\documentclass{article}
\usepackage{amssymb,graphicx,cancel}
\usepackage[margin=2cm]{geometry}
\usepackage{slashed}
\usepackage{hepunits}
\usepackage{color}
\usepackage[table]{xcolor}
\usepackage{graphicx}
\usepackage{multirow}
\usepackage{enumitem}
\usepackage{pgf}
\usepackage{pgfplots}
\usepackage{url}
\usepackage{bm}
\usepackage{gensymb}

 \usepackage{jheppub}

\usepackage{color}

\preprint{KCL-2020-27, FTUV-20-0625.4735, IFIC/20-33}
\title{Relaxing Cosmological Neutrino Mass Bounds with Unstable Neutrinos}

\author[a]{Miguel Escudero,${}^1$\note{ORCID: \href{http://orcid.org/0000-0002-4487-8742}{0000-0002-4487-8742}}}
\author[b]{Jacobo Lopez-Pavon,${}^2$\note{ORCID: \href{https://orcid.org/0000-0002-9554-5075}{0000-0002-9554-5075}}}
\author[b]{Nuria Rius,${}^3$\note{ORCID: \href{http://orcid.org/0000-0002-0606-4297}{0000-0002-0606-4297}}}
\author[b]{and Stefan Sandner,${}^4$\note{ORCID: \href{http://orcid.org/0000-0002-1802-9018}{0000-0002-1802-9018}}}

\affiliation[a]{Theoretical Particle Physics and Cosmology Group, Department of Physics,  \\
King's College London, Strand, London WC2R 2LS, UK}

\affiliation[b]{Departamento de F\'isica Te\'orica and IFIC, Universidad de Valencia-CSIC\\
C/ Catedr\'atico Jos\'e Beltr\'an, 2, E-46980 Paterna, Spain}

\abstract{At present, cosmological observations set the most stringent bound on the neutrino mass scale. Within the standard cosmological model ($\Lambda$CDM), the Planck collaboration reports $\sum m_\nu < 0.12\,\text{eV}$ at 95 \% CL. This bound, taken at face value, excludes many neutrino mass models. However, unstable neutrinos, with lifetimes shorter than the age of the universe $\tau_\nu \lesssim t_U$, represent a particle physics avenue to relax this constraint. 
Motivated by this fact, we present a taxonomy of neutrino decay modes, categorizing them in terms of particle content and final decay products. Taking into account the relevant phenomenological bounds, our analysis shows that 2-body decaying neutrinos into BSM particles are a promising option to relax cosmological neutrino mass bounds.
We then build a simple extension of the type I seesaw scenario by adding one sterile state $\nu_4$ and 
a Goldstone boson $\phi$, in which $\nu_i \to \nu_4 \, \phi$ decays can loosen the neutrino mass bounds up to $\sum m_\nu \sim 1\,\text{eV}$, without spoiling the light neutrino mass generation mechanism. Remarkably, this is possible for a large range of the right-handed neutrino masses, from the electroweak up to the GUT scale. We successfully implement this idea in the context of minimal neutrino mass models based on a $U(1)_{\mu-\tau}$ flavor symmetry, which are otherwise in tension with the current bound on 
$\sum m_\nu$.

}

\emailAdd{miguel.escudero@kcl.ac.uk}
\emailAdd{jlpavon@ific.uv.es}
\emailAdd{nuria.rius@ific.uv.es}
\emailAdd{stefan.sandner@ific.uv.es}


\begin{document}
\maketitle
\flushbottom

\vspace{-0.9cm}
\section{Introduction}
\label{sec:intro}
\vspace{-0.15cm}

Neutrino oscillation experiments show that neutrinos are massive and there is flavor 
mixing not only in the hadronic sector but also in the leptonic sector. At present, three out of the four parameters of the three family leptonic mixing matrix, $U$, are experimentally determined ($\theta_{12}$, $\theta_{23}$ and $\theta_{13}$) and there is a $\sim 3\,\sigma$ hint pointing to a CP-violating value of the Dirac CP phase $\delta$~\cite{Abe:2019vii,Acero:2019ksn}. Regarding neutrino masses, neutrino oscillation measurements can provide information only about the neutrino mass squared differences $\Delta m_{ij}^2 = m_{\nu_i}^2 -m_{\nu_j}^2$. In particular, we currently know that $\sqrt{\Delta m_{21}^2} \simeq 0.0086 \,\text{eV}$ and $|\sqrt{\Delta m_{32}^2}| \simeq |\sqrt{\Delta m_{31}^2}| \simeq 0.05\,\text{eV}$~\cite{Esteban:2018azc,deSalas:2020pgw,Capozzi:2020qhw}. Thus, concerning light neutrino masses, there are still two open fundamental questions: \textit{i)} what is the neutrino mass ordering~\cite{deSalas:2018bym}, i.e. $m_{\nu_3} > m_{\nu_1}$ (Normal Ordering, NO) or $m_{\nu_3} < m_{\nu_1}$ (Inverted Ordering, IO), and \textit{ii)} what is the absolute neutrino mass scale or, equivalently, $\sum m_\nu \equiv m_{\nu_1}+m_{\nu_2}+m_{\nu_3}$. Near future neutrino oscillation facilities will be able to definitely answer \textit{i)} but question \textit{ii)} can not be addressed in this type of experiments.

The best laboratory constraint on the absolute neutrino mass scale comes from the KATRIN experiment that reports the following Feldman-Cousins upper limit: $m_{\bar{\nu}_e}\equiv\sqrt{\sum_{i} |U_{ei}|^2 m_{\nu_i}^2} < 0.9\,\text{eV}$ at 95\% CL~\cite{Aker:2019uuj}. Using the information from neutrino oscillation data on $U$, this bound can be translated into the subsequent upper limit on the sum of the light neutrino masses~\cite{Tanabashi:2018oca}: $\sum m_{\nu} \lesssim 2.7\,\text{eV}$ at 95\% CL.
This should be compared with the substantially stronger constraints arising from Cosmology~\cite{Lesgourgues:2006nd,Wong:2011ip,Lattanzi:2017ubx}. In particular, by using Cosmic Microwave Background (CMB) measurements in conjunction with Baryon Acoustic Oscillations (BAO) data, and within the framework of the standard cosmological model $\Lambda$CDM, the Planck collaboration reports~\cite{Aghanim:2018eyx}: $\sum m_\nu < 0.12\,\text{eV}$ at 95\% CL. Thus, at face value, the cosmological bound from Planck is more than one order of magnitude stronger than the one reported by KATRIN.
From the experimental viewpoint, the CMB bound is very robust since systematic effects in Planck CMB measurements have been shown to be very small~\cite{Aghanim:2018eyx,Akrami:2018vks,Aghanim:2019ame,Efstathiou:2019mdh} and because, given a cosmological model, the theoretical error in predicting CMB spectra is negligible~\cite{Ma:1995ey,Lewis:1999bs,Howlett:2012mh,Lesgourgues:2011re,Blas:2011rf}. 
However, all cosmological bounds on $\sum m_\nu$ are cosmological model dependent and $\sum m_\nu < 0.12\,\text{eV}$ (at 95\% CL) only applies if $\Lambda$CDM is the actual model describing our Universe.
 Observationally, $\Lambda$CDM is an extremely successful cosmological model~\cite{Aghanim:2018eyx} but, of course, when comparing cosmological and laboratory bounds on $\sum m_\nu$, one wonders to what extent the cosmological ones depend upon the data set and cosmological model under consideration. In order for the reader to have a feeling of this matter, we summarize in Table~\ref{tab:mnu_constraints} a suite of cosmological constraints on $\sum m_\nu$ arising from analyzing various data sets, and by using the same data set but within different cosmological models. From Table~\ref{tab:mnu_constraints} we can draw some important conclusions:
\vspace{-0.1cm}
\begin{enumerate}
\item Within the framework of $\Lambda$CDM current cosmological bounds on $\sum m_\nu$ are driven by Planck. This is relevant since, as argued before, Planck constraints are very robust.
\item Cosmological bounds on $\sum m_\nu$ are not substantially altered in standard extensions of $\Lambda$CDM. \\
In particular, when the modifications of $\Lambda$CDM entail non-standard dark energy, non-standard inflationary perturbation spectra, curvature, extra dark radiation, or some of these together.
\item Neutrino mass bounds \textit{can} considerably be alleviated if neutrinos possess non-standard properties such as a time dependent $m_\nu$ or if neutrinos decay on cosmological timescales.
\end{enumerate}
\vspace{-0.1cm}

In particular, the analysis performed in~\cite{Chacko:2019nej} which includes Planck+BAO+SNIa+SDSS large scale structure (LSS) data, yields $\sum m_\nu < 0.9\,\text{eV}$ at 95\% CL, provided that neutrinos decay invisibly to massless BSM states with lifetimes in the range $10^{-4}\lesssim \tau_\nu/t_U \lesssim 0.1$\footnote{Note, however, that CMB observations also bound neutrinos to have a lifetime~\cite{Escudero:2019gfk}: $\tau_\nu > 1.3\times 10^9\,\text{s} \,(m_\nu/0.05\,\text{eV})^3\simeq 10^{-9}\,t_U\,(m_\nu/0.05\,\text{eV})^3$ at 95\% CL. This constraint is discussed in detail in Section~\ref{sec:parameterpsace}. }, where $t_U = 13.8\,\text{Gyr} = 4.35 \times 10^{17}\,\text{s}$~\cite{Aghanim:2018eyx} is the age of the Universe. 
This bound is to be compared with the one obtained within the $\Lambda$CDM model in the stable neutrino framework, $\sum m_\nu < 0.12\,\text{eV}$ at 95\% CL~\cite{Aghanim:2018eyx}. This remarkable relaxation has important implications for neutrino mass models since many of them predict neutrinos with masses $\sum m_{\nu} \gtrsim 0.12\,\text{eV}$. For instance, most of the two-zero neutrino mass textures, and therefore the flavor models realizing these textures, lead to $\sum m_\nu > 0.12\,\text{eV}$~\cite{Lavoura:2004tu,Verma:2011kz,Alcaide:2018vni}. Along these lines, we refer to~\cite{Lattanzi:2020iik} for a very recent discussion of the  implications of cosmological neutrino mass bounds for models leading to quasi-degenerate neutrinos.

 \begin{table*}
\begin{center}
{\def\arraystretch{1.2}
\begin{tabular}{c|l|cc|l}
\hline\hline
\multicolumn{5}{c}{Cosmological and Laboratory bounds on $\sum m_\nu/\text{eV}$ and $m_\nu^{\rm lightest}/\text{meV}$ at 95\% CL}   \\ \hline\hline
\textbf{Scenario}  &Ref and Scenario/Data Set & $ \sum m_\nu $ & $m_{\nu}^{\rm lightest}$   & Comment \\ \hline
  \multirow{7}{*}{ \rotatebox[origin=c]{90}{$\Lambda$CDM }   $ \,$  \rotatebox[origin=c]{90}{  \texttt{Various Data Sets}}}  $ \,$
& \cite{Aghanim:2018eyx}~Planck		& 0.24   & 75/69    & Linear Cosmology \\
&\cite{Aghanim:2018eyx}~Planck+BAO 		& 0.12   &  30/16     & Linear Cosmology \\
&\cite{Ivanov:2019pdj}~BOSS $P(k)$ 		& 0.86   & 280    &Mildly non-Linear Cosmology \\
&\cite{Ivanov:2019hqk}~Planck+BOSS $P(k)$ & 0.16   & 46/37    & -\\
&\cite{Palanque-Delabrouille:2019iyz}~Lyman-$\alpha$+$H_0$ prior & 0.71   &  230   & Non-Linear Cosmology \\
&\cite{Palanque-Delabrouille:2019iyz}~Planck+Lyman-$\alpha$  & 0.10   & 22/0    & -\\
&\cite{RoyChoudhury:2019hls}~Planck+BAO+$H_0$ & 0.08   & 12/*    & Combines data sets in tension \\
  \hline

  \multirow{6}{*}{ \rotatebox[origin=c]{90}{$\Lambda$CDM Extensions}   $ \,$  \rotatebox[origin=c]{90}{  \texttt{Planck+BAO}}}  $ \,$
&\cite{Aghanim:2018eyx}~$\Lambda$CDM 			& \textbf{0.12}   	& 30/16   & Standard Cosmological Model\\
&\cite{RoyChoudhury:2019hls}~CDM+$\omega_0$+$\omega_a$& 0.25   	& 78/73     & Dynamical Dark Energy  \\
&\cite{RoyChoudhury:2019hls}~$\Lambda$CDM+$\Omega_k$& 0.15   	& 42/32    & Including Curvature   \\
&\cite{Aghanim:2018eyx}~$\Lambda$CDM+$N_{\rm eff}$ & 0.23  	 & 71/66     & Including Dark Radiation \\
&\cite{DiValentino:2019dzu}~CDM+$N_{\rm eff}$+$\omega$+$\alpha_s$ & 0.17   & 49/41 & Including some of the above  \\
&\cite{Yang:2020tax}~CDM+$N_{\rm eff}$+DM-DE int & 0.19   & 57/50 & Including DM-DE interactions  \\
  \hline
   \multirow{4}{*}{ \rotatebox[origin=c]{90}{ Exotic     }   $ \,$  \rotatebox[origin=c]{90}{ Neutrinos }}  $ \,$
&\cite{Lorenz:2018fzb}~$\Lambda$CDM+$m_\nu(t)$	 	&\multirow{2}{*}{4.8}   	& \multirow{2}{*}{1600}    & Time dependent $m_\nu$ \\
& Data: Planck+BAO+SNIa	&    	&    & from a phase transition \\
&\cite{Chacko:2019nej}~$\Lambda$CDM+Decaying Neutrinos	&\multirow{2}{*}{\textbf{0.9}}   	& \multirow{2}{*}{300}    & Invisible Neutrino Decays \\
& Data: Planck+BAO+SNIa+LSS	&    	&    & only if $10^{-4}\lesssim \tau_\nu/t_U \lesssim 0.1$  \\
  \hline
\texttt{KATRIN} &\cite{Aker:2019uuj}~Tritium $\beta$ decay spectrum 	&\textbf{2.7}   	&   900 & Absolute neutrino mass bound \\

  \hline \hline
\end{tabular}
}
\end{center}\vspace{-0.5cm}
\caption{Bounds on $\sum m_\nu/\text{eV}$ and $m_{\nu}^{\rm lightest}/\text{meV}$ at 95\% CL from various cosmological data sets and within various cosmological models. In the $m_{\nu}^{\rm lightest}$ column, the value quoted on the left (right) corresponds to NO (IO). The cosmological bounds listed here assume 
that neutrinos are degenerate, but relaxing this assumption has a negligible impact for these bounds, see e.g.~\cite{RoyChoudhury:2019hls,Vagnozzi:2017ovm}. In the last row we show the constraint on $\sum m_\nu$ from the KATRIN experiment, obtained from precision measurements of the electron spectrum from Tritium $\beta$ decay~\cite{Aker:2019uuj}.  *IO excluded at 95\% CL. 
 }\label{tab:mnu_constraints}
\end{table*}

Moreover, upcoming galaxy surveys such as \texttt{DESI}~\cite{Aghamousa:2016zmz} and \texttt{Euclid}~\cite{Amendola:2016saw} will have a $1\sigma$-sensitivity of $\sigma (\sum m_\nu) \simeq 0.02\,\text{eV}$ and are expected to detect neutrino masses within the next $\sim$5-10 years if the cosmological model describing our Universe is $\Lambda$CDM, see e.g.~\cite{Hu:1997mj,Font-Ribera:2013rwa,Allison:2015qca,Brinckmann:2018owf,Xu:2020fyg}. A neutrino mass detection by \texttt{DESI}/\texttt{Euclid} would provide us with extremely valuable information about the neutrino lifetime~\cite{Serpico:2007pt,Chacko:2020hmh}, implying $\tau_\nu \gtrsim t_U$ and, thus, excluding cosmologically relevant neutrino decay scenarios. 
Perhaps even more interestingly, if neutrino masses are not detected by \texttt{DESI}/\texttt{Euclid} and only upper bounds on $\sum m_\nu$ are reported, neutrino decays will become a prime scenario to explain the non-detection of neutrino masses in cosmological observations. Notice that this would have also a crucial impact in the context of neutrinoless double beta decay experiments,
since the theoretical prediction for the decay rate strongly depends on both the absolute neutrino mass scale and the neutrino ordering (for two recent reviews on the topic see for instance~\cite{DellOro:2016tmg,Dolinski:2019nrj}).

In this work, motivated by the potential relaxation of cosmological neutrino mass bounds, we set up a comprehensive study of the models that can lead to neutrino decays with lifetimes $\tau_\nu \lesssim  t_U$.
We focus on invisible decay channels because the ones with final states that can interact electromagnetically are bounded to have rates $\tau_\nu > (10^2-10^{10}) \,t_U $~\cite{Fujikawa:1980yx,Beda:2013mta,Borexino:2017fbd,Mirizzi:2007jd,Aalberts:2018obr,Raffelt:1990pj,ARCEODIAZ20151,Raffelt:1999gv}. 
 The possibility of neutrino decays was considered already in the 70's~\cite{Bahcall:1972my}, and in fact, two neutrino mass eigenstates decay within the Standard Model~\cite{Petcov:1976ff,Hosotani:1981mq,Pal:1981rm} albeit at a very slow rate, $\tau_\nu^{\rm SM} > (G_F^2 m_\nu^5)^{-1} \sim 10^{23}\,\text{yr}\,(\text{eV}/m_\nu)^5 \gg t_U$. We note that several invisible neutrino decay models have been put forward, see e.g.~\cite{Schechter:1981cv,Wilczek:1982rv,Valle:1983ua,Gelmini:1983ea,Joshipura:1992vn,Akhmedov:1995wd}. However, to our knowledge, the majority of theoretical investigations of invisible neutrino decays in cosmology were developed prior to the discovery of neutrino oscillations. Therefore, we find timely to develop new models for invisible neutrino decay and to perform an exhaustive identification of the required particle content, coupling strengths, and mass scales of the particles needed to trigger invisible neutrino decays at a rate $\tau_\nu \lesssim t_U$.
 
\vspace{0.4cm}
 
The structure of this work is the following: In Section~\ref{sec:Taxonomy}, at the phenomenological level, we categorize the different decay modes of invisible neutrino decays. In particular, in terms of 2-body and 3-body neutrino decay final states as mediated by neutrinophilic scalars and vector bosons. In Section~\ref{sec:Results}, we identify the relevant neutrino decay channels and lifetimes that can potentially lead to the alleviation of the $\sum m_\nu$ bounds and briefly review the most stringent constraints on invisible neutrino decays. We then map this information to the properties of the particles involved in the decay, identifying the region of the coupling and mass parameter space compatible with $\tau_\nu \lesssim t_U$ and all the relevant present constraints. In Section~\ref{sec:ModelBuilding}, we propose a simple extension of the seesaw mechanism in which neutrinos naturally can decay loosening the neutrino mass bounds up to $\sum m_\nu \sim 1\,\text{eV}$,  without spoiling the light neutrino mass generation mechanism. Then, we embed this extension within a minimal model based on a spontaneously broken $U(1)_{\mu-\tau}$ flavor symmetry~\cite{Choubey:2004hn,Araki:2012ip}, which predicts  $\sum m_\nu > 0.12\,\text{eV}$. Finally, in Section~\ref{sec:Summary} we summarize the main results obtained in this study and draw our conclusions.

\vspace{-0.2cm}
\section{Taxonomy of Invisible Neutrino Decays}
\label{sec:Taxonomy}

\vspace{-0.2cm}
In this section we categorize the various possibilities for invisible neutrino decays, namely, those in which the decay products do not interact electromagnetically. We then write the most general effective Lagrangians parametrizing them. 

\subsection{Categories}
\label{subsec:category}

We classify invisible neutrino decays according to two criteria: \textit{i)} nature of the decay products and \textit{ii)} number of particles in the final state. The categorization for each case goes as follows:

\begin{enumerate}[leftmargin=0.5cm,itemsep=0.4pt]
\item[\textit{i)}]  \textit{Nature of the decay products}
\begin{enumerate}[]

\item[ \textit{a)}] At least another active neutrino mass eigenstate $\nu_j$.  \\
Since energy and angular momentum conservation ensure the lightest neutrino state to be stable, the cosmological constraint on $\sum m_\nu$ cannot be relaxed by more than $ 0.06\,\text{eV}$ and $0.1\,\text{eV}$ for NO and IO, respectively. We refer the reader to Section~\ref{sec:Results} for more details.

\item[\textit{s)}] 
Only BSM species.

In this case, the constraint on $\sum m_\nu$ can be significantly relaxed to the level of $\sum m_\nu \lesssim 1\,\text{eV}$ at 95\% CL, provided that the BSM particles are massless, see Section~\ref{sec:Results}.
\end{enumerate}

\item[\textit{ii)}] \textit{Number of particles in the final state}
\begin{enumerate}[]
\item[\textit{2)}] 2-body decays.

Angular momentum conservation requires the decay products to be one fermion and one boson with spin~$0$ (scalar $\phi$) or~$1$ (vector $Z'$). Regarding the fermion, we will consider two possibilities: a light neutrino mass eigenstate $\nu_i$ or a sterile neutrino $\nu_4$.

\item[\textit{3)}] 3-body decays.

For sufficiently massive bosons, i.e. $m_{\phi,\,Z'} > m_{\nu_i}$, any 2-body decay is kinematically closed. However, such a boson can mediate off-shell a 3-body decay, which becomes the dominant channel.  

\item[\textit{4)}] 4-body decays and beyond.

We will not consider this possibility here since, as we will show in Section~\ref{sec:results}, already the 3-body decay channels are only capable of rendering $\tau_\nu < t_U$ across a narrow window of the parameter space.
\end{enumerate}
\end{enumerate}

\begin{figure}[t]
\centering
\includegraphics[width=1\textwidth]{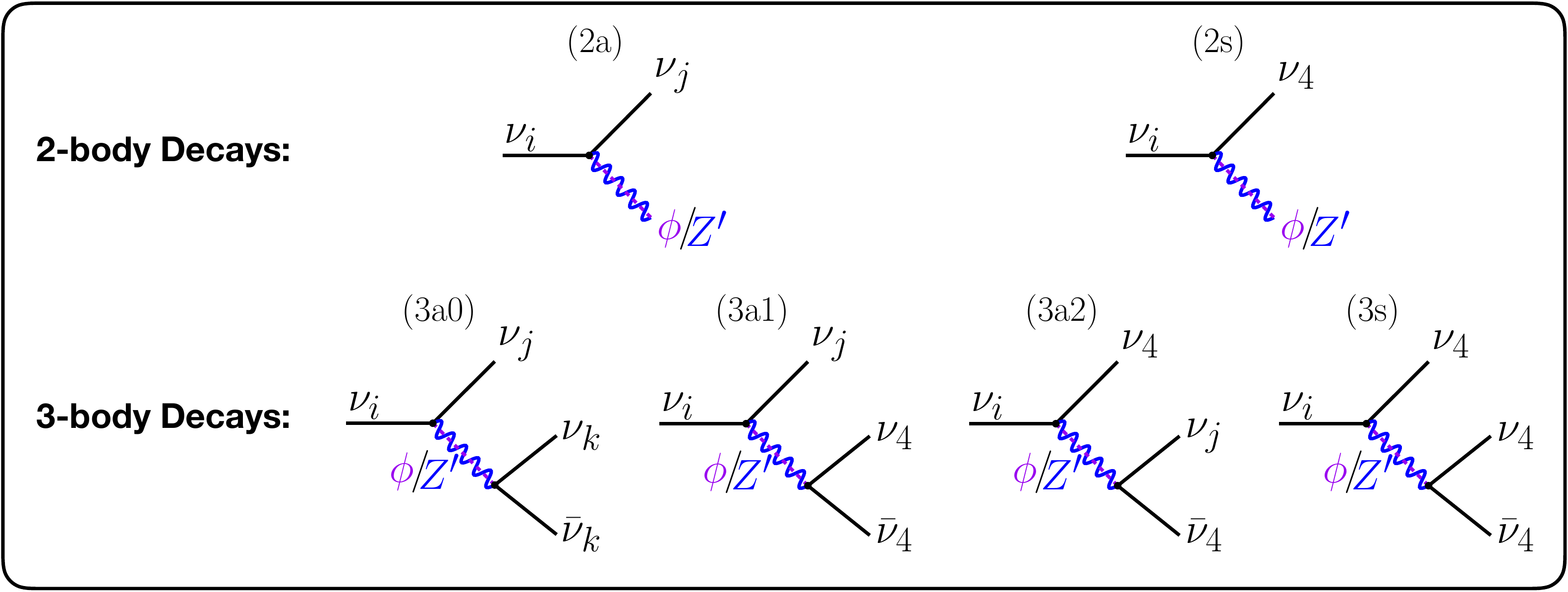}
\vspace{-0.7cm}
\caption{Possible diagrams describing invisible neutrino decays of the mainly active neutrino mass eigenstate $\nu_i$. Here the indices $i,\,j,\,k$ take values from $1$ to $3$. $\nu_4$ refers to a mainly sterile mass eigenstate with small mixing with the active neutrinos $\nu_\alpha$. $\phi$ and $Z'$ are neutrinophilic scalars and vector bosons, respectively.}\label{fig:NeutrinoDecayDiagrams}
\end{figure}

Therefore, according to this classification, there are six different possible topologies for the neutrino decays that we show in Figure~\ref{fig:NeutrinoDecayDiagrams}. The diagrams are labelled by their number of final state particles $N = 2,\,3$ and by whether there are active neutrinos in the final state \textit{a} or only sterile species \textit{s}.

\subsection{Effective Lagrangians}\label{sec:effLag}

\begin{table}[t]
\centering
{\def\arraystretch{1.1}
\begin{tabular}{l|l|c|l}
 \hline\hline
  \multicolumn{4}{c}{\textbf{Invisible Neutrino Decay Rates}} \\ \hline\hline
& \textbf{Decay Channel} & $\!\!\!\!\!$ \textbf{Case} $\!\!\!\!\!$  & \textbf{Decay Rate} \\  \hline\hline
  \multirow{4}{*}{ \rotatebox[origin=c]{90}{\textbf{2-Body Decays} $\qquad$}   }
& $ \nu_{i} \to \nu_{j}\,\phi$  & \textit{2a} &  $  \begin{aligned} \Gamma &= \lambda_{ij}^2/(16\pi) \,   (m_{\nu_i}-m_{\nu_j})^3(m_{\nu_i}+m_{\nu_j})/m_{\nu_i}^3 \\ &\simeq   t_{U}^{-1} \left(\frac{m_{\nu_i}}{0.05~\text{eV}}\right) \left(\frac{\lambda_{ij}}{1.2\times 10^{-15}}\right)^2  ~\text{for}~m_{\nu_j} \ll m_{\nu_i} \end{aligned}$ \\ 
\cline{2-4} 
& $ \nu_{i} \to \nu_{4}\,\phi$ & \textit{2s} &  $  \begin{aligned} \Gamma &= \frac{\lambda_{i4}^2}{16\pi} \, m_{\nu_i} \simeq  t_{U}^{-1} \left(\frac{m_{\nu_i}}{0.3~\text{eV}}\right) \left(\frac{\lambda_{i4}}{5\times 10^{-16}}\right)^2  \end{aligned}  $\\
\cline{2-4} 
& $ \nu_{i} \to \nu_{j} \, Z'$ & \textit{2a} &  $\begin{aligned} \!\!\Gamma &\simeq g^{L}_{ij} {}^2/(32\pi)  \, (m_{\nu_i}^2-m_{\nu_j}^2)^3/(m_{Z'}^2m_{\nu_i}^3)  \\ &\simeq t_{U}^{-1}\left(\frac{m_{\nu_i}}{0.05~\text{eV}} \! \right)^3 \! \left( \frac{29\, \text{TeV}}{m_{Z'}/g_{ij}^L}\right)^2  ~\text{for}~m_{\nu_j} \! \ll \! m_{\nu_i}\end{aligned} $   \\

 \cline{2-4} 
&$ \nu_{i} \to \nu_{4}\, Z'$ & \textit{2s} &  $  \begin{aligned} \Gamma &\simeq \frac{g^{L}_{i4} {}^2}{16\pi} \, \frac{m_{\nu_i}^3}{m_{Z'}^2} \simeq t_{U}^{-1}\left(\frac{m_{\nu_i}}{0.3~\text{eV}}\right)^3  \left( \frac{400 \,\text{TeV}}{m_{Z'}/g_{ij}^L}\right)^2  \!\!\!\!\!\!\!\!\!  \end{aligned}$\\ 
 \hline\hline
  \multirow{4}{*}{ \rotatebox[origin=c]{90}{\textbf{3-Body Decays} $\,\,\,\,\,\,\,\,\,$}}   
& $ \nu_{i} \to \nu_{j}\phi^{*} \to \nu_{j}\nu_{4}\nu_{4}$ & \textit{3a1} &  $  \begin{aligned}  \Gamma & \simeq  t_{U}^{-1}  \left(\frac{m_{\nu_ i}}{0.05~\text{eV}}\right)^5 \! \left(\frac{\lambda_{44}}{4\pi}\right)^2 \!\left(\frac{\lambda_{ij}}{10^{-12}}\right)^2 \! \left(\frac{1~\text{eV}}{m_\phi}\right)^4 \text{for}~m_{\nu_j} \! \ll \! m_{\nu_i} \!\!\!\!\!\!\! \end{aligned} $\\ 
 \cline{2-4}
& $ \nu_{i} \to \nu_{4}\phi^{*} \to \nu_{4}\nu_{4}\nu_{4}$ & \textit{3s} &  $  \begin{aligned}   \Gamma  &\simeq  t_{U}^{-1}  \left(\frac{m_{\nu_i}}{0.3~\text{eV}}\right)^5 \left(\frac{\lambda_{44}}{4\pi}\right)^2 \left(\frac{\lambda_{i4}}{10^{-6}}\right)^2 \left(\frac{10~\text{keV}}{m_\phi}\right)^4  \end{aligned} $\\
 \cline{2-4}
& $ \nu_{i} \to \nu_{j}Z'^{*} \to \nu_{j}\nu_{4}\nu_{4}$ & \textit{3a1}&  $ \begin{aligned}  \Gamma   &\simeq t_{U}^{-1}  \left(\frac{m_{\nu_i}}{0.05~\text{eV}}\right)^5  \! \left(\frac{(|g_{44}^{L}|^2+|g_{44}^{R}|^2)^{1/2}}{4\pi}\right)^2 \! \left( \frac{g_{ij}^{L}}{10^{-12}}\right)^2 \! \left(\frac{1~\text{eV}}{m_{Z'}}\right)^4  \!\!\!\!\!\!\!     \end{aligned} $\\
 \cline{2-4} 
& $ \nu_{i} \to \nu_{4}Z'^{*} \to \nu_{4}\nu_{4}\nu_{4}$ & \textit{3s} &  $  \begin{aligned} \Gamma  &\simeq  t_{U}^{-1}   \left(\frac{m_{\nu_i}}{0.3~\text{eV}}\right)^5 \!   \left(\frac{(|g_{44}^{L}|^2+|g_{44}^{R}|^2)^{1/2}}{4\pi}\right)^2 \!  \left( \frac{g_{i4}^{L}}{10^{-6}}\right)^2 \! \left(\frac{8~\text{keV}}{m_{Z'}}\right)^4 \!\!\!\!\!\!\! \end{aligned}$\\
\hline\hline

\end{tabular}
}\vspace{-0.4cm}
\caption{Rates for invisible neutrino decays. For neutrinos coupled to a scalar we have considered pseudo-scalar couplings only for concreteness. We have simplified the 3-body decay rates by integrating out the mediating boson and $\phi/Z'^*$ means off-shell mediators. See Appendix~\ref{sec:appendix_rates} for detailed formulae. The reference numbers for each decay are chosen so as to match relevant scales/parameters, see Section~\ref{sec:results}. }\label{tab:rates}
\end{table}

Now that we have clarified the particles that can participate in the decays, in this section we will introduce the most general effective Lagrangians that parametrize their interactions. The details about the computation of the decay rates required for the phenomenological analysis are given in Appendix~\ref{sec:appendix_rates}. Approximated expressions of the decay rates for all the channels under consideration in the relevant limits are presented in Table~\ref{tab:rates}. 

On the one hand, the most general renormalizable Lagrangian that describes the interactions between the light neutrinos, a sterile state, and a scalar field $\phi$ in the neutrino mass basis is given by:
\begin{align}\label{eq:Lag_scalar}
\mathcal{L}^{\phi} & \supset - \frac{\phi}{2} \, \left[ \overline{\nu_i} \left(h_{ij} + i \lambda_{ij} \gamma_5  \right) \nu_j   + \overline{\nu_i} \left(h_{i4} + i \lambda_{i4} \gamma_5  \right) \nu_4 + \overline{\nu_4} \left(h_{4i} + i \lambda_{4i} \gamma_5  \right) \nu_i +\overline{\nu_4} \left(h_{44} + i \lambda_{44} \gamma_5  \right) \nu_4 \right] +  \text{h.c.}\,,
\end{align}
where $\nu_i$ and $\nu_4$ are the mainly active and mainly sterile neutrino mass eigenstates, respectively. Here $h$ and $\lambda$ parametrize scalar and pseudo-scalar Yukawa couplings between active/sterile neutrinos and $\phi$. For the sake of concreteness, we shall assume that neutrinos are Majorana particles. The formulae for the decay rates shown in Appendix~\ref{sec:appendix_rates} and Table~\ref{tab:rates} can be applied to the Dirac neutrino case doing the following mapping: $(h,\,\lambda)_{ij}^{\rm Dirac} \leftrightarrow 2 (h,\,\lambda)_{ij}^{\rm Majorana}$. 

On the other hand, the interaction Lagrangian parametrizing active/sterile neutrino interactions with a $Z'$ is:
\begin{align}\label{eq:Lag_vector}
\mathcal{L}^{Z'} & \supset  -\frac{Z'_\mu}{2} \left[ g_{ij}^{L}\, \bar{\nu}_i \gamma^\mu P_L  \nu_j+ g_{i4}^{L} \,\bar{\nu}_i  \gamma^\mu P_L \nu_4 + g_{4i}^{L} \,\bar{\nu}_4  \gamma^\mu P_L \nu_i+ \bar{\nu}_4   \gamma^\mu\left(g^L_{44} P_L +g^R_{44} P_R   \right) \nu_4 \right]+ \text{h.c.}\,, 
\end{align}
where $P_{L(R)}$ is the left (right) handed projection operator and $g^{L(R)}$ represent left (right) handed couplings. 

Notice that, in order to translate Equations~(\ref{eq:Lag_scalar}) and (\ref{eq:Lag_vector}) to the neutrino flavor basis, active ($\nu_\alpha$) and sterile ($\nu_s$), we just need to perform the correspondent rotation: $\nu_\alpha=U_{\alpha i}\,\nu_i+\theta_{\alpha 4}\,\nu_4$, $\nu_s=\theta_{s i}\,\nu_i+\theta_{s 4}\,\nu_4$. Since the mixing between the sterile and active neutrinos $\theta_{\alpha 4}$ and $\theta_{s i}$ should be small, $U_{\alpha i}$ with $\alpha=e,\mu,\tau$ and $i=1,2,3$ is given by the PMNS matrix, up to subleading corrections driven by the active-sterile neutrino mixing.

Given the phenomenological approach of this section, we will assume that the coupling constants present in the effective Lagrangians introduced in Equations~(\ref{eq:Lag_scalar}) and (\ref{eq:Lag_vector}) are independent parameters. However, in UV complete models some couplings may be absent and, if all are present, they are expected to present correlations among them. In Section~\ref{sec:ModelBuilding} we will map the effective Lagrangians presented here into UV complete models able to realize invisible neutrino decays in the region of the parameter space that can relax cosmological bounds on $\sum m_\nu$. In any case, from the UV perspective, one would typically expect the following possible relationships between $\lambda_{ij}: \lambda_{i4} : \lambda_{44}$~\footnote{Similarly for $h_{ij}: h_{i4} : h_{44}$, and $g_{ij}^L: g_{i4}^L : g_{44}^L$.}: 
\begin{itemize}
\item $\lambda_{i4} = 0,\,\lambda_{44} = 0$: Scenario with no light sterile neutrinos. This is the simplest possibility and we will consider neutrino mass models realizing invisible neutrino decay of this sort in Section~\ref{sec:Minimalmutau}.  
\item $(\lambda_{ij}: \lambda_{i4} : \lambda_{44})  \simeq (1:1:1)$: Scenarios in which both the SM lepton doublets and the sterile neutrino are charged under a gauge lepton flavor symmetry. This case is expected to generically suffer from stringent constraints since the bosons mediating this interaction will couple to both charged leptons and neutrinos with couplings of the same order.  
\item $(\lambda_{ij}: \lambda_{i4} : \lambda_{44})  \simeq (1:\theta:\theta^2)$: Only active neutrinos interacting directly with a new force carrier and $\nu_4-\phi/Z'$ interaction arising via active-sterile neutrino mixing $\theta$. 
\item $(\lambda_{ij}: \lambda_{i4} : \lambda_{44})  \simeq (\theta^2:\theta:1)$: Sterile neutrinos interacting directly with a new force carrier while active neutrinos interact only via the mixing $\theta$.
\end{itemize}

This should not be considered as a complete list of all possible scenarios but rather as a sample of typical cases. Indeed, in Section~\ref{sec:Guidelines} we will propose a neutrino decay model with $\lambda_{i4} \gg \lambda_{ij}$ and $\lambda_{44}=0$, that does not belong to any of the above items in the list.

\section{Relaxing Neutrino Mass Bounds with Invisible Neutrino Decays}
\label{sec:Results}

The aim of this section is twofold. Firstly, in~\ref{sec:parameterpsace}, we discuss the present bounds on invisible neutrino decays and their impact on the extraction of cosmological neutrino mass bounds. We also show the viable region of the $\tau_\nu-\sum m_\nu$ parameter space where the cosmological constraint on $\sum m_\nu$ can be relaxed. Secondly, in~\ref{sec:results}, we highlight the range of values for the coupling constants and masses involved in the possible invisible decays (classified in Figure~\ref{fig:NeutrinoDecayDiagrams}) in which this goal can be achieved.

\subsection{Effects on cosmological neutrino mass bounds}\label{sec:parameterpsace}

\begin{figure}[t]
\hspace{-0.5cm}  \includegraphics[width=1.05\textwidth]{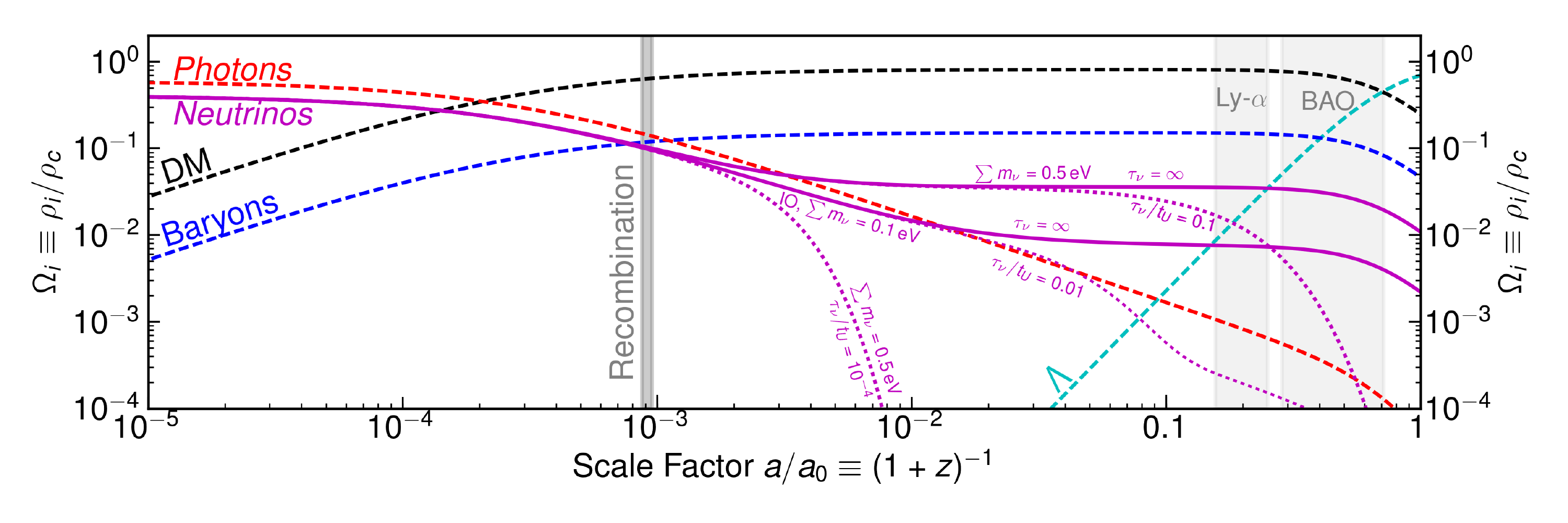}  
 \vspace{-1cm}
\caption{Energy density evolution for several neutrino decay scenarios. The period in which recombination takes place and the ones relevant for the Lyman-$\alpha$ forest and Baryon Acoustic Oscillations data are indicated by the grey regions. We show the contribution to the energy density from photons (red), neutrinos (magenta), dark matter (black), baryons (blue), and dark energy (cyan). Note that for the sake of simplicity we do not show the energy density of the neutrino decay products. 
}\label{fig:nuiDecaysCosmo}
\end{figure}

In order to understand how invisible neutrino decays can potentially relax cosmological neutrino mass bounds, we will first briefly review the thermal history of neutrinos in the Standard Model and how it is modified if neutrinos decay. This is illustrated in Figure~\ref{fig:nuiDecaysCosmo} where we show the evolution of neutrino energy density in the Standard Model framework and within several invisible neutrino decay scenarios. The energy density evolution has been computed by solving the relevant Boltzmann equations as in~\cite{Chacko:2019nej}. 
Note that, for simplicity, in Figure~\ref{fig:nuiDecaysCosmo} we only show the contribution from the neutrino energy density.
However, this does not change our conclusions since even though the contribution from the neutrino decay products is not negligible for $t\sim \tau_\nu$, it is diluted as $(1+z)^4$. For example, for $\sum m_\nu = 0.5\,\text{eV}$ and $\tau_\nu = 10^{-4}\,t_U$ the energy density of the decay products today is similar to that of the photons and represents only 0.5\% of the energy density that stable neutrinos with $\sum m_\nu =0.5\,\text{eV}$ would encode. The energy density of the decay products becomes relevant for lifetimes that are somewhat similar to $t_U$ and for $\sum m_\nu \gtrsim 0.2\,\text{eV}$. However, as we discuss below, such regions of parameter space are excluded by current cosmological observations.

Within the standard cosmological picture, neutrinos decouple from the SM plasma at $T\sim 2\,\text{MeV}$ ($t\sim 0.2\,\text{s}$) and ever since they simply free stream without interacting with any species in the Universe. For temperatures $T\ll m_e$, after electrons and positrons annihilate, the ratio between the neutrino and photon energy densities is fixed and keeps being fixed as long as neutrinos are relativistic, as it can be seen in Figure~\ref{fig:nuiDecaysCosmo}. In particular, $T_\gamma/T_\nu \simeq  1.4$, which corresponds to $N_{\rm eff}^{\rm SM} \equiv 8/7 \,(11/4)^{4/3}\,\rho_\nu/\rho_\gamma \simeq 3$. Eventually, when the average momentum of relativistic neutrinos, given by the Fermi-Dirac distribution, drops below its rest mass, $T_\nu < m_\nu/3.15$, neutrinos become non-relativistic and enhance their contribution to the energy density with respect to the photons. This corresponds to a redshift $z \simeq 1100 \, m_\nu/0.58\,\text{eV}$ (see Figure~\ref{fig:nuiDecaysCosmo}).
Therefore, neutrinos become non-relativistic after recombination ($z_\star \simeq 1090$) provided that $m_\nu < 0.58\,\text{eV}$. As it can be observed in Figure~\ref{fig:nuiDecaysCosmo}, while neutrinos are still relativistic their contribution to the energy density is independent of their mass. 
However, after becoming non-relativistic, the total matter energy density gets an additional non-negligible contribution arising from stable neutrinos, which today is given by $\Omega_\nu h^2 = \sum m_\nu/93.14\,\text{eV}$. 
Essentially, CMB observations are sensitive to $\sum m_\nu$ via the contribution of massive neutrinos to the total energy density after recombination. 

The impact of massive neutrinos on the CMB has been reviewed with great level of detail in the literature, see e.g.~\cite{Tanabashi:2018oca,Lesgourgues:2006nd,Wong:2011ip,Lattanzi:2017ubx}\footnote{We particularly recommend the \textit{Neutrinos in Cosmology} review of the PDG~\cite{Tanabashi:2018oca} for readers with a particle physics background.}. The position of the acoustic peaks in the CMB spectra is extremely well measured and it depends both upon \textit{i)} the expansion history of the Universe prior to recombination (which is essentially independent of  $\sum m_\nu$ provided $\sum m_\nu \lesssim 1.5\,\text{eV}$ as can be observed in Figure~\ref{fig:nuiDecaysCosmo}) and \textit{ii)} the angular diameter distance to recombination, that can be modified via the contribution of massive neutrinos to the matter energy density between recombination and today. 
Thus, perhaps the most direct effect of massive neutrinos on CMB observations, encoded in $\Omega_\nu h^2 = \sum m_\nu/93.14\,\text{eV}$, is due to their contribution to the angular diameter distance to recombination. 
This effect is, however, strongly degenerate with the value of the Hubble constant. 
Using data from BAO or local measurements of the Hubble constant this parameter degeneracy can be broken.
In addition, when CMB observations are considered in isolation, the most relevant effect of neutrinos that become non-relativistic after recombination ($m_\nu < 0.58\,\text{eV}$) is to affect the lensing of the CMB power spectra via their contribution to the gravitational matter potential between recombination and today. We note that there are other additional effects of $m_\nu \neq 0$~\cite{Tanabashi:2018oca,Lesgourgues:2006nd,Wong:2011ip,Lattanzi:2017ubx} but they are sub-leading, given current data, as compared to the effects mentioned above.

\begin{figure}[t]
\centering
\begin{tabular}{cc}
\hspace{-0.5cm}
\includegraphics[width=0.5\textwidth]{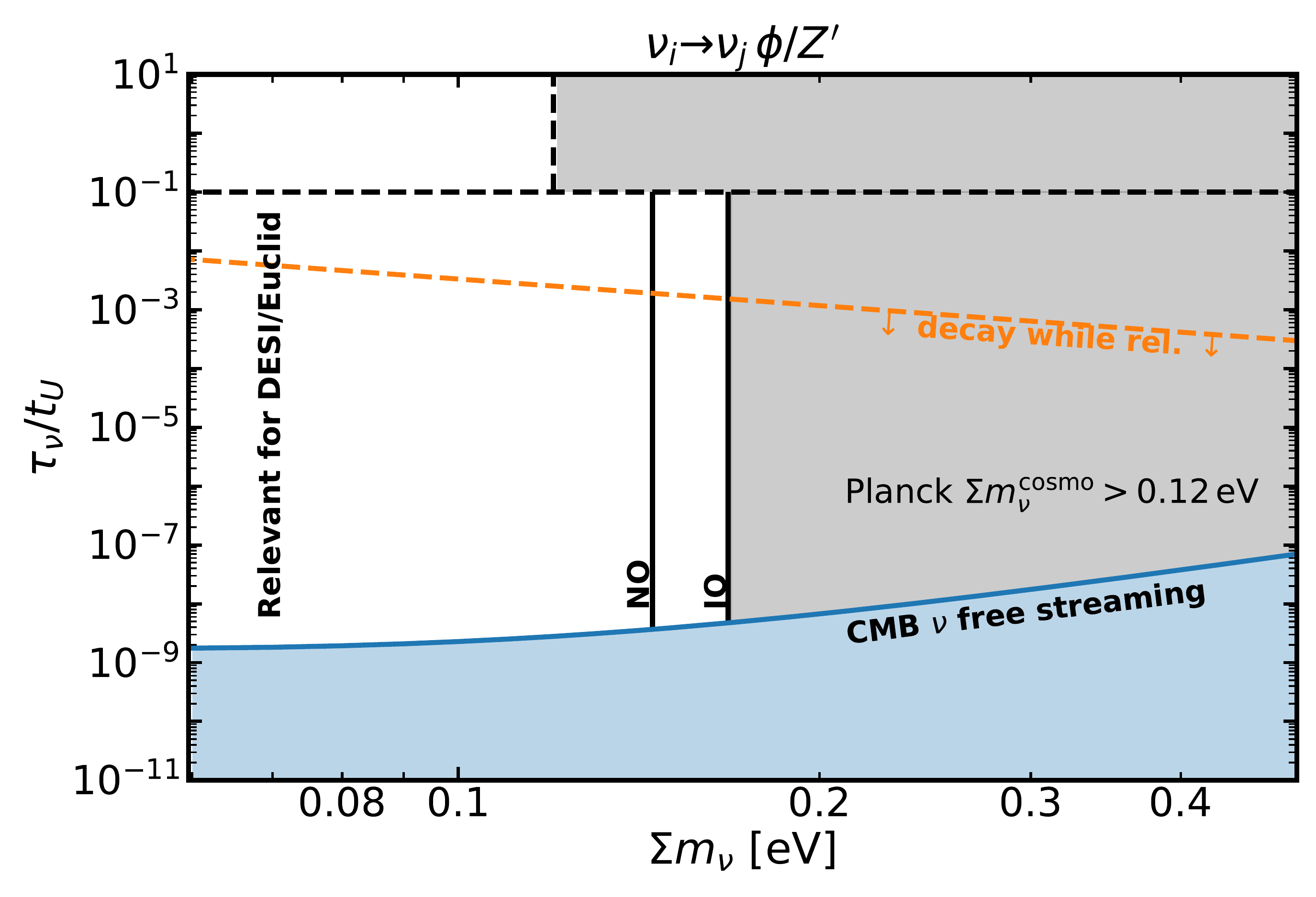} & \hspace{-0.3cm}  \includegraphics[width=0.5\textwidth]{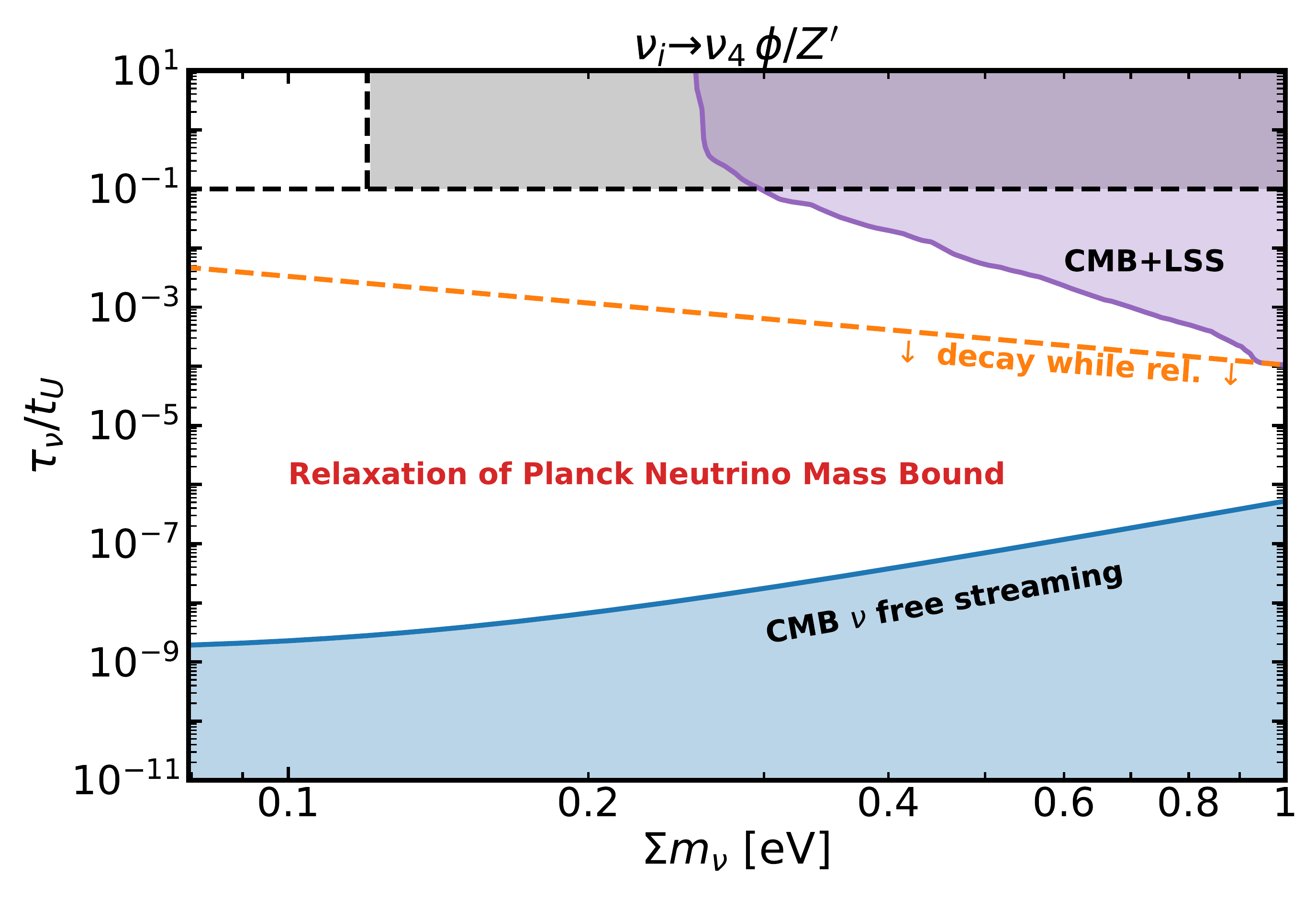}     \\
\end{tabular}
\vspace{-0.4cm}
\caption{Constraints on invisible neutrino decays. The blue regions are excluded by Planck legacy observations analyzed in the relativistically decaying neutrino framework~\cite{Escudero:2019gfk}. The magenta region is excluded by the analysis of late time matter clustering and CMB lensing from~\cite{Chacko:2019nej}. In orange we show the line separating relativistically and non-relativistically neutrino decays given by~\cite{Chacko:2019nej}: $1/\tau_\nu \simeq H_{0} \sqrt{\Omega_{m}} \left( {\sum m_{\nu}}/{(9T_{\nu,0})}\right)^{3/2}$. \textit{Left panel:} bounds on decays of the type $\nu_i \to \nu_j\,\phi/Z'$ with $m_{\phi,\,Z'} =0$. \textit{Right panel:}. bounds on decays of the type $\nu_i \to \nu_4\,\phi/Z'$ with $m_{\nu_4} = m_{\phi,\,Z'}=0$ (the same bounds apply to $\nu_i \to \nu_4 \bar{\nu}_4 \nu_4$ decays). Note that the impact of $\nu_i \to \nu_j\,\phi/Z'$ in late time cosmological observables has not yet been studied in the literature. Such study may yield additional constraints. Therefore, the region of parameter space excluded by the Planck bound on $\sum m^{\text{cosmo}}_\nu$ (see Figure~\ref{fig:numassamelioration}) shown in the left panel constitutes an estimated constraint.
}
\label{fig:lifetimes}
\end{figure}

If neutrinos decay while or soon after becoming non-relativistic, their energy contribution to matter is reduced with respect to the stable scenario and can even become completely negligible as it is shown in Figure~\ref{fig:nuiDecaysCosmo}. This would thus relax their main impact on the CMB spectra\footnote{Of course, decaying neutrinos can have additional effects: i) alter the neutrino perturbations~\cite{Escudero:2019gfk,Archidiacono:2013dua,Hannestad:2005ex}; ii) potentially modify the equation of state of the Universe and thus the CMB spectra~\cite{Hannestad:2004qu}; iii) change the matter clustering relevant for the Lyman-$\alpha$ forest and the distribution of the large scale structures of the Universe~\cite{Chacko:2019nej,Lopez:1998jt,Hannestad:1998cv,Lopez:1999ur,Kaplinghat:1999xy,Hannestad:1999xy,Hannestad:2004qu}.}. In such a case, the cosmological bounds on $\sum m_\nu$ can be relaxed. Nonetheless, CMB and galaxy survey observations can still be used to constrain the neutrino mass and lifetime in the regime in which neutrinos decay non-relativistically~\cite{Lopez:1998jt,Hannestad:1998cv,Lopez:1999ur,Kaplinghat:1999xy,Hannestad:1999xy}. In this regime, the mass and lifetime of neutrinos is mainly constrained by the shape of the matter power spectrum as relevant for galaxy surveys and also, although to a less extent, by the CMB lensing spectrum. Very recently, the alleviation of the cosmological $\sum m_\nu$ bound via neutrino decay has been analyzed in~\cite{Chacko:2019nej}. There, it has been shown that, if neutrinos decay while non-relativistic into invisible massless dark radiation with $10^{-4}\lesssim \tau_\nu/t_U \lesssim 0.1$, the cosmological $\sum m_\nu$ bound can be relaxed up to $\sum m_\nu \lesssim 0.9\,\text{eV}$. This is illustrated in the right panel of Figure~\ref{fig:lifetimes} where the bounds from~\cite{Chacko:2019nej} are labeled as ``CMB+LSS". We note that there is no study of the impact of relativistically decaying neutrinos on late time cosmological observables such as the matter power spectrum. However, since the energy density evolution in the relativistic decay scenario renders always a smaller energy density than in the non-relativistic decay case, we expect that neutrinos decaying into massless dark radiation would lead to a similar maximum alleviation of the $\sum m_\nu$ bound in both cases.

Neutrinos with lifetimes $\tau_\nu \lesssim 10^{12}\,\text{s}\,\text{eV}/m_\nu$ decay while relativistic prior to recombination. In this regime, neutrino decays alter the free streaming property of neutrinos in the early Universe. Given the large energy density encoded in neutrinos prior to recombination, this effect has a strong impact on neutrino perturbations and thus on the CMB spectra which is broadly independent of the cosmological model~\cite{Bashinsky:2003tk}. Therefore, CMB data can also be used to constrain relativistic invisible neutrino decays~\cite{Archidiacono:2013dua,Hannestad:2005ex}. The strongest current bound comes from the analysis of Planck legacy CMB observations performed in~\cite{Escudero:2019gfk}. Provided all neutrinos decay, and irrespectively of the neutrino decay final state, Ref.~\cite{Escudero:2019gfk} reports the following bound on the neutrino lifetime: $\tau_\nu > 1.3 \times  10^{9}\,\text{s}\,(m_\nu/0.05\,\text{eV})^3$. This constraint is shown in Figure~\ref{fig:lifetimes} labeled as ``CMB $\nu$ free streaming". Complementary bounds on the neutrino lifetime from laboratory experiments and astrophysics are summarized in Appendix~\ref{sec:appendix_astrolab}.

\begin{figure}[t]
\centering
\begin{tabular}{cc}
 \includegraphics[width=0.5\textwidth]{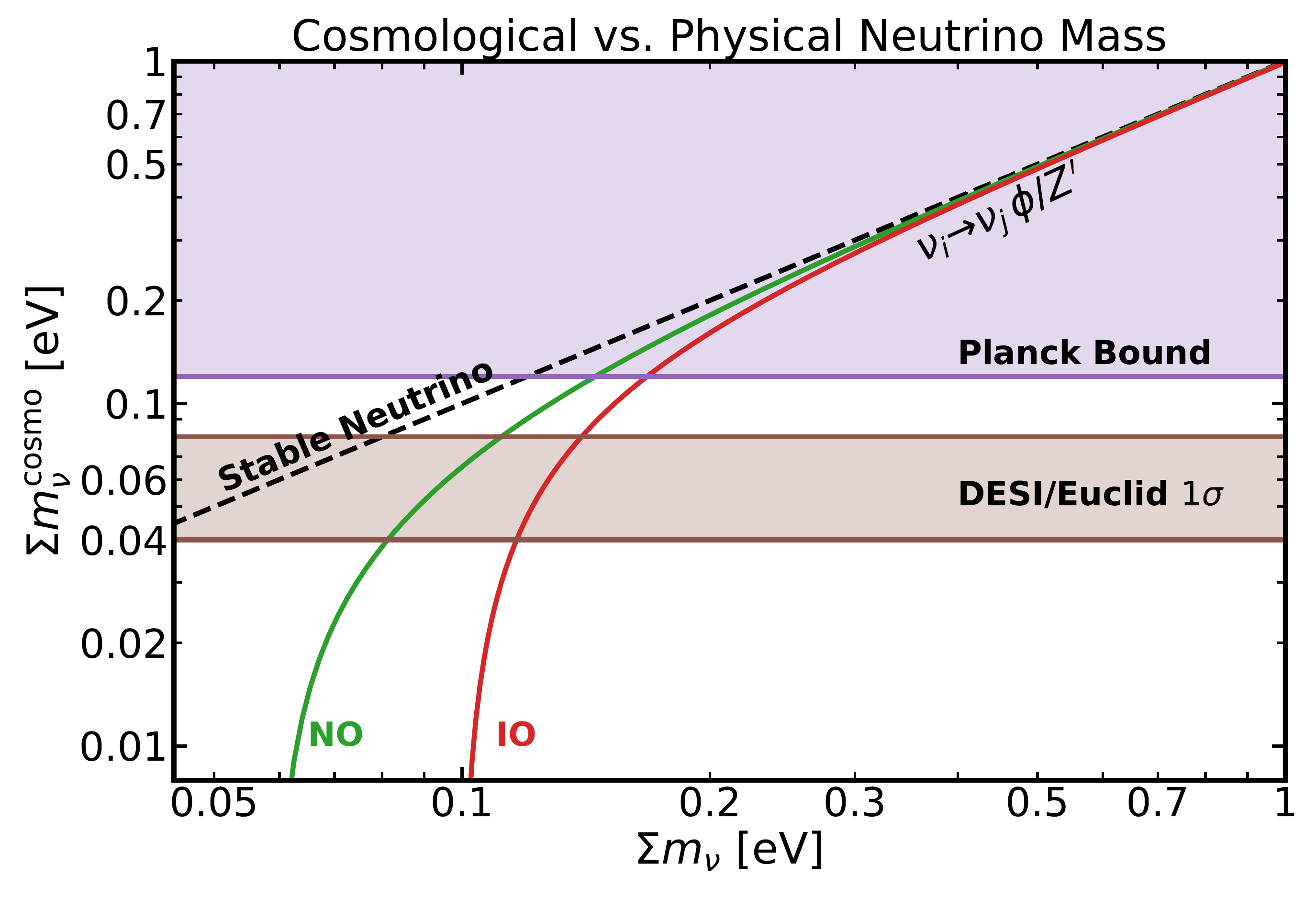}   
\hspace{-0.5cm}
 \end{tabular}
\vspace{-0.4cm}
\caption{Estimation for the maximum cosmological neutrino mass bound relaxation induced by $\nu_i \to \nu_j \,\phi/Z'$ decays with $\tau_\nu \lesssim t_U/10$. The y-axis corresponds to $\sum m_\nu^{\rm cosmo}$, namely, to the effective sum of neutrino masses inferred from the neutrino energy density today $\Omega_\nu h^2 \equiv \sum m_\nu^{\rm cosmo}/93.14\,\text{eV}$, which corresponds to $\sum m_\nu^{\rm cosmo}=3\,m_{\rm lightest}$ when all neutrinos except the lightest one decay. The x-axis is the actual sum of neutrino masses. In brown we highlight the $1\sigma$ sensitivity from \texttt{DESI}/\texttt{Euclid} for stable neutrinos with $\sum m_\nu =0.06\,\text{eV}$.}
\label{fig:numassamelioration}
\end{figure}

Finally, we note that a detailed analysis of the implications for late time observables of $\nu_i \to \nu_j \,\phi/Z'$ decays is absent from the literature. Such an analysis is beyond the scope of this paper. However, based on fermion number conservation and the previous discussion on the neutrino contribution to matter energy density, we estimate that the effective sum of neutrino
masses inferred from the neutrino energy density today, $\Omega_\nu h^2 \equiv \sum m_\nu^{\rm cosmo}/93.14\,\text{eV}$, will be given by $\sum m_\nu^{\rm cosmo}=3\,m_{\rm lightest}$ provided that all neutrinos except the lightest one decay at a rate $\tau_\nu \lesssim t_U/10$. In other words, we estimate that the potential maximum relaxation of the bound on the actual sum of neutrino masses, $\sum m_\nu$, is given by a $m_{\nu_3}-m_{\nu_1} + m_{\nu_2}-m_{\nu_1}$ ($m_{\nu_1}-m_{\nu_3} + m_{\nu_2}-m_{\nu_3}$) shift for NO (IO). This is illustrated in Figure~\ref{fig:numassamelioration} where we show the estimated value of $\sum m_\nu^{\rm cosmo}$ as a function of the actual sum of neutrino masses $\sum m_\nu$. On the one hand, given the Planck bound $\sum m_\nu^{\rm cosmo} < 0.12\,\text{eV}$, if two neutrinos decay via the $\nu_i \to\nu_j \,\phi/Z'$ channel with $\tau_\nu \lesssim t_U/10$, the region of the actual neutrino masses parameter space that would be excluded is $\sum m_\nu >  0.15\,\text{eV}$ ($\sum m_\nu > 0.17\,\text{eV}$) for NO (IO). In such a case, neutrino decays can only slightly relax the current bound. We note, however, that the constrain we report here should  be  taken  as  an estimate since a detailed analysis could lead to a slightly different result\footnote{For non-relativistically decaying neutrinos with $\sum m_\nu \lesssim 0.2\,\text{eV}$, the phase space is such that  that the neutrino in the final state could be relativistic thus weakening even further the constraints on $\sum m_\nu^{\rm cosmo}$. On the other hand, for $\sum m_\nu \gtrsim 0.2\,\text{eV}$ the neutrino in the final state will be non-relativistic and we do not expect our estimate to be altered.}. On the other hand, notice that the relaxation of the neutrino mass bound is larger for small values of $\sum m_\nu$ with a maximum alleviation given by a $0.06\,\text{eV}$ ($0.1\,\text{eV}$) shift for NO (IO) in the massless lightest neutrino limit. This might be very relevant for upcoming galaxy surveys as \texttt{DESI} or \texttt{Euclid} since if, lets say, a bound $\sum m_\nu^{\rm cosmo} <0.04\,\text{eV}$ is eventually reported it would not necessarily exclude NO or IO, being compatible with both NO and IO in the neutrino decay scenario, with $\sum m_\nu < 0.08\,\text{eV}$ ($\sum m_\nu < 0.12\,\text{eV}$) for NO (IO). 

We can conclude from this section that decays into purely massless dark radiation represent the best avenue to relax current cosmological constraint on $\sum m_\nu$. Neutrino decays including active neutrinos in the final state can potentially play a relevant role in the near future with upcoming data from \texttt{DESI}/\texttt{Euclid}. This can be clearly observed already in Figure~\ref{fig:lifetimes}, where we show (in white color) the viable region of the $\tau_\nu-\sum m_\nu$ parameter space where the cosmological bound on $\sum m_\nu$ can be relaxed. In the next section we will explore the range of values for the coupling constants and masses of the particles involved in the decays for which this can be realized.

\vspace{-0.1cm}

\subsection{Allowed parameter space}\label{sec:results}
\vspace{-0.1cm}

Using the effective Lagrangians considered in Subsection~\ref{sec:effLag}, in Appendix~\ref{sec:appendix_rates} we have computed the invisible neutrino decay rates for all the relevant channels summarized in Figure~\ref{fig:NeutrinoDecayDiagrams}. In this section, we will determine the region of couplings and masses parameter space that can lead to $\tau_\nu < t_U$, including the most constraining model independent bounds on neutrinophilic bosons and invisible neutrino decays. According to the categorization presented in Section~\ref{subsec:category}, and following a model independent approach, this will be done separately for each decay channel using the decay rates computed in Appendix~\ref{sec:appendix_rates}. In order to have a feeling of the main dependence of these channels on the different parameters we display in Table~\ref{tab:rates} a summary of approximated expressions for the neutrino decay rates in some relevant phase space limits. Table~\ref{tab:rates} has been built so as to highlight relevant scalings with masses and couplings for interesting cases, i.e., for neutrino masses $\cal{O}$(0.3 eV) when the decay channel allows $\sum m_\nu \sim 1$ eV and $\cal{O}$(0.05 eV) in the case of decay to one active neutrino, which could be relevant for future surveys. We note that there is no interference between the different decay diagrams and the rates only depend upon the modulus of the individual coupling constants. Therefore, to simplify the notation, throughout this section we shall use $\lambda$, $h$ and $g$ to denote their absolute value.  

\noindent\textbf{2-Body Decays:}

\textbf{\textit{Case2a.}} In a realistic scenario, once the possibility of a neutrino decay is given, we expect all possible decay channels to be opened, i.e., $\nu_{3}\to\nu_{1,2}\,\phi/Z'$ and $\nu_{2}\to\nu_{1}\,\phi/Z'$ ($\nu_{2}\to\nu_{1,3}\,\phi/Z'$ and $\nu_{1}\to\nu_{3}\,\phi/Z'$) for NO (IO). Furthermore, theoretically one would expect that once couplings are turned on they are of similar strength, given the observed mixing among active neutrinos. The resulting allowed parameter space for a pseudo-scalar $\phi$ and $Z'$ in the final state is shown in Figure~\ref{fig:2acase}. In Appendix~\ref{sec:individual} we show the result for each individual mode, including also the case of $\phi$ being a scalar boson, assuming that the rest of the channels are switched off. Notice that, as we have discussed in the previous section, the maximum relaxation of the $\sum m_\nu$ bound is obtained for the case in which two neutrinos decay (as it is shown in Figures.~\ref{fig:numassamelioration} and~\ref{fig:2acase}). The present Planck constraint can only be very slightly alleviated by $0.03$ eV ($0.05$) for NO (IO).

\textit{Case 2a: Pseudo-scalar couplings in $\nu_i\to\nu_j \, \phi$ decays}. From the left panels of Figure~\ref{fig:2acase} (see also the detailed plots for each individual mode in Appendix~\ref{sec:individual}), we can extract the following two pieces of information: 
\textit{i)} CMB bounds on invisible neutrino decays restrict pseudo-scalar couplings to be:  $\lambda_{31}, \lambda_{32}\lesssim 10^{-10}$, and $\lambda_{21}\lesssim 10^{-8}$; \textit{ii)} the present cosmological neutrino mass bound can only marginally be alleviated for coupling strengths in the range $10^{-15} \lesssim \lambda_{31},\lambda_{32} \lesssim 10^{-10}$ and $10^{-13} \lesssim \lambda_{21} \lesssim 10^{-8}$, which give rise to decays occurring on timescales shorter than the age of the universe without substantially altering neutrino free streaming in the early Universe.

\begin{figure}[t]
\centering
\begin{tabular}{cc}
\hspace{-0.5cm}
\includegraphics[width=0.5\textwidth]{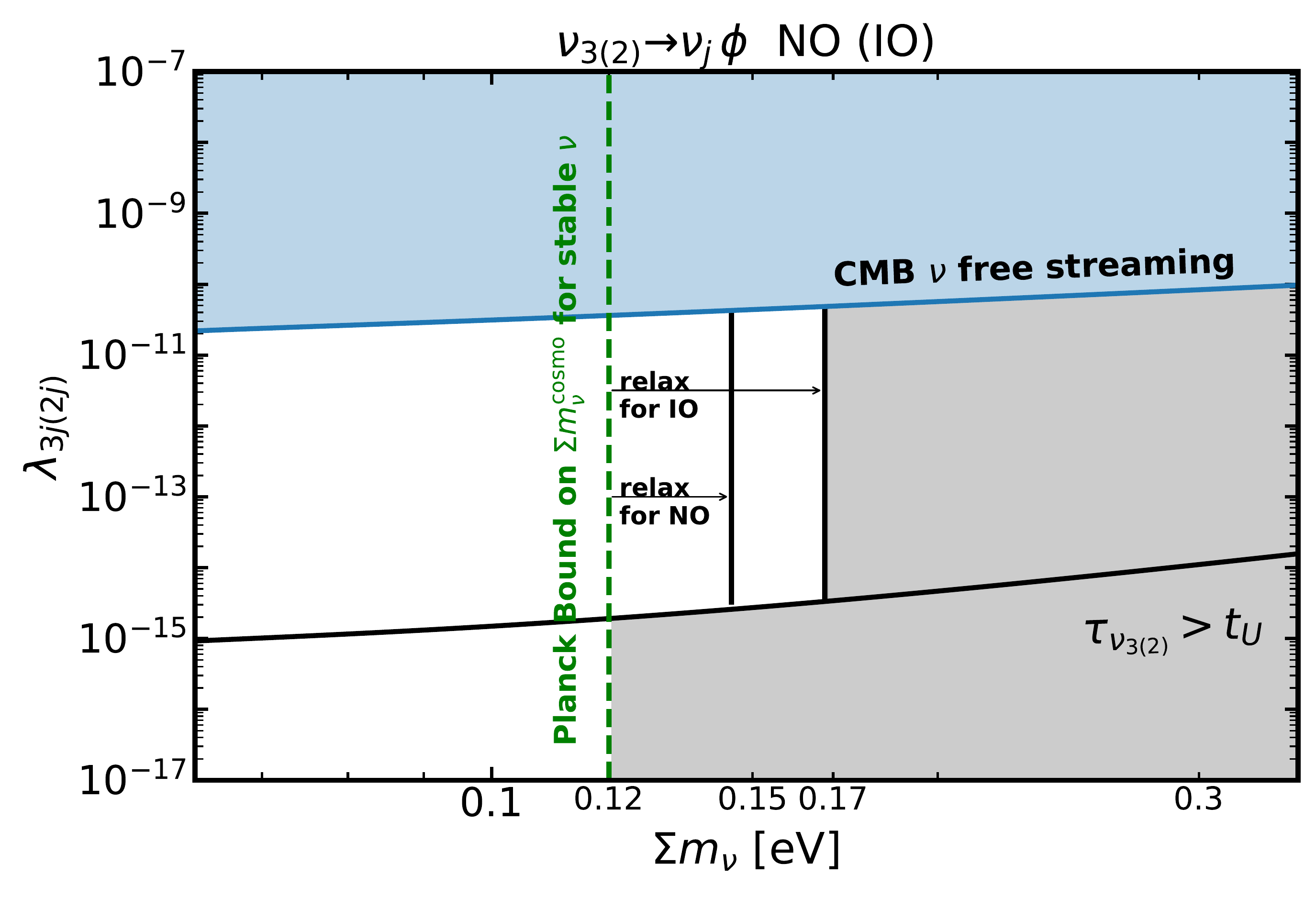} & \hspace{-0.3cm}  \includegraphics[width=0.5\textwidth]{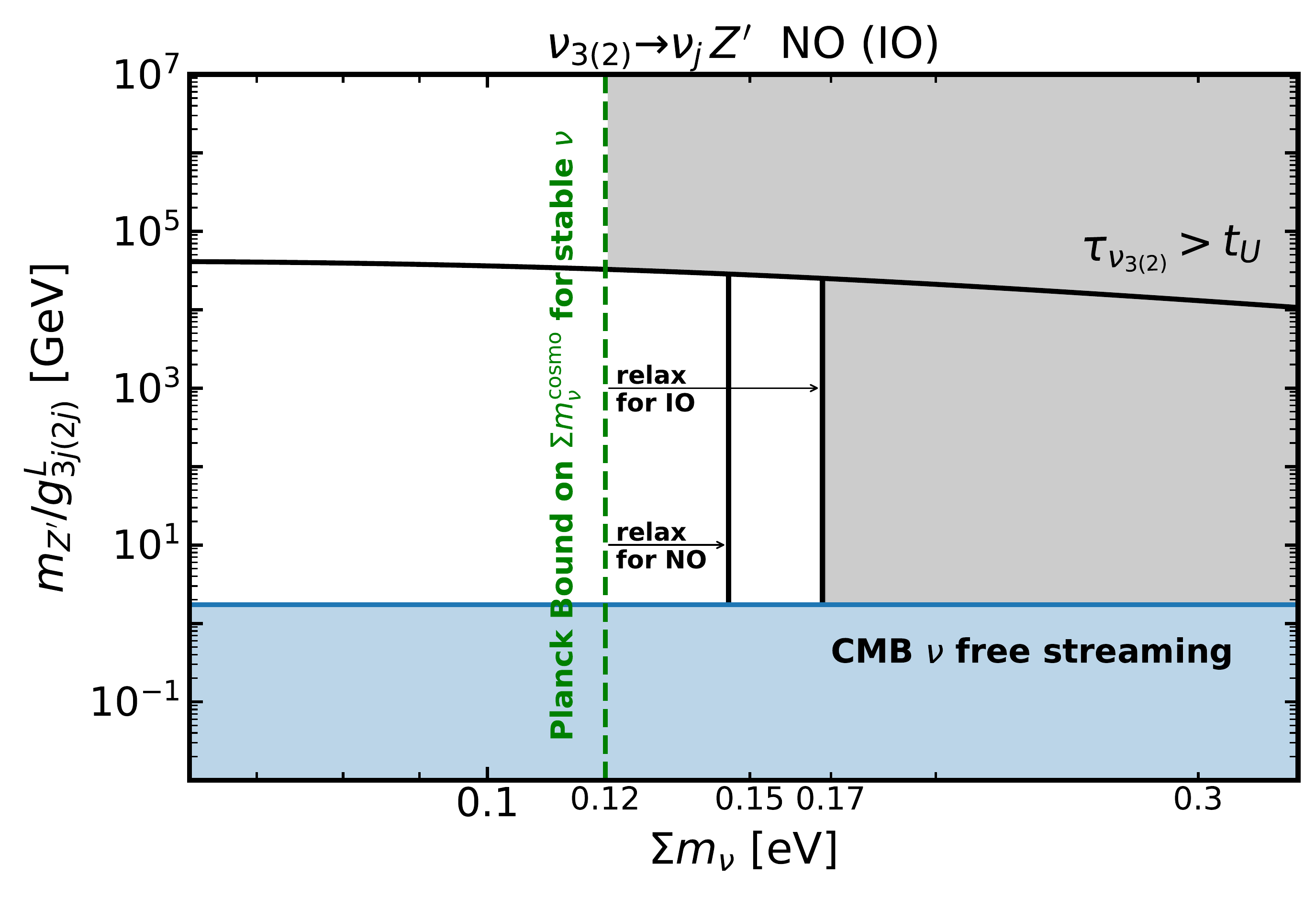} \\     
\hspace{-0.5cm}
\includegraphics[width=0.5\textwidth]{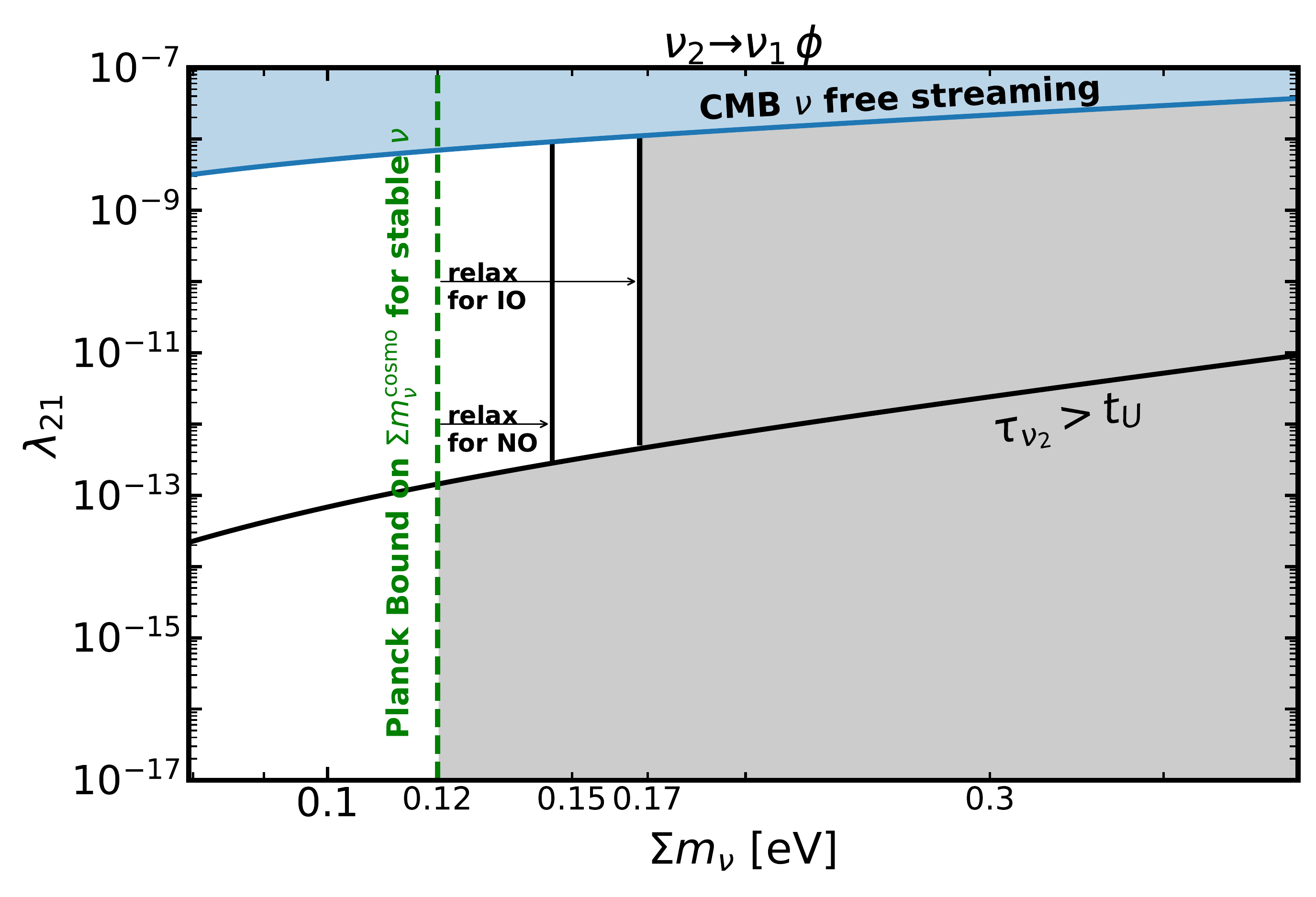} & \hspace{-0.3cm}  \includegraphics[width=0.5\textwidth]{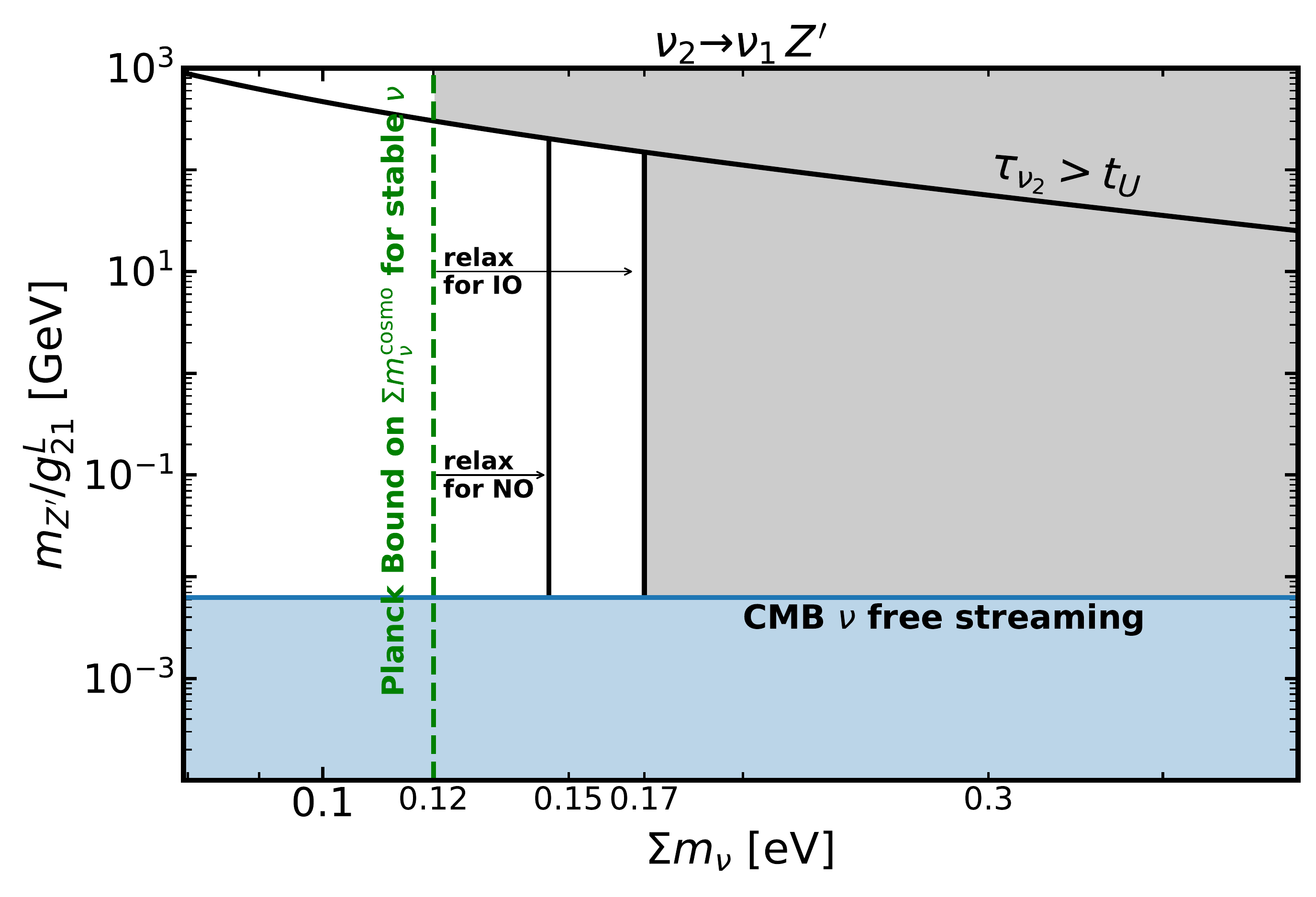} \
 \end{tabular}
\vspace{-0.4cm}
\caption{Parameter space for case \textit{2a} (see Figure~\ref{fig:NeutrinoDecayDiagrams}). \textit{Left panel:} $\nu_i \to\nu_j\,\phi$ decays with $\phi$ being a pseudoscalar boson. \textit{Right panel:} $\nu_i\to \nu_j \,Z'$ decays. Here we have assumed that all the couplings are different from zero and all neutrinos decay except the lightest one. The blue region is excluded by the constraint on invisible neutrino decays from Planck CMB observations~\cite{Escudero:2019gfk}. In dashed green we highlight the current bound on $\sum m_\nu$ within the $\Lambda$CDM~\cite{Aghanim:2018eyx} and stable neutrino framework. The arrows indicate to what extent it can be relaxed, see also Figure~\ref{fig:numassamelioration}. The grey region is excluded by the bound $\sum m_\nu^{\rm cosmo} < 0.12\,\text{eV}$ reinterpreted in the neutrino decay scenario.  
}
\label{fig:2acase}
\end{figure}

\textit{Case 2a: Scalar couplings in $\nu_i\to\nu_j\,\phi$ decays}. 
As far as the mass of the active neutrino in the final state can be neglected, the results are very similar to the pseudo-scalar case, becoming well differentiated when the decaying neutrino mass is close to the final state neutrino mass, due to the extra suppression for degenerate neutrinos in the case of pseudo-scalar couplings, see Equation~(\ref{eq:rate2a}). We find that CMB constraints on invisible neutrino decays bound the couplings responsible for neutrino decay to be $h_{31} ,\,h_{32}\lesssim 10^{-11} \, (0.05\,\text{eV}/m_{\nu})$ and $h_{21} \lesssim  5\! \times \! 10^{-11} \, (0.05\,\text{eV}/m_{\nu})$ (see Figure~\ref{appfig:2acase}). We also find that couplings in the range $10^{-15} \lesssim h_{31}, h_{32} \lesssim 10^{-11} \, (0.05\,\text{eV}/m_{\nu})$ allow a slight relaxation of the cosmological $\sum m_\nu$ bound, in the same way as we discussed in the previous scenario.

\textit{Case 2a: $\nu_i\to\nu_j\,Z'$ decays}. From the right panels of Figure~\ref{fig:2acase}  we can see that CMB bounds on invisible neutrino decays imply $m_{Z'}/g^L_{31},  m_{Z'}/g^L_{32} \gtrsim 1\,\text{GeV}$ and $m_{Z'}/g^L_{21} \gtrsim 10\,\text{MeV}$. We find that the cosmological bounds on $\sum m_\nu$ could be slightly relaxed as in the scalar cases, provided $1\,\text{GeV} \lesssim m_{Z'}/g^L_{31}, m_{Z'}/g^L_{32} \lesssim 25\,\text{TeV}$. In order to plot our results as a function of $m_{Z'}/g^L_{ij}$, we have used the leading order expanded expression for the decay rate as given in Table~\ref{tab:rates}, where $m_{Z'}/m_{\nu} \sim 0$ is the expansion parameter. This will allow us to easily make the connection between the effective coupling $m_{Z'}/g^{L}_{ij}$ and a vacuum expectation value within a concrete model realization. The parameter spaces arising when this approximation or the full formula are considered only differ in a tiny region of $m_{Z'}$ (where $m_{Z'}$ has a significant impact in the phase space). Our approximation is thus in excellent agreement with the full formula given in Equation~\eqref{appeq:twobodyzprime}.

\begin{figure}[t]
\centering
\begin{tabular}{cc}
\hspace{-0.5cm}
 \includegraphics[width=0.5\textwidth]{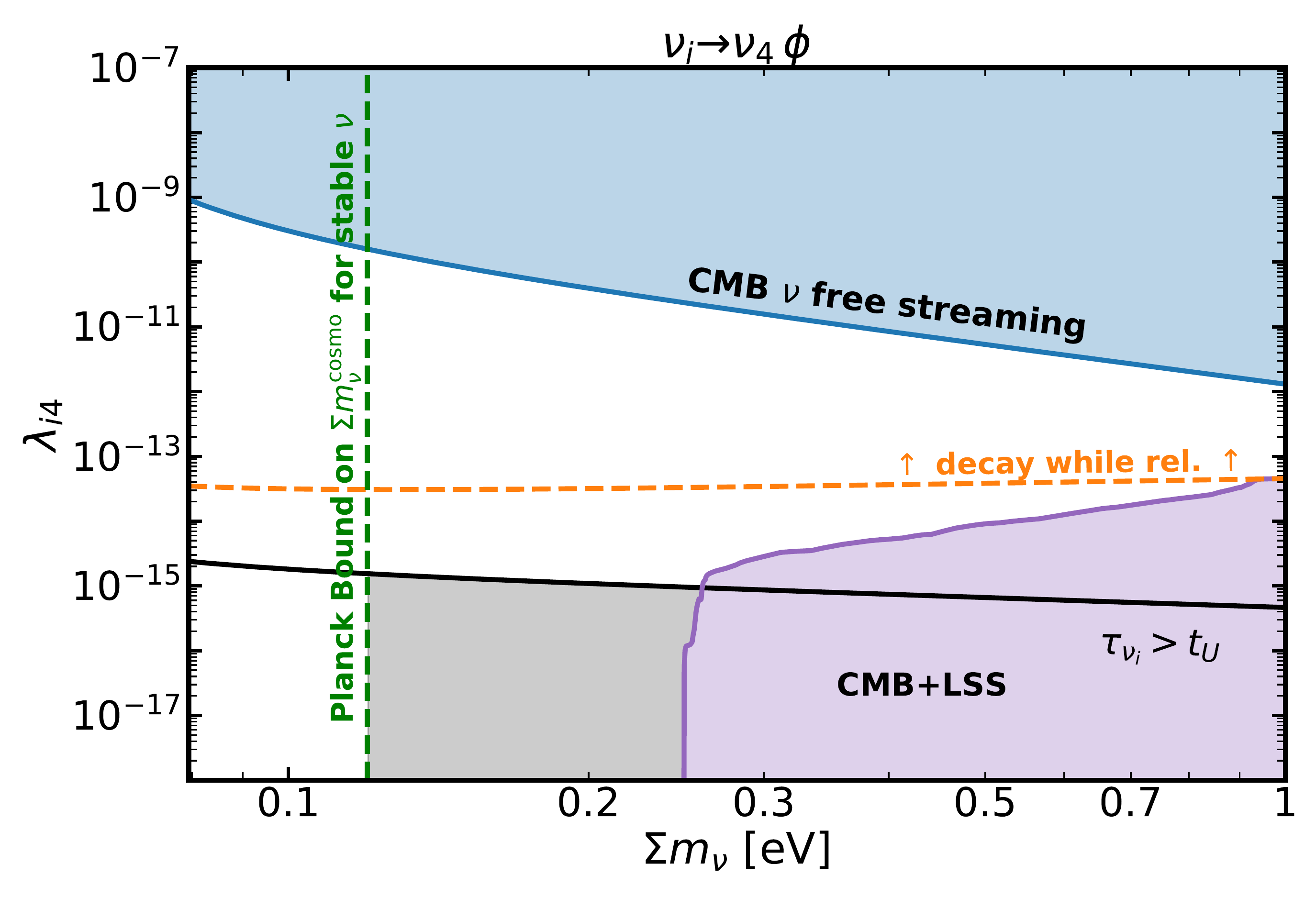} & \hspace{-0.3cm}  \includegraphics[width=0.5\textwidth]{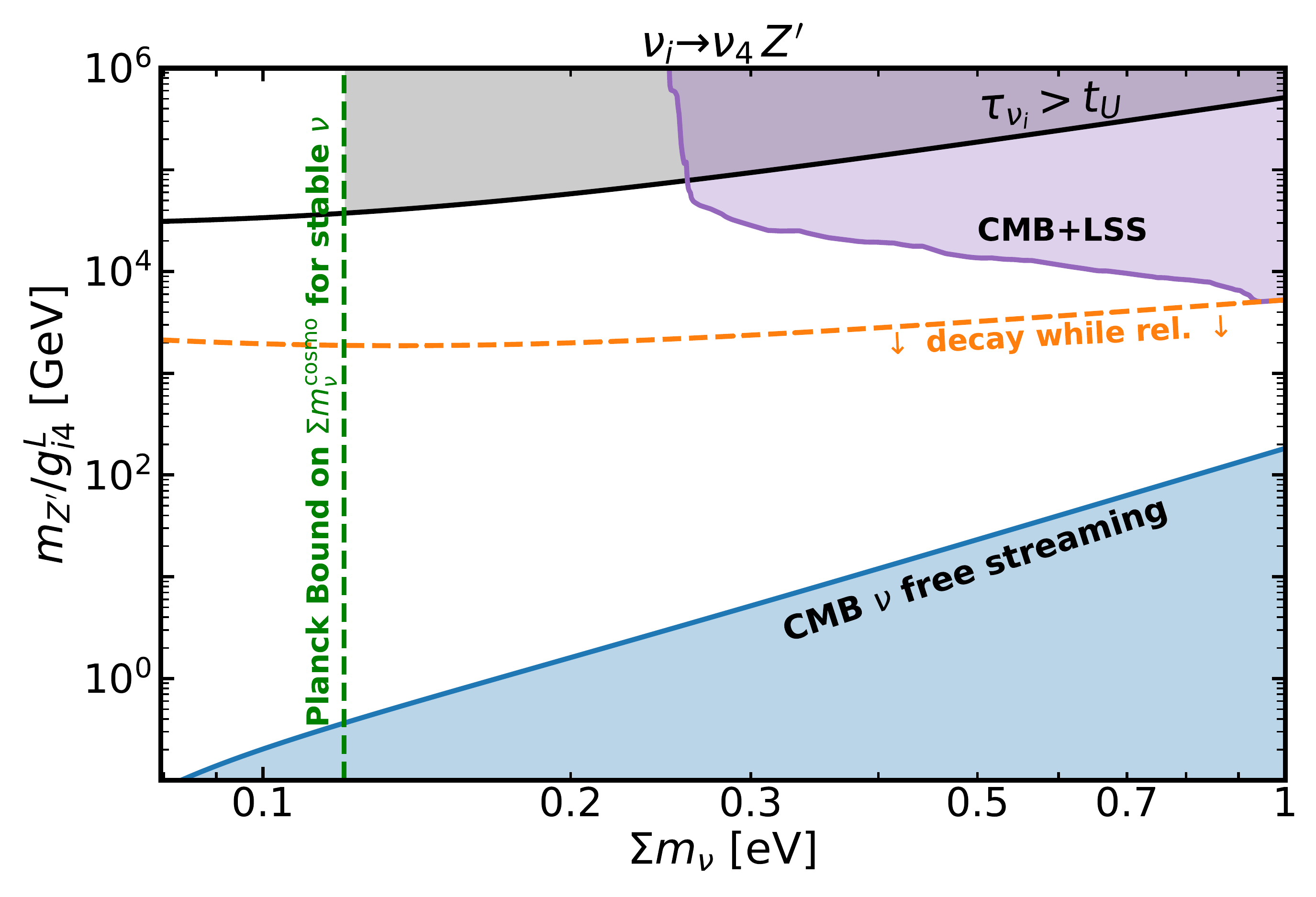}    
\hspace{-0.5cm}
 \end{tabular}
\vspace{-0.4cm}
\caption{Parameter space for case \textit{2s} (see Figure~\ref{fig:NeutrinoDecayDiagrams}). \textit{Left panel:} parameter space for $\nu_i \to \nu_4 \, \phi$ decays. \textit{Right panel:} parameter space for $\nu_i \to \nu_4 \, Z'$ decays. Both correspond to the \textit{2s} case in Figure~\ref{fig:NeutrinoDecayDiagrams}. We have chosen $m_{\phi,\,Z'} \ll m_{\nu_i}$.  For $\nu_i \to \nu_4 \, \phi$ we consider only pseudo-scalar couplings but note that the rates are identical for scalar couplings. In orange we show the line separating relativistically and non-relativistically neutrino decays given by~\cite{Chacko:2019nej}: $1/\tau_\nu \simeq H_{0} \sqrt{\Omega_{m}} \left( {\sum m_{\nu}}/{(9T_{\nu,0})}\right)^{3/2}$. The blue region is excluded by CMB constraints on relativistically decaying neutrinos~\cite{Escudero:2019gfk}. The magenta region is excluded by cosmological bounds on non-relativistically decaying neutrinos~\cite{Chacko:2019nej}.  The Planck bound on $\sum m_\nu$ extracted within $\Lambda$CDM is indicated by the dashed green line. The constraint from~\cite{Chacko:2019nej} does not saturate the Planck bound for $\tau_\nu \gg t_U$ simply because different data sets were considered in the analysis of~\cite{Chacko:2019nej}. 
}
\label{fig:2scase}
\end{figure}

\vspace{0.1cm}
\textbf{\textit{Case2s:}} This case corresponds to $\nu_{i}\to\nu_{4}\,\phi/Z'$, with $\phi$ being a scalar or pseudo-scalar and is shown in Figure~\ref{fig:2scase}. Both, scalar and pseudo-scalar, give rise to the same decay rate if $m_{\nu_4},m_{\phi,Z'} \ll m_{\nu_i}$ and $h = \lambda $.

\textit{Case 2s. $\nu_i\to\nu_4\,\phi$ decays}. We display the relevant parameter space for this scenario in the left panel of Figure~\ref{fig:2scase} for the case $m_{\nu_4} = m_\phi = 0$. The upper bound is driven by the modification of the CMB due to neutrino free streaming when relativistic neutrino decays are included, and is given by $\lambda_{i4} \lesssim 10^{-12}(m_{\nu}/0.3\,\text{eV})$. CMB and LSS bounds on non-relativistic invisible neutrino decays exclude the region of the parameter space given by $\sum m_\nu \gtrsim 0.25\,\text{eV}$ and $\lambda_{i4} \lesssim  4\! \times \! 10^{-14}(m_\nu/0.3\,\text{eV})$. Further requiring the decay to happen faster than the age of the universe, we identify a large plateau in the range $10^{-15} \lesssim \lambda_{4i}  \lesssim 10^{-12} \, (0.3\,\text{eV}/m_\nu) $ where the cosmological neutrino mass bound can be alleviated up to $\sum m_\nu \sim 1\,\text{eV}$ (see also Figure~\ref{fig:lifetimes}). 

\textit{Case 2s: $\nu_i\to\nu_4\,Z'$ decays}. The region of parameter space $m_{Z'}/g_{i4}^L < 100\,\text{GeV} (m_\nu/0.3\,\text{eV})^3$ is excluded by the CMB constraints on relativistically decaying neutrinos, as it can be observed in the right panel of Figure~\ref{fig:2scase}. Cosmological bounds on non-relativistically decaying neutrinos exclude the region given by $\sum m_\nu > 0.25\,\text{eV}$ and $m_{Z'}/g_{i4}^L \gtrsim 10 \,\text{TeV}$. In summary, we find that for $100\,\text{GeV} (m_\nu/0.3\,\text{eV})^3 \lesssim m_{Z'}/g_{i4}^L \lesssim 10\,\text{TeV}$, this decay mode can lead to an amelioration of the cosmological neutrino mass bounds up to $\sum m_\nu \lesssim 1\,\text{eV}$.

\noindent \textbf{3-Body Decays:}

The 3-body decays rates are fairly independent upon the nature of the bosonic mediator. 
In particular, the case of $Z'$ mediated decays and $\phi$ mediated decays with scalar couplings are fully analogous to the $\phi$ mediated decay with pseudo-scalar couplings, given the following approximate mapping: $\left(g^L_{ij},\, g^{L}_{i4},\,\sqrt{g^{L}_{44}{}^2+g^{R}_{44}{}^2} \right)\to \left(\lambda_{ij},\,\lambda_{i4},\,\lambda_{44}\right)$ and  $\left(h_{ij},\,h_{i4},\,h_{44}\right) \to \left( \lambda_{ij},\,\lambda_{i4},\,\lambda_{44} \right)$, respectively. Therefore, we only discuss decays solely realized via pseudo-scalar couplings ($\lambda \neq 0\,,h=0$). But the same conclusions that we will draw here for the pseudo-scalar case apply to the other scenarios.

Several bounds apply to sub-MeV neutrinophilic bosons as relevant for 3-body invisible neutrino decays. The most relevant ones are: \textit{i)} CMB constraints on neutrino free streaming that exclude $\nu_i-\phi/Z'$ couplings as small as $10^{-13}$ for $0.1\,\text{eV}<m_{\phi/Z'} < 300\,\text{eV}$~\cite{Escudero:2019gvw}, \textit{ii)} BBN constraints, and \textit{iii)} supernova cooling constraints~\cite{Raffelt:1996wa}. There are several studies dealing with BBN and SN1987A bounds on neutrinophilic bosons, see e.g.~\cite{Farzan:2018gtr,Escudero:2019gfk,Escudero:2019gzq} and~\cite{Choi:1989hi,Kachelriess:2000qc,Farzan:2002wx,Heurtier:2016otg,Brune:2018sab,Croon:2020lrf}, respectively. However, given the wide range of masses and types of couplings we consider in our analysis, we have rederived the bounds using our effective Lagrangians, see Equations~(\ref{eq:Lag_scalar}) and (\ref{eq:Lag_vector}). The derivation is described in Appendix~\ref{sec:Bosonsconstraints}.

\begin{figure}[t]
\centering
\begin{tabular}{cc}
\hspace{-0.5cm}
 \includegraphics[width=0.5\textwidth]{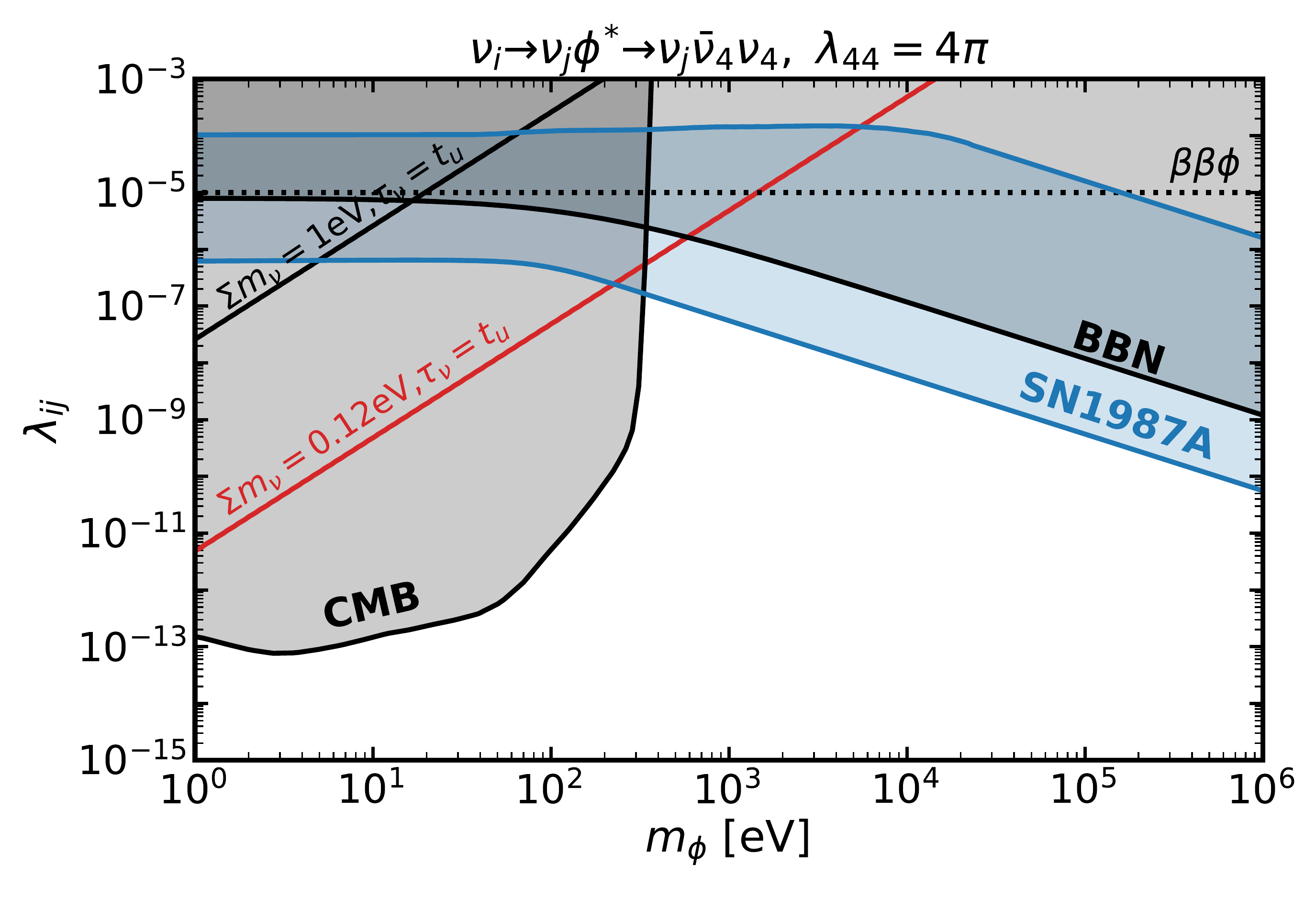} & \hspace{-0.3cm}  \includegraphics[width=0.5\textwidth]{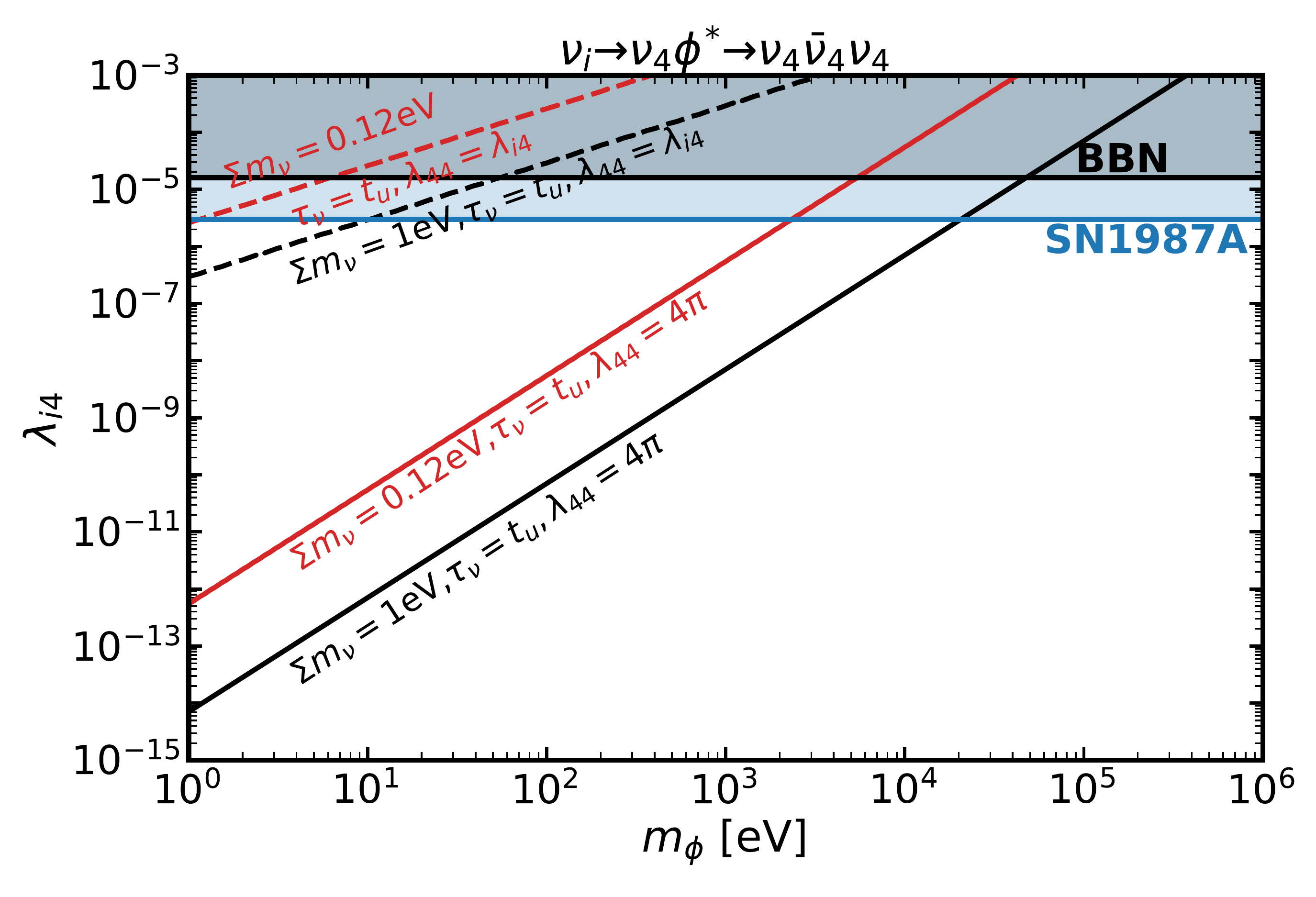}   
\hspace{-0.5cm}
 \end{tabular}
\vspace{-0.4cm}
\caption{\textit{Left panel:} parameter space for $\nu_i \to \nu_j \phi^\star \to \nu_j \bar{\nu}_4 \nu_4$ decays (case \textit{3a1} in Figure~\ref{fig:NeutrinoDecayDiagrams}). We have assumed that $m_{\nu_4} = 0$ and that active neutrinos have universal couplings to $\phi$. 
\textit{Right panel:} parameter space for $\nu_i \to \nu_j \phi^\star \to \nu_4 \bar{\nu}_4 \nu_4$ decays (case \textit{3s} in Figure~\ref{fig:NeutrinoDecayDiagrams}). We show laboratory bounds from KamLAND-Zen~\cite{Gando:2012pj}, CMB constraints on $\nu$-$\phi$ interactions from Planck CMB observations~\cite{Escudero:2019gvw}, and SN1987A and BBN constraints on neutrinophilic bosons -- see Appendix~\ref{sec:Bosonsconstraints} for the derivation of the latter two. Note that for decays happening at the same rate, the difference in the final phase space enforces larger couplings for increasing active neutrino mass if one active neutrino is in the final state, whereas the opposite occurs in the case with only sterile final states. Identical conclusions can be drawn for both scenarios but mediated by a $Z'$.}
\label{fig:3a1}
\end{figure}

\newpage
\textit{Case 3a0: $\nu_i\to\nu_j\bar{\nu}_k\nu_k$}. In this case, $\tau_\nu < t_U$ requires a light and not very weakly coupled neutrinophilic boson with $\lambda_{ij} \gtrsim 0.6 \,{m_{\phi}}/{10\,\text{keV}}$ which is already excluded by meson decays, double beta decay searches, and various astrophysical and cosmological constraints (see left panel of Figure~\ref{fig:3a1}), and thus this channel is not phenomenologically viable. 

\textit{Case 3a1: $\nu_i\to\nu_j\bar{\nu}_4\nu_4$}. This decay mode is controlled by $\lambda_{ij}$ and $\lambda_{44}$, see Table~\ref{tab:rates}.
 $\lambda_{ij}$ is constrained by various laboratory experiments, astrophysics and cosmology. However, $\lambda_{44}$ is largely unconstrained since it parametrizes the interaction between a sterile neutrino and a neutrinophilic boson. In the left panel of Figure~\ref{fig:3a1} we show the most favorable region in which $\tau_\nu < t_U$ can be realized. As it can be seen in the figure, even saturating $\lambda_{44} = 4\pi$, the bounds from the CMB, BBN and SN1987A on sub-MeV neutrinophilic mediators exclude $\tau_\nu < t_U$ for this type of topology.

\textit{Case 3a2 : $\nu_i\to\nu_j\bar{\nu}_4\nu_4$}. This decay mode is solely driven by $\lambda_{i4}$. We have found that $\tau_\nu < t_U$ requires $\lambda_{i4} > 10^{-5} \,m_{\phi/Z'}/1\,\text{eV}$, which is already excluded by BBN and SN1987A data.

\textit{Case 3s $\nu_i\to\nu_4\bar{\nu}_4\nu_4$}. The right panel of Figure~\ref{fig:3a1} shows the relevant parameter space for this decay channel. 
The couplings controlling this decay mode ($\lambda_{i4}$ and $\lambda_{44}$) are not particularly well constrained. The strongest bounds on $\lambda_{i4}$ arise from BBN data and supernova cooling, as shown in Figure~\ref{fig:3a1}, while $\lambda_{44}$ is essentially unbounded. In the right panel of Figure~\ref{fig:3a1} we show the most favorable case with $\lambda_{44} = 4\pi$. We can clearly appreciate that for $\nu_i \to \nu_4 \bar{\nu}_4 \nu_4$ decays to be cosmologically relevant, mediators  with $m_{\phi/Z'} \lesssim 10\,\text{keV}$ and coupling strengths $\lambda_{i4}  > 10^{-14} \frac{4\pi}{\lambda_{44}} \left({m_{\phi/Z'}}/{\text{eV}}\right)^2 \sqrt{0.3\,\text{eV}/m_\nu }$ are required. For $\lambda_{i4} = \lambda_{44}$, however, only a tiny window with ${m_{\phi/Z'}}\lesssim 10\,{\text{eV}}$ would be viable.

\noindent Therefore, at the phenomenological level, $\nu_i\to\nu_4\bar{\nu}_4\nu_4$ decays can still satisfy $\tau_\nu < t_U$ for $m_{\phi/Z'} \lesssim 10\,\text{keV}$. It is important to remark that, even tough the decay rate is independent of $\lambda_{ij}$ at the model independent level, in a UV complete model $\lambda_{ij}$ is expected to be present and correlated via mixing with the parameters driving the decay. Since this coupling is subject to stringent cosmological and astrophysical constraints (see left panel of Figure~\ref{fig:3a1}), $\nu_i\to\nu_4\bar{\nu}_4\nu_4$ decays can only render $\tau_\nu < t_U$ provided $\lambda_{44} \,, \lambda_{i4}\gg \lambda_{ij} $. 

\noindent \textbf{Summary:}  We have identified the cosmologically relevant regions of parameter space in our effective Lagrangians parametrizing invisible neutrino decay, see Equations~\eqref{eq:Lag_scalar} and~\eqref{eq:Lag_vector}. 
We have shown that a wide region of parameter space is excluded by current CMB, BBN and SN1987A constraints.
Furthermore, we have also highlighted the range of values for the relevant couplings and masses that can lead to $\tau_\nu < t_U$ and in which, therefore, invisible neutrino decays can relax cosmological bounds on $\sum m_\nu$. We find that 2-body decays into new light BSM species can alleviate such bounds in a substantial region of the parameter space.
Finally, our results show that among all possible 3-body decays, only those of the type $\nu_i \to \nu_4 \bar{\nu}_4 \nu_4$ can lead to $\tau_\nu < t_U$ and only in a narrow window of the parameter space. 

It is worth mentioning that although 2-body decays with one active neutrino in the final state can only very slightly relax the current Planck bound, such decays may become relevant if future galaxy surveys such as \texttt{DESI} or \texttt{Euclid} do not detect neutrino masses at all, or point towards a NO spectrum. From Figure~\ref{fig:numassamelioration}, we see that the absence of a neutrino mass signal could be a hint of unstable neutrinos. Moreover, a measurement of  $\sum m_\nu^{\rm cosmo} = 0.06$ eV, which would be interpreted as a signal of NO in the standard framework, could instead correspond to IO if the active neutrinos decay to the lightest one. This would be particularly relevant in the context of neutrinoless double beta decay experiments since the theoretical prediction for the decay rate crucially depends on both the neutrino ordering and the absolute neutrino mass scale.

\section{Models for Neutrino Masses and Invisible Neutrino Decays}\label{sec:ModelBuilding}

A large number of models generating the neutrino mass and mixing pattern observed in neutrino oscillation experiments are in strong tension with cosmological bounds on $\sum m_\nu$. For instance, 5 out of 7 potentially viable two-zero neutrino mass textures predict $\sum m_\nu > 0.12\,\text{eV}$~\cite{Alcaide:2018vni}. This tension with cosmology generically occurs also for inverse Majorana neutrino mass matrices with two zeros textures~\cite{Lavoura:2004tu,Verma:2011kz}.    
Any model generating these textures would thus be disfavored by cosmology as it is the case, for instance, for minimal $U(1)_{\mu-\tau}$ neutrino mass models~\cite{Asai:2017ryy,Asai:2018ocx,Asai:2019ciz,Araki:2019rmw}. Furthermore, plenty of generic models do not necessarily strictly predict $\sum m_\nu > 0.12\,\text{eV}$, but accommodate $\sum m_\nu > 0.12\,\text{eV}$ across large regions of the parameter space (see~\cite{GonzalezGarcia:2002dz,King:2003jb,Altarelli:2004za,Mohapatra:2005wg,Mohapatra:2006gs,Altarelli:2010gt,King:2013eh,King:2014nza,Cai:2017jrq} for reviews on neutrino mass models).

In this section, we first propose a simple extension of the seesaw mechanism  with an extra scalar field and a sterile neutrino state, both charged under an additional $U(1)_X$ symmetry, in which $\nu_i \to \nu_4 \, \phi$ decays with $m_{\nu_i} \gg m_{\nu_4} \simeq m_\phi = 0$ can naturally occur, alleviating the neutrino mass bounds up to 
$\sum m_\nu \sim 1\,\text{eV}$. Then, we embed this extension within a minimal neutrino mass model based on a spontaneously broken $U(1)_{\mu-\tau}$ flavor symmetry~\cite{Choubey:2004hn,Araki:2012ip}, which predicts  $\sum m_\nu > 0.12\,\text{eV}$~\cite{Asai:2017ryy,Asai:2018ocx,Asai:2019ciz,Araki:2019rmw} and therefore is in strong tension with the current Planck bound.

\subsection{A simple model for $\nu_i \to \nu_4 \, \phi$ decays within the seesaw framework}\label{sec:Guidelines}

Here we will study a simple extension of the seesaw  mechanism able to open new active neutrino decay channels. In particular the \textit{2s} channel ($\nu_i \to \nu_4\, \phi)$ that can lead to a substantial relaxation of the cosmological neutrino mass bounds, see Figure~\ref{fig:2scase}. The presence of this decay mode necessarily requires to introduce new fields in order to account for the extra light sterile neutrino and the light boson in the decay final state. Of course, the new sector should not spoil the light neutrino mass generation mechanism and, in particular, the new sterile state can not present a large mixing with the active neutrinos in order to satisfy bounds on light sterile neutrinos. From the theoretical point of view, this means that a new extremely light (or massless) sterile neutrino state, essentially decoupled from the light neutrino masses, needs to be generated.

Our proposal is to enlarge the usual content of the seesaw model (i.e., three right-handed neutrinos, $N_R$) by adding an extra fermion singlet $S_L$,  and one complex scalar singlet $\Phi$. To avoid large couplings with the active sector and generate a massless sterile state, we also extend the symmetry of the model with a new global $U(1)_{X}$. Charging $S_L$ and $\Phi$ with opposite $U(1)_{X}$ charges and leaving the rest of the fields uncharged,
the only new terms allowed by the symmetry are of the form $y \Phi\overline{N}_{R}S_L$. Notice that, at tree level, the Majorana mass term $\overline{S^c_L}S_L$ and Yukawa coupling between $S_L$ and the active neutrinos are forced to be zero by the symmetry. $U(1)_{X}$ is dynamically broken when $\Phi$ takes a vacuum expectation value (vev) $v_\Phi$ and, since we chose $U(1)_{X}$ to be global, one of the two components of $\Phi$ will remain massless (the Majoron) while the other will have a mass of the order of $v_\Phi$. After symmetry breaking, the complete $7\times7$ neutrino mass matrix in the flavor basis is given by
\begin{equation}
 M_\nu=\begin{pmatrix}
0 & m_D & 0\\
m_D^t & M_R & \Lambda\\
0 & \Lambda^t &0
\end{pmatrix},
\label{eq:Mnu}
\end{equation}
where $m_D$ and $M_R$ are the $3\times 3$ Dirac and Majorana matrices already present in the seesaw model, while $\Lambda$ is a $3\times1$ matrix whose elements are 
given by  $\Lambda_a=y_a v_{\Phi}$. 
Diagonalizing the mass matrix under the assumption $\Lambda \ll m_D\ll M_R$, we obtain 
\begin{eqnarray}
 m_i&\simeq& m_D^2/M_R,\;\;\;\nu_\alpha\simeq \nu_i+\left(m_D/M_R\right)N_i-\left(\Lambda/m_D\right)\nu_4;
\nonumber\\
M_i&\simeq& M_R,\;\;\;\;\;\;\;\;\;N_R^c\simeq -(m_D/M_R)\nu_i+N_i;
\nonumber\\
m_4&=& 0,\;\;\;\;\;\;\;\;\;\;\;\;\;\;S_L\simeq (\Lambda/m_D)\nu_i+(\Lambda/M_R)N_i+\nu_4;
\end{eqnarray}
Notice that for the sake of simplicity, we have ignored the dependence on the PMNS leptonic mixing matrix $U$. The light neutrino masses and mixing are generated through the seesaw mechanism, and the mixing among the active and heavy states remains the same as in the standard seesaw. There is a zero mode associated to the new sterile state that presents a mixing with $\nu_i$ given by $\Lambda/m_D$ and with the heavy $N_i$ of $\Lambda/M_R$. The corrections from the new sector to the active neutrino mass matrix are remarkably suppressed, $\mathcal{O}(\Lambda^2/M_R)$, and thus negligible as long as $\Lambda \ll m_D$. Obviously, the same requirement leads to a small mixing between the sterile state and the active neutrinos, avoiding any potential tension with neutrino oscillations or BBN data. 
In particular the sterile-active neutrino mixing, $\theta_{\alpha 4}\sim \Lambda/m_D$, is best constrained by BBN data that roughly requires $|\theta_{\alpha 4}|^2 \lesssim 4\times 10^{-6} \,\text{eV}/\sqrt{|\Delta m_{4i}^2|}$ (see Appendix~\ref{sec:BBN_steriles}) and neutrino oscillation experiments that set $|\theta_{\alpha 4}|^2 \lesssim \mathcal{O}(0.01)$~\cite{Gariazzo:2015rra,Diaz:2019fwt,Boser:2019rta}. Remarkably, since in our model $m_{\nu_{4}}^2-m_{\nu_{i}}^2 < 0$, the sterile neutrino will resonantly be produced via mixing in the early universe, analogously to the MSW-effect and contrary to the widely studied case with $m_{\nu_{4}}^2-m_{\nu_{i}}^2 > 0$. Consequently, the BBN constraint that we have derived in our scenario is roughly two orders of magnitude more stringent than the one derived for the $m_{\nu_{4}}^2-m_{\nu_{i}}^2 > 0$ case~\cite{Gariazzo:2019gyi,Hagstotz:2020ukm,Hasegawa:2020ctq}\footnote{We thank Matheus Hostert for pointing this out and refer the reader to Appendix~\ref{sec:BBN_steriles} for further details.}. Given that $\sqrt{|\Delta m_{4i}^2|}=m_{\nu_{i}} \lesssim 1\,\text{eV}$, BBN thus gives a stronger bound than neutrino oscillation data: $\theta_{\alpha 4}\sim \Lambda/m_D \lesssim  10^{-3}$. Therefore, considering a mild hierarchy $\Lambda/m_D \lesssim 10^{-3}$ the model is in agreement with the constraints from the active neutrino sector.

The  structure of the mass matrix given in Equation~(\ref{eq:Mnu}), also called Minimal Extended Seesaw (MES),  has been proposed in order to  address the hints of light sterile neutrinos (but still heavier than the active ones) in some oscillation experiments, and may arise as a consequence of some flavour symmetries 
\cite{Chun:1995js, Barry:2011wb, Zhang:2011vh, Heeck:2012bz}, or a dark $U(1)$ \cite{Ballett:2019cqp}. Notice that the rank of the matrix is $6$, and therefore there is always a 
massless neutrino at tree level.
However typically the hierarchy $\Lambda \gg m_D $ is assumed, so the massless state is mainly active, while in our scenario $\Lambda \ll m_D $ leads to a mainly sterile massless neutrino.

In order to do a mapping to the parameters of the phenomenological Lagrangian, we write now the coupling between the $N_R$ and the new fields in the mass basis
\begin{eqnarray}
 \Delta  \mathcal{L}&=& \frac{y}{2}\Phi\overline{N_R}S_L+h.c \supset -\frac{y}{2}\frac{\Lambda}{M_R}\overline{\nu_i^c}(\sigma-i\gamma_5\phi)\nu_j-\frac{y}{2}\frac{m_D}{M_R}\overline{\nu_i^c}(\sigma-i\gamma_5\phi)\nu_4
 \nonumber\\
 &+&\frac{y}{2}\overline{N_i}(\sigma-i\gamma_5\phi)\nu_4 -\frac{y}{2}\frac{\Lambda}{m_D}\overline{N_i}(\sigma-i\gamma_5\phi)\nu_j+h.c,
  \label{eq:DeltaL}
\end{eqnarray}
where $\phi$ ($\sigma$) is the massless (massive) component of $\Phi$ and, for the sake of simplicity, we are again ignoring the dependence on the PMNS mixing matrix. The first line of the above equation give us the mapping to the phenomenological parameters from~\eqref{eq:Lag_scalar}:
\begin{align}\label{eq:mapping}
\lambda_{ij} \simeq  y \,\Lambda/M_R, \,\,\,\, 
\lambda_{i4}\simeq y\, m_D/M_R,\,\,\,\,\lambda_{44}\simeq0\,.
\end{align}
Notice that the hierarchy among the $\lambda$ couplings does not coincide with the naive expectation $\lambda_{ij}\sim\lambda_{i4}\theta\sim\lambda_{44}\theta^2$. In our model, the dominant coupling of the light neutrino (active and sterile) sector with the massless scalar field is given by $\lambda_{i4}$, and is controlled by the mixing between the active and heavy neutrinos. Thus, if the standard seesaw mechanism is at work, we have

\begin{equation}
\lambda_{i4}\simeq y\cdot 10^{-13}\sqrt{\left(\frac{m_{\nu}}{0.1\rm{eV}}\right)\left(\frac{10^{16}\rm{GeV}}{M_R}\right)}\,,
\label{eq:li4}
\end{equation}
matching the required range of values for the \textit{2s} channel ($\nu_i \to \nu_4\, \phi)$ to relax the $\sum m_\nu$ bound (see Figure~\ref{fig:2scase}), as we will discuss in more detail in a concrete neutrino mass model below. Note that, if the massive scalar $\sigma$ has a mass $ m_\sigma \gtrsim m_\nu$ it could potentially mediate 3-body decays. In Section~\ref{sec:results} we concluded that only the \textit{3s} channel is phenomenologically viable, and only provided that $h_{ij} \ll h_{44},\,h_{i4}$. However, in our neutrino decay model $h_{44} \simeq 0$ which precludes the possibility of realizing the \textit{3s} channel.

Our proposal provides an extra massless sterile neutrino, avoiding at the same time large corrections to the light neutrino sector, thanks to the addition of an extra $U(1)_{X}$ global symmetry. This prevents $S_L$ to couple to the active neutrinos and forbids the Majorana $S_L$ self coupling. A pertinent question arises in this context: is the model radiatively stable? The most relevant 1-loop effect is driven by the finite correction to the Majorana mass term $\overline{S^c_L}S_L$
\begin{equation}
 \delta M_{LL}\sim \frac{y^2}{(4\pi)^2} \frac{M_R}{M_R^2/m_{\sigma}^2-1} \log\left(\frac{M_R^2}{m_{\sigma}^2}\right)
 \simeq \frac{y^2}{(4\pi)^2}\frac{m_\sigma^2}{M_R}\log\left(\frac{M_R^2}{m_{\sigma}^2}\right),
\end{equation}
which is completely negligible for the parameter space that can alleviate the $\sum m_{\nu}$ cosmological bound tension via active neutrino decay. The model is thus stable at the radiative level, and can be easily embedded in any extension of the standard seesaw scenario.

Finally, it is worth to point out that although we have considered a global $U(1)_X$ symmetry, the ideas presented here can also be applied  to a feebly coupled 
gauged $U(1)_X$. In such a case, at least two extra fermion singlets $S_L$ with opposite $U(1)_X$ charges are required to cancel anomalies, while the corresponding gauge boson $X$ associated to the  $U(1)_X$ symmetry should be extremely light to allow for the 2-body decay $\nu_i \to \nu_4 \,X$.

\subsection{Neutrino masses and decays within a $U(1)_{\mu-\tau}$ symmetry}\label{sec:Minimalmutau}

In order to illustrate how the model independent results presented in Section~\ref{sec:parameterpsace} can be mapped to UV complete models, in this section we consider neutrino mass models based on a spontaneously broken $U(1)_{\mu-\tau}$ flavor symmetry~\cite{Choubey:2004hn,Araki:2012ip} which are potentially in tension with current Planck data. This type of flavor symmetries are easily anomaly free, given the Standard Model particle content~\cite{He:1990pn,He:1991qd}, and considering the right choice of charges for the right-handed neutrinos introduced in order to account for light neutrino masses. Neutrino mass models based on a $U(1)_{\mu-\tau}$ flavor symmetry correctly fit the light neutrino mass and mixing pattern observed in neutrino oscillation experiments. However, they generically predict $\sum m_\nu > 0.12\,\text{eV}$~\cite{Asai:2017ryy,Asai:2018ocx,Asai:2019ciz,Araki:2019rmw}. For the sake of concretness and illustration, we shall concentrate on the most minimal $U(1)_{\mu-\tau}$ SM extension featuring a SM singlet scalar $\varphi$ and 3 heavy sterile neutrinos with $U(1)_{\mu-\tau}$ charges $Q(\varphi) = +1$, $Q(N_e) = 0$, $Q(N_\mu) =+1$, and $Q(N_\tau) = -1$. In this set up, after electroweak and  $U(1)_{\mu-\tau}$ symmetry breaking, active neutrinos get a mass via the seesaw mechanism. The resulting pattern of masses and mixings within this framework has been analysed in detail in~\cite{Asai:2017ryy,Asai:2018ocx,Asai:2019ciz,Araki:2019rmw}. For this particular setting, only NO is consistent with the measured neutrino sector parameters, and the prediction for $\sum m_\nu $ is mainly dependent upon the value of $\theta_{23}$~\cite{Asai:2018ocx}. For values of $\theta_{23} = 48.3\degree {}^{+1.1\degree}_{-1.9\degree}$ (corresponding to the best fit $\pm$ $1\,\sigma$ from \texttt{NuFIT 4.1} without SK data\footnote{Inlcuding SK data does not change any of the conclusions and in particular one finds $0.16\,\text{eV} \! \! < \sum m_\nu < 0.30\,\text{eV}$ $(0.13\,\text{eV} \!\! < \sum m_\nu < 2.7\,\text{eV})$ at $1\sigma\,(3\sigma)$.}~\cite{Esteban:2018azc}) one finds that $0.17\,\text{eV}\!\! < \sum m_\nu < 0.47\,\text{eV}$, while for $\theta_{23} =  40.8\degree \! \to \! 51.3\degree$ (corresponding to the $3\,\sigma$ range) $ 0.13 < \sum m_\nu < 2.7\,\text{eV}$, where the upper bound is given by KATRIN. Therefore, the model is in strong tension with the Planck bound on $\sum m_\nu$. 

In what follows,  we will consider a weakly coupled realization of this minimal $U(1)_{\mu-\tau}$ neutrino mass model and highlight the allowed regions of parameter space for which neutrinos decay within cosmological timescales being able to relax the $\sum m_\nu$ bound from cosmology. We will first study the case in which $U(1)_{\mu-\tau}$ is promoted to a gauge symmetry where the $\nu_i \to \nu_j \, Z'$ decays become the most relevant channels. We then consider the case in which $U(1)_{\mu-\tau}$ is a global symmetry  (\textit{\`{a} la}~\cite{Valle:1983ua}) and the main decay channels are of the type $\nu_i \to \nu_j \, \phi$, where $\phi$ is the Goldstone mode associated to $\varphi$ upon $U(1)_{\mu-\tau}$ spontaneous symmetry breaking. 
Although from the  discussion in Section~\ref{sec:results} we expect that for these scenarios the neutrino mass bound can only be slightly alleviated, since the decay products contain always an active neutrino and they are quite heavy, we use them as an illustration of the matching of a UV completed model to our effective Lagrangian parameters.
Moreover, some of our conclusions are easily applicable to other realizations of the  $U(1)_{\mu-\tau}$ symmetry which lead to slightly lighter active neutrinos, where such decays could relax the tension with the Planck bound more significantly \cite{Araki:2019rmw}.
Finally, we will implement the simple extension described in the previous section in the context of the minimal $U(1)_{\mu-\tau}$ model, showing that  in this case the neutrino decay to only BSM species can naturally occur, ameliorating the neutrino mass bounds up to $\sum m_\nu \sim 1\,\text{eV}$.

\textbf{Gauge Case:} The $U(1)_{\mu-\tau}$ interaction is mediated by a $Z'$ boson. The corresponding $Z'$-neutrino interaction arises through the coupling with the $SU(2)$ lepton doublets and is given by
\begin{align}
\label{eq:Lflavorgauge}
\mathcal{L} \supset -Z'_{\mu} g_{\mu-\tau} \left( \bar{\nu}_{\mu} \gamma^{\mu} P_{L} \nu_{\mu} -  \bar{\nu}_{\tau} \gamma^{\alpha} P_{L} \nu_{\tau} \right) = -\frac{1}{2} Z'_{\mu} g_{\mu-\tau} \left[U_{\mu j}^\star U_{\mu i} - U_{\tau j}^\star U_{\tau i}\right]\, \bar{\nu}_i \gamma^\mu P_L  \nu_j+ \text{h.c.}  \,,
\end{align}
where in the second term we have explicitly written the Lagrangian in the mass basis as in~\eqref{eq:Lag_vector}. 
In order to illustrate the strength of the couplings that play a role in the relevant decay channels, we map the parameters in Equation~(\ref{eq:Lflavorgauge}) to the phenomenological parameters of Equation~(\ref{eq:Lag_vector}), introducing the bestfit values of the PMNS mixing parameters from \texttt{NuFIT 4.1}~\cite{Esteban:2018azc} for NO:
 \begin{align}
\label{eq:Lflavorgauge_map}
|g_{12}^L|/g_{\mu-\tau} &\simeq 0.14 \,,\qquad \qquad \, |g_{13}^L|/g_{\mu-\tau} \simeq  0.54 \,,\qquad \qquad \,\,\,\, |g_{23}^L|/g_{\mu-\tau} \simeq 0.82 \,,\\
g_{11}^L/g_{\mu-\tau} &\simeq -0.14\,,\qquad \qquad g_{22}^L/g_{\mu-\tau} \simeq 0.023 \,,\qquad \qquad \,\,\, \,g_{33}^L/g_{\mu-\tau} \simeq 0.11\,, \nonumber
\end{align}
where the $g_{ij}^L$ couplings with $i\neq j$ lead to invisible neutrino decays of the type $\nu_i \to \nu_j \, Z'$ provided $m_{Z'} < m_{\nu_i} - m_{\nu_j}$. In the model under consideration, neutrino oscillation data can only be fitted for NO and approximately degenerate light neutrinos. Thus, given the values of the couplings above, $\nu_3 \to \nu_2 \, Z'$ and  $\nu_3 \to \nu_1 \, Z'$ decays occur at roughly the same rate. 

As can be seen from Table~\ref{tab:rates}, the total rate of invisible neutrino decays depends only on the light neutrino masses, $g_{\mu-\tau}$ and $m_{Z'}$ for which we assume $m_{Z'} \ll m_{\nu}$ (see e.g.~\cite{Williams:2011qb} for theoretical arguments motivating why this could be the case). Using Equation~(\ref{eq:Lflavorgauge_map}), the results shown in Figure~\ref{fig:2acase} (right panels) can be easily mapped to the present model. As also recently pointed out in~\cite{Dror:2020fbh}, we find that bounds on relativistically decaying invisible neutrino decays represent the most stringent constraint on the considered parameter space $10^{-10}\,\text{eV}\lesssim m_{Z'} \lesssim m_\nu$\footnote{See~\cite{Dror:2020fbh} for bounds applicable in the region $m_{Z'}< 10^{-10}\,\text{eV}$, and \cite{Escudero:2019gzq} for constraints within the $m_{Z'} > m_\nu$ case.}. In particular, the vacuum expectation value of the scalar field should fulfil $v_{\mu-\tau} \equiv m_{Z'}/g_{\mu-\tau} > 1\,\text{GeV}$ for $m_{Z'} \ll m_\nu$. We also find that for $v_{\mu-\tau} < 25\,\text{TeV}$  neutrinos have a lifetime $\tau_\nu < t_U$.
Therefore, for $1\,\text{GeV}<v_{\mu-\tau} < 25\,\text{TeV}$ and $m_{Z' }\ll m_\nu$, in this model neutrino decays are able to slightly relax the tension between the model prediction for $\sum m_{\nu}$ and the Planck bound.

\textbf{Global Case:} If $U(1)_{\mu-\tau}$ is a global symmetry, the CP-odd component of the $\varphi$ scalar field becomes one of the final states in the 2-body decay channel \textit{2a}\footnote{The CP-even component obtains a mass $\sim v_{\mu-\tau}$ and hence could only participate in $3-$body decays which are, however, not phenomenologically viable, see Section~\ref{sec:results}. }. 
Upon spontaneous symmetry breaking of $U(1)_{\mu-\tau}$, the CP-odd component of the scalar field, i.e. $\phi$, becomes a Goldstone boson. This scalar $\phi$ can be seen as a flavorful massless majoron with $m_\phi = 0$. Quantum gravity is expected to break all global symmetries~\cite{Kallosh:1995hi} leading to  $m_\phi \neq 0$. However, given that  the exact way in which gravity breaks global symmetries is highly dependent upon the unknown nature of the gravitational theory at the Planck scale,
we shall assume that $m_\phi \ll m_\nu$. 

In this setting, $h_{ij} =0$ in the effective Lagrangangian given in Equation~(\ref{eq:Lag_scalar}) since $\phi$ is a CP-odd scalar. Furthermore, the effective coupling $\lambda_{ij}$, which controls the $\nu_i \to \nu_j \, \phi$ decay, is generated via mixing between the heavy sterile and active neutrinos through the interaction $\mathcal{L} \supset -\mathcal{Y}_{e \mu} \varphi \overline{N_{e}^c} N_{\mu} -  \mathcal{Y}_{e \tau} \varphi^{\dagger} \overline{N_{e}^c} N_{\tau}$, where $N_\alpha$ are the heavy right handed neutrinos and $\mathcal{Y}_{e \mu}$ and $\mathcal{Y}_{e \tau}$ are Yukawa couplings. Taking into account the canonical seesaw scaling of the mixing between the heavy states and the active neutrinos $\theta_{N\nu} \simeq \sqrt{m_\nu/M_N}$, and assuming that $\mathcal{Y}_{e \mu}\sim \mathcal{Y}_{e \tau}$, we can estimate
\begin{align}\label{eq:Global_mutau_rel}
\lambda_{ij} \simeq \mathcal{Y}_{e \mu} \theta_{N_{e} \nu_i} \theta_{N_\mu \nu_j} + \mathcal{Y}_{e \tau} \theta_{N_{e} \nu_i} \theta_{N_\tau \nu_j} \simeq  \frac{\sqrt{m_{\nu_{i}}m_{\nu_{j}}}}{v_{\mu-\tau}} \simeq \frac{m_{\nu_{i}}}{v_{\mu-\tau}} = 10^{-11} \frac{m_{\nu_{i}}}{0.1~\text{eV}} \frac{10~\text{GeV}}{v_{\mu-\tau}}\,.
\end{align}
To be in accordance with the observed light neutrino mass spectrum and mixing,
the contribution to the right-handed neutrino Majorana masses from the coupling to the scalar should be of the same order of the mass terms allowed by the 
$U(1)_{\mu-\tau}$ symmetry, $M_{ee}  \overline{N_{e}^c} N_{e}$ and $M_{\mu\tau}  \overline{N_{\mu}^c} N_{\tau}$, as we have considered in the above equation. 

Using Equation~\eqref{eq:Global_mutau_rel}, the allowed parameter space shown in the left panels of Figure~\ref{fig:2acase}, maps to $v_{\mu-\tau} > 1\,\text{GeV}$ (neutrinos not spoiling Planck CMB observations) and corresponds to $v_{\mu-\tau} \lesssim 10\,\text{TeV}$ (active neutrino lifetime satisfying $\tau_\nu < t_U$). Interestingly, the relevant range for the energy scale of the model lies around the electroweak scale, $v_{\mu-\tau} \simeq v_H$. Hence, similarly to the gauge case, such energy scales point towards a low-scale realization of the seesaw mechanism at $\mathcal{O}(\text{TeV})$.

\textbf{Decay to sterile neutrinos:} 
Now we propose a particular realization of the scenario described in Section~\ref{sec:Guidelines} within the $U(1)_{\mu-\tau}$ minimal model (with either gauge or global symmetry), 
that opens the possibility of active neutrinos decaying to a lighter, mainly sterile state.
Recall that the new species are one extra fermion singlet $S_L$ and one complex scalar singlet $\Phi$, which  have opposite charges under a global $U(1)_{X}$ symmetry.
We assign zero $U(1)_{\mu-\tau}$ charge to $S_L$, in order to keep the anomaly cancellation, while $\Phi$ could carry $U(1)_{\mu-\tau}$ charge. 
 There are only three viable choices for the $U(1)_{\mu-\tau}$ charge of $\Phi$ that allow the required  interaction terms $y \Phi\overline{N}_{R}S_L$:  $Q(\Phi) = 0,\,\pm 1$. For $Q(\Phi) = 0$, we have $\Lambda^t=\left\{\Lambda_e,0,0\right\}$ while for $Q(\Phi)=\pm1$, $\Lambda^t=\left\{0,\Lambda_\mu,\Lambda_\tau\right\}$, 
where  $\Lambda_\alpha=y_\alpha v_{\Phi}$.

\begin{figure}[t]
\centering
\hspace{-0.4cm} \includegraphics[width=0.75\textwidth]{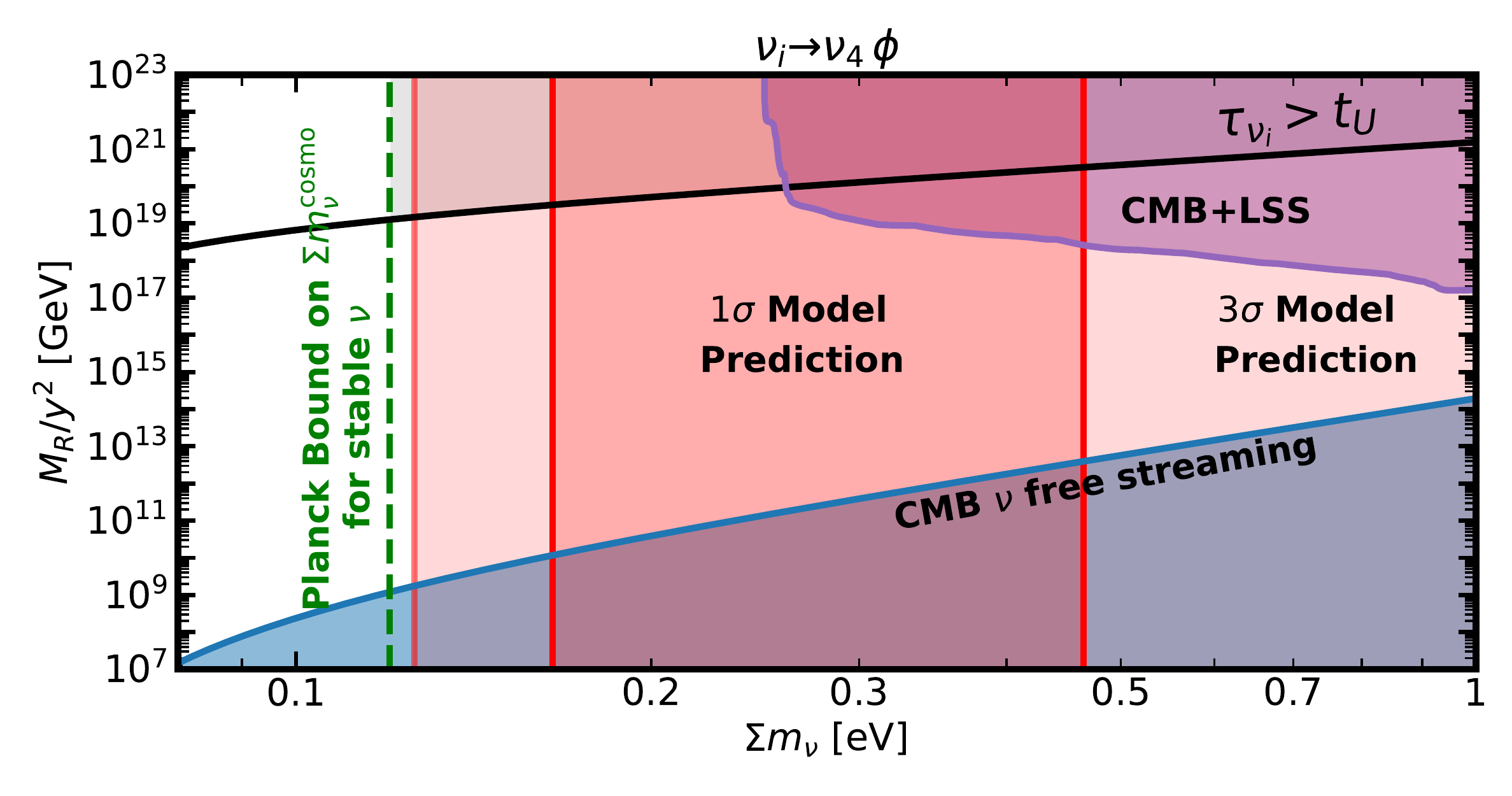}
\vspace{-0.5cm}
\caption{Parameter space for our proposed model realizing $\nu_i \to \nu_4 \,\phi$ decays within a $U(1)_{\mu-\tau}$ symmetry. We highlight in red the prediction from the $U(1)_{\mu-\tau}$ mass model required in order to fit at $1\sigma$ and $3\sigma$ neutrino oscillation data. 
}
\label{fig:case2smutau}
\end{figure}

From the mapping given in Equation~(\ref{eq:mapping}), we see that the coupling driving the $\nu_i \to \nu_4 \,\phi$ decay is given by $\lambda_{i4}\simeq y\, m_D/M_R$, which is linked to the light neutrino mass scale via Equation~(\ref{eq:li4}) since the seesaw mechanism is at work. In Figure~\ref{fig:case2smutau} we show the relevant parameter space in our scenario, as well as the theoretical prediction for $\sum m_\nu$. We see that our extension can evade the bound on $\sum m_\nu$ from cosmology across a large region of the parameter space. In particular, for $y\sim 1$, our model can naturally evade the bound on $\sum m_{\nu}$ for
$10^{10}\,\text{GeV}\lesssim M_R \lesssim 10^{19}\, \text{GeV}$, which points to the canonical seesaw. Such a high scale for $M_R$ avoids any potential constrain from the early universe on the new Lagrangian terms given in the second line of Equation~(\ref{eq:DeltaL}). Any sizeable contribution from the heavy $N_R$ to low energy particle physics experiments would also be suppressed. Notice that lighter $M_R$ scales can also be easily realized by simply considering smaller values of the $y$ coupling. 

In summary, our model provides a natural framework to alleviate the $\sum m_{\nu}$ cosmological bound via $\nu_i \to \nu_4 \,\phi$ decays (\textit{2s}). In this scenario, light neutrino masses are generated via the canonical seesaw mechanism and, remarkably, are not spoiled by the presence of the massless sterile neutrino that mixes with the active sector. This is possible just invoking a hierarchy between the vev of the new scalar field and the Higgs vev of $v_\Phi/v_H \lesssim 10^{-3}$.

\section{Summary, Conclusions and Outlook}\label{sec:Summary}
The cosmological bounds on the absolute neutrino mass scale within $\Lambda$CDM are very stringent. In particular, the Planck collaboration reports $\sum m_\nu < 0.12\,\text{eV}$ at 95\% CL~\cite{Aghanim:2018eyx}. This bound, taken at face value, is more than one order of magnitude stronger than the one reported by the KATRIN experiment~\cite{Aker:2019uuj}, which can be reinterpreted as $\sum m_\nu < 2.7\,\text{eV}$ at 95 \% CL. Even though cosmological neutrino mass bounds are cosmological model dependent, in typical extensions of $\Lambda$CDM the bound is only slightly relaxed to $\sum m_\nu \lesssim 0.2\,\text{eV}$, see Table~\ref{tab:mnu_constraints}. Furthermore, the constraints depend on the data set under consideration, but the ones driven by Planck CMB observations are quite robust (see Section~\ref{sec:intro}). This puts several neutrino mass models in trouble since a large number of them predict $\sum m_\nu \gtrsim 0.12\,\text{eV}$.

In this context, invisible neutrino decays -- namely, those in which the decay products do not interact electromagnetically -- represent a particle physics avenue to relax the potential tension between theoretical model predictions and cosmology. Indeed, the $\sum m_\nu$ bound can be loosen up to $\sum m_\nu \lesssim 1\,\text{eV}$ for neutrinos decaying on cosmological timescales~\cite{Chacko:2019nej}.
Motivated by this, in this paper we have carried out a comprehensive analysis of models that can accomplish neutrinos decaying invisibly within cosmological timescales. For this purpose, in Section~\ref{sec:Taxonomy}, we have presented a taxonomy of all possible invisible neutrino decays as highlighted in Figure~\ref{fig:NeutrinoDecayDiagrams}.
At the phenomenological level, we have then written down general effective Lagrangians which describe these decay channels in Equations~\eqref{eq:Lag_scalar} and~\eqref{eq:Lag_vector}. In Section~\ref{sec:Results}, after reviewing the most relevant constraints on invisible neutrino decays,
we have identified the parameter space compatible with current limits that can lead to a
relaxation of the cosmological $\sum m_\nu$ bound for each possible decay channel.

\noindent The main findings of our phenomenological study are:\vspace{-0.1cm}
\begin{itemize}[leftmargin=0.5cm,itemsep=0.4pt]

\item \textit{2s} invisible decay channels (see Figure~\ref{fig:NeutrinoDecayDiagrams}) of the type of $\nu_i \to \nu_4  \,  \phi$ with $m_\phi = m_{\nu_4} = 0$, can produce a significant relaxation of the neutrino mass bounds up to $\sum m_\nu \lesssim 1\,\text{eV}$ while satisfying all known present constraints, for $\nu_i\!-\!\nu_4\!-\!\phi$ interaction strengths given by $10^{-15} \lesssim \lambda \lesssim 10^{-10}$ -- see Figure~\ref{fig:2scase}. An analogous result is obtained for the $\nu_i \to \nu_4\,{Z'}$ decay scenario, provided that $1 \,\text{GeV} \lesssim m_{Z'}/g_{i4}^{L} \lesssim 25 \,\text{TeV} $ and $m_{Z'} \ll m_{\nu_i}$.

\item \textit{2a} neutrino decays driven by the $\nu_i \to \nu_j \, \phi$ modes, with $m_\phi \ll m_{\nu_i}$, can only be able to slightly relax current cosmological bounds while simultaneously being compatible with CMB constraints on invisible neutrino decays -- see Figure~\ref{fig:2acase}. This requires $\nu_i\!-\!\nu_j\!-\!\phi$ coupling strengths in the $10^{-15} \lesssim \lambda \lesssim 10^{-10}$ range. We reach a similar conclusion for $\nu_i \to \nu_4 \, {Z'}$ decays provided that $1 \,\text{GeV} \lesssim m_{Z'}/g_{ij}^{L} < 25 \,\text{TeV} $ with $m_{Z'} \ll m_{\nu_i}$.

\item Due to a combination of terrestrial, astrophysical, and cosmological constraints on sub-MeV neutrinophilic bosons, 3-body neutrino decays involving at least one active neutrino in the final state (cases \textit{3a0}, \textit{3a1} and \textit{3a2} in Figure~\ref{fig:NeutrinoDecayDiagrams}) cannot lead to $\tau_\nu < t_U$.
The only topology that remains phenomenologically viable involves only sterile neutrinos as final states and is given by $\nu_i \to \nu_4 \bar{\nu}_4 \nu_4$ (\textit{3s}  in Figure~\ref{fig:NeutrinoDecayDiagrams}). However, this requires a particular hierarchy among the active-sterile-mediator couplings, $\lambda_{44} \,, \lambda_{i4} \gg \lambda_{ij}$, and the mediating particle to be lighter than $\sim10\,\text{keV}$.

\end{itemize}

In light of the above, we believe that near future experiments provide further motivation to study the phenomenology of neutrinos decaying within cosmological timescales. Indeed, the upcoming galaxy surveys \texttt{DESI}~\cite{Aghamousa:2016zmz} and \texttt{Euclid}~\cite{Amendola:2016saw} will have a $1\sigma$-sensitivity of $\sigma (\sum m_\nu) \simeq 0.02\,\text{eV}$ and are therefore expected to detect neutrino masses in cosmology within the next 5-10 years if the cosmological model describing our Universe is $\Lambda$CDM.
If such a detection is made, it will lead to a bound on the neutrino lifetime $\tau_\nu \gtrsim t_U$~\cite{Serpico:2007pt,Chacko:2020hmh} and our study shows that wide regions of parameter space in decaying neutrino models would be excluded.
On the other hand, the possibility that both \texttt{DESI} and \texttt{Euclid} will not measure $\sum m_\nu$, reporting only upper bounds, is even more interesting.
Were that to be the case, invisible neutrino decays with $\tau_\nu < t_U$ will become a prime scenario to understand such measurements. In our analysis we have already identified the minimal particle content and relevant parameter space required to implement such decays.
Furthermore, in the next $\sim$5 years, the KATRIN experiment is expected to reach a 90\% CL sensitivity to $m_{\bar{\nu}_e}$ of 0.2 eV~\cite{Osipowicz:2001sq,Aker:2019uuj}. If KATRIN reports a neutrino detection\footnote{Note that the data for the second neutrino mass campaign has already been taken and its analysis is expected to yield a sensitivity on $m_{\bar{\nu}_e}$ of 0.7 eV at 90\% CL, see the \href{https://indico.fnal.gov/event/43209/contributions/187858/attachments/129656/158582/Day1_Talk9_Mertens.m4v}{talk} of Susanne Mertens at Neutrino 2020.} and only upper bounds on $\sum m_\nu$ arise from cosmology, invisible neutrino decays will become a very well motivated possibility to reconcile both sets of data.

From the model building perspective,  in Section~\ref{sec:ModelBuilding}, we have proposed a simple extension of the seesaw scenario which contains a (mainly) sterile neutrino lighter than the (mainly) active ones, together with a massless Goldstone boson $\phi$ from a spontaneously broken $U(1)_X$ global symmetry, so that invisible neutrino decays of the type $\nu_i \to \nu_4 \, \phi$ can be cosmologically relevant.
Remarkably, in this set up the mixing between active neutrinos and $\nu_4$ are naturally small and the new sterile state does not spoil the standard seesaw mechanism for generating active neutrino masses, so it could be easily embedded in more complicated models within this framework.

As an illustration, we have considered minimal neutrino mass models based on a spontaneously broken $U(1)_{\mu-\tau}$ flavor symmetry, which  generically predict
$\sum m_\nu > 0.12 \,\text{eV}$ and are thus  in tension with the current cosmological bound on neutrino masses. We have found that in feebly coupled realizations of such models
 neutrino decays are possible, and are in somewhat better agreement with cosmological observations. Our analysis shows that the cosmological
$\sum m_\nu$ bounds are slightly relaxed for $U(1)_{\mu-\tau}$ symmetry breaking scales $v_{\mu-\tau} \lesssim 25\,\text{TeV}$, while $v_{\mu-\tau} > 1\,\text{GeV}$ is required in order to satisfy current CMB constraints on invisible neutrino decays.

Furthermore, we have implemented the minimal extension of the seesaw model described above within the context of the $U(1)_{\mu-\tau}$ flavor symmetry. In this case the tension between the prediction for $\sum m_\nu$ from $U(1)_{\mu-\tau}$ models and cosmology can be fully evaded, as it is shown in Figure~\ref{fig:case2smutau}. Within this scenario, $\tau_\nu < t_U$ can be associated to energy scales $1\,\text{GeV}\lesssim v_{\mu-\tau} \lesssim 10^{19}\,\text{GeV}$. Such energy scales  suggest that UV complete models featuring invisible neutrino decay modes could be linked to phenomena such as baryogenesis, leptogenesis, and dark matter. Although beyond the scope of this paper, we think that it would be very interesting to pursue such avenues and we leave it for future work.

\textbf{Note Added in Proof:}
When this paper was accepted for publication and in proof, Ref.~\cite{Barenboim:2020vrr} appeared on the ArXiv, revisiting the process of invisible neutrino decay in the early Universe. Even though the new analysis from Ref.~\cite{Barenboim:2020vrr} does not affect significantly our results and conclusions, a detailed discussion of the potential implications for our study can be found in Appendix~\ref{sec:comments}.

\vspace{-0.1cm}

\section*{Acknowledgments}
We thank Pilar Hern\'andez, Matheus Hostert, and Sergio Palomares-Ruiz for useful discussions. JLP thanks Nemo for illuminating discussions. 
ME is supported by the European Research Council under the European Union's Horizon 2020 program (ERC Grant Agreement No 648680 DARKHORIZONS).
The work of SS received the support of a fellowship from ”la Caixa” Foundation (ID 100010434) with fellowship code LCF/BQ/DI19/11730034.
This project has received funding from the European Union's Horizon 2020 research and innovation programme under the Marie Sk\l{}odowska-Curie grant agreements 674896 and 690575, from the Spanish MINECO under Grant FPA2017-85985-P, and from Generalitat Valenciana through the ``plan GenT'' program (CIDEGENT/2018/019) and the grant PROMETEO/2019/083.

\bibliographystyle{JHEP}
\bibliography{refs}{}

 \appendix
\section{Appendices}
\label{sec:appendix}
\subsection{Invisible neutrino decay rates}\label{sec:appendix_rates}

We devote this appendix to the calculation of the decay rates that we have considered in the phenomenological analysis performed in Section~\ref{sec:results}. 
The neutrino decay rates are dictated by the two effective Lagrangians presented in Equations~(\ref{eq:Lag_scalar}) and (\ref{eq:Lag_vector}).
We will assume the sterile neutrino to be (approximately) massless ($m_{\nu_4} \simeq 0$). A non-negligible $m_{\nu_4}$ close to the mass of the decaying neutrino would further reduce the parameter space due to a phase space suppression similar to the one associated to the decay channels involving active neutrinos in the final state. In any case, the formulae for the decay rates involving active neutrinos in the final state are applicable to the sterile neutrino case with $m_{\nu_{4}} \lesssim m_{\nu_{i}}$ doing a straightforward mapping of the coupling constants. In such a case, the sterile neutrino mass would be a free parameter at the phenomenological level, contrary to the light neutrino masses which are correlated because the light neutrino squared mass differences have been measured in neutrino oscillation experiments.

\subsubsection*{Two Body}

\textbf{Case $2a$:}
The $\nu_{i} \to \nu_{j} \, \phi$  decay rate is given by
\begin{equation}
\begin{split}
    \Gamma_{\nu_{i}\to\nu_{j}\,\phi} &= \frac{1}{16\pi m_{\nu_{i}}} \sqrt{\left[m_{\nu_{i}}^2-\left(m_{\nu_{j}}+m_{\phi}\right)^2\right] \left[m_{\nu_{i}}^2 - \left(m_{\nu_{j}}-m_{\phi}\right)^2\right]} \\
    &\times \left\{ 
    |h_{ij}|^2 \left[\left(1+\frac{m_{\nu_{j}}}{m_{\nu_{i}}}\right)^2 - \frac{m_{\phi}^2}{m_{\nu_{i}}^2}\right] + 
    |\lambda_{ij}|^2 \left[ \left(1-\frac{m_{\nu_{j}}}{m_{\nu_{i}}}\right)^2 - \frac{m_{\phi}^2}{m_{\nu_{i}}^2}\right] 
    \right\} \,.
\end{split}
\label{eq:rate2a}
\end{equation}
The $m_{\phi} \ll m_{\nu_{i}}$ limit can be trivially obtained by setting $m_{\phi} = 0$.

For a massive $Z'$ the decay rate is given by
\begin{equation}
\label{appeq:twobodyzprime}
    \begin{split}
        \Gamma_{\nu_{i}\to\nu_{j}\,Z'} &=\frac{|g^{L}_{ij}|^2 m_{\nu_{i}}}{32\pi}
        \left\{1+\frac{m_{\nu_{j}}^2}{m_{\nu_{i}}^2}-2\frac{m_{Z'}^2}{m_{\nu_{i}}^2}+\frac{(m_{\nu_{i}}^2-m_{\nu_{j}}^2)^2}{m_{\nu_{i}}^2m_{Z'}^2}\right\}  \\
        &\times \sqrt{\left(1-\frac{(m_{\nu_{j}}-m_{Z'})^2}{m_{\nu_{i}}^2}\right) \left(1-\frac{(m_{\nu_{j}}+m_{Z'})^2)}{m_{\nu_{i}}^2}\right)}\,.
    \end{split}
\end{equation}
\\\textbf{Case $2s$:}
The corresponding two decay rates can be simply obtained by setting $m_{\nu_{j}}\to 0$ in case $2a$ and performing the following replacement
\begin{align}
(h_{ij},\lambda_{ij}) \to (h_{i4},\lambda_{i4})\,,~~g^{L}_{ij} \to g^{L}_{i4}\,,
\end{align}
in the above equations.

\subsubsection*{Three Body}
\textbf{Case 3a0}:
No simple analytic expression neither for scalar nor vector boson mediator can be obtained. However, this case is phenomenologically irrelevant.\\
\textbf{Case 3a1}:
The result for the decay $\nu_{i} \to \nu_{j} \, \phi^{*} \to \nu_{j} \bar{\nu_{4}} \nu_{4}$ is
\begin{equation}
\begin{split}
    \Gamma &= \frac{|h_{44}|^2+|\lambda_{44}|^2}{1536 \pi ^3m_{\phi}^{4}} m_{\nu_{i}}^5 
\left\{\left(|h_{ij}|^2+|\lambda_{ij}|^2\right)
    \left[1- 8\frac{m_{\nu_{j}}^2}{m_{\nu_{i}}^2}+24\frac{m_{\nu_{j}}^4}{m_{\nu_{i}}^4}\log\left(\frac{m_{\nu_{i}}}{m_{\nu_{j}}}\right)+ 8\frac{m_{\nu_{j}}^6}{m_{\nu_{i}}^6}-\frac{m_{\nu_{j}}^8}{m_{\nu_{i}}^8} \right]\right.
    \\
    &\left. +4\left(|h_{ij}|^2-|\lambda_{ij}|^2\right)\frac{m_{\nu_{j}}}{m_{\nu_{i}}}\left[
    1+9\frac{m_{\nu_{j}}^2}{m_{\nu_{i}}^2}-12\frac{m_{\nu_{j}}^2}{m_{\nu_{i}}^2}\left(1+\frac{m_{\nu_{j}}^2}{m_{\nu_{i}}^2}\right)\log\left(\frac{m_{\nu_{i}}}{m_{\nu_{j}}}\right)-9\frac{m_{\nu_{j}}^4}{m_{\nu_{i}}^4}-\frac{m_{\nu_{j}}^6}{m_{\nu_{i}}^6}\right]\right\}.    
    \end{split}
\end{equation}

\noindent The rate for the same decay but mediated by a $Z'$ reads
\begin{equation}
\label{appeq:case3a1z}
    \begin{split}
        \Gamma &= \frac{m_{\nu_{i}}^5}{1536 \pi^3m_{Z'}^4}|g_{ij}^{L}|^2 \left(|g_{44}^{L}|^2+|g_{44}^{R}|^2\right) 
        \left\{1- 8\frac{m_{\nu_{j}}^2}{m_{\nu_{i}}^2}+24\frac{m_{\nu_{j}}^4}{m_{\nu_{i}}^4}\log\left(\frac{m_{\nu_{i}}}{m_{\nu_{j}}}\right)+ 8\frac{m_{\nu_{j}}^6}{m_{\nu_{i}}^6}-\frac{m_{\nu_{j}}^8}{m_{\nu_{i}}^8} \right\}\,.
    \end{split}
\end{equation}
\\\textbf{Case 3a2}:
The $\nu_{i} \to \nu_{4} \, \phi^{*} \to \nu_{4} \bar{\nu}_4 \nu_{j}$ decay rate is given by
\begin{align}
  \Gamma =  \frac{ \left(|h_{i4}|^2+|\lambda_{i4}|^2 \right) \left(|h_{j4}|^2+|\lambda_{j4}|^2 \right) }{1536 \pi ^3m_{\phi}^{4}} m_{\nu_{i}}^5   \left[1- 8\frac{m_{\nu_{j}}^2}{m_{\nu_{i}}^2}+24\frac{m_{\nu_{j}}^4}{m_{\nu_{i}}^4}\log\left(\frac{m_{\nu_{i}}}{m_{\nu_{j}}}\right)+ 8\frac{m_{\nu_{j}}^6}{m_{\nu_{i}}^6}-\frac{m_{\nu_{j}}^8}{m_{\nu_{i}}^8} \right]\  \,.
\end{align}
The expression for the $\nu_{i} \to \nu_{4} \, Z'^{*} \to \nu_{4} \bar{\nu}_4 \nu_{j}$ decay rate is given by Equation~\eqref{appeq:case3a1z} performing the following mapping
\begin{align}
    g_{ij}^{L} \to g_{i4}^{L}, ~|g_{44}^{L}|^2+|g_{44}^{R}|^2 \to |g_{j4}^{L}|^2\,.
\end{align}
\\\textbf{Case 3s}:
The result for the decay $\nu_{i} \to \nu_{4} \, \phi^{*} \to \nu_{4} \bar{\nu}_4 \nu_{4}$ is
\begin{align}
    \Gamma = \frac{m_{\nu_{i}}^5}{1536 \pi^3 m_{\phi}^4} (|h_{i4}|^2+|\lambda_{i4}|^2)(|h_{44}|^2+|\lambda_{44}|^2)\,,
\end{align}
and for the $Z'$ mediated case
\begin{align}
  \Gamma=  \frac{m_{\nu_{i}}^5}{1536\pi^3m_{Z'}^4} |g_{ij}^{L}|^2(|g_{44}^{L}|^2+|g_{44}^{R}|^2)\,.
\end{align}

\newpage
\subsection{Complementary bounds on invisible neutrino decay rates}\label{sec:appendix_astrolab}
In this Appendix we will briefly review the bounds on the invisible neutrino decay lifetime that can be extracted from astrophysics and laboratory experiments, which are complementary to the stronger constraints from cosmology presented in Section~\ref{sec:parameterpsace}. 

\textbf{Astrophysical Bounds.} The observation of neutrinos from the supernova SN1987A constrains the neutrino lifetime to be $\tau_\nu/m_\nu > 5.7\times 10^{5}\,\text{s}\,\text{eV}^{-1} $~\cite{Frieman:1987as}, provided all neutrinos decay into very light and non-interacting BSM species. Given the low number of observed neutrinos from SN1987A, there are no solid bounds from SN1987A on decay modes that involve only one neutrino decay or that count with active neutrinos in the final state~\cite{Ando:2003ie,Ando:2004qe}. However, given the sensitivity of current neutrino detectors, data from the next galactic supernova is expected to probe decay modes of the type $\nu_i \to \nu_j \phi$ with rates $\tau_\nu \lesssim 10^{7}\,\text{s}\,\text{eV}^{-1}$~\cite{Ando:2003ie,Ando:2004qe,deGouvea:2019goq}. Similarly, neutrino decays can be searched for with neutrino telescopes~\cite{Bustamante:2016ciw,Denton:2018aml,Abdullahi:2020rge}. A very recent analysis of IceCube data~\cite{Bustamante:2020niz} shows that $\tau_{\nu_{1,\,2}} > (1-3)\times 10^{-3}\,\text{s}\,\text{eV}^{-1}$ at 90\% CL for IO. For NO no bound has been reported yet, but similar lifetimes are expected to be probed in the years to come~\cite{Bustamante:2020niz}.
Furthermore, neutrino telescopes are expected to reach sensitivities to neutrino decays up to $\tau_\nu/m_\nu \lesssim 10^{3}\,\text{s}\,\text{eV}^{-1} $ from astrophysical sources~\cite{Bustamante:2016ciw,Denton:2018aml,Abdullahi:2020rge,Bustamante:2020niz}.

\textbf{Laboratory Bounds.}  Neutrino oscillations are also sensitive to neutrino decays. The strongest bounds arise from Solar neutrino experiments that limit the neutrino lifetime to be $\tau_{\nu_i}/m_{\nu_i} \gtrsim (10^{-5}-10^{-3})\,\text{s}\,\text{eV}^{-1}$~\cite{Beacom:2002cb,Aharmim:2018fme,Berryman:2014qha,Funcke:2019grs}. Atmospheric and long-baseline neutrino experiments lead to weaker but complementary constraints~\cite{GonzalezGarcia:2008ru,Gomes:2014yua,Gago:2017zzy,Choubey:2018cfz}. In particular, the combined analysis of Super-Kamiokande, K2K and MINOS data provides the bound: $\tau_{\nu_3}/m_{\nu_3} \gtrsim 2.9\times 10^{-10}\,\text{s} \, \text{eV}^{-1} $~\cite{GonzalezGarcia:2008ru}. Upcoming neutrino oscillation experiments as DUNE, JUNO or ORCA, are expected to slightly improve the present bound on $\tau_{\nu_3}$~\cite{Abrahao:2015rba,Coloma:2017zpg,Choubey:2017dyu,Choubey:2017eyg,deSalas:2018kri,Ascencio-Sosa:2018lbk,Tang:2018rer,Ghoshal:2020hyo}.

Even though laboratory and astrophysical bounds on the neutrino lifetime are orders of magnitude weaker than those derived from cosmological observations, the latter can be avoided if neutrinos decay today but not in the early Universe~\cite{Dvali:2016uhn}. This can potentially occur if the neutrino masses and/or couplings with additional BSM species are time dependent. We consider that this possibility is not generic and we have not contemplated it in this work. 

\subsection{Allowed parameter space for individual $\nu_{i}\to\nu_{j}\,\phi/Z'$ decays}\label{sec:individual}

\begin{figure}[t]
\centering
\begin{tabular}{cc}
\hspace{-0.5cm}
\includegraphics[width=0.5\textwidth]{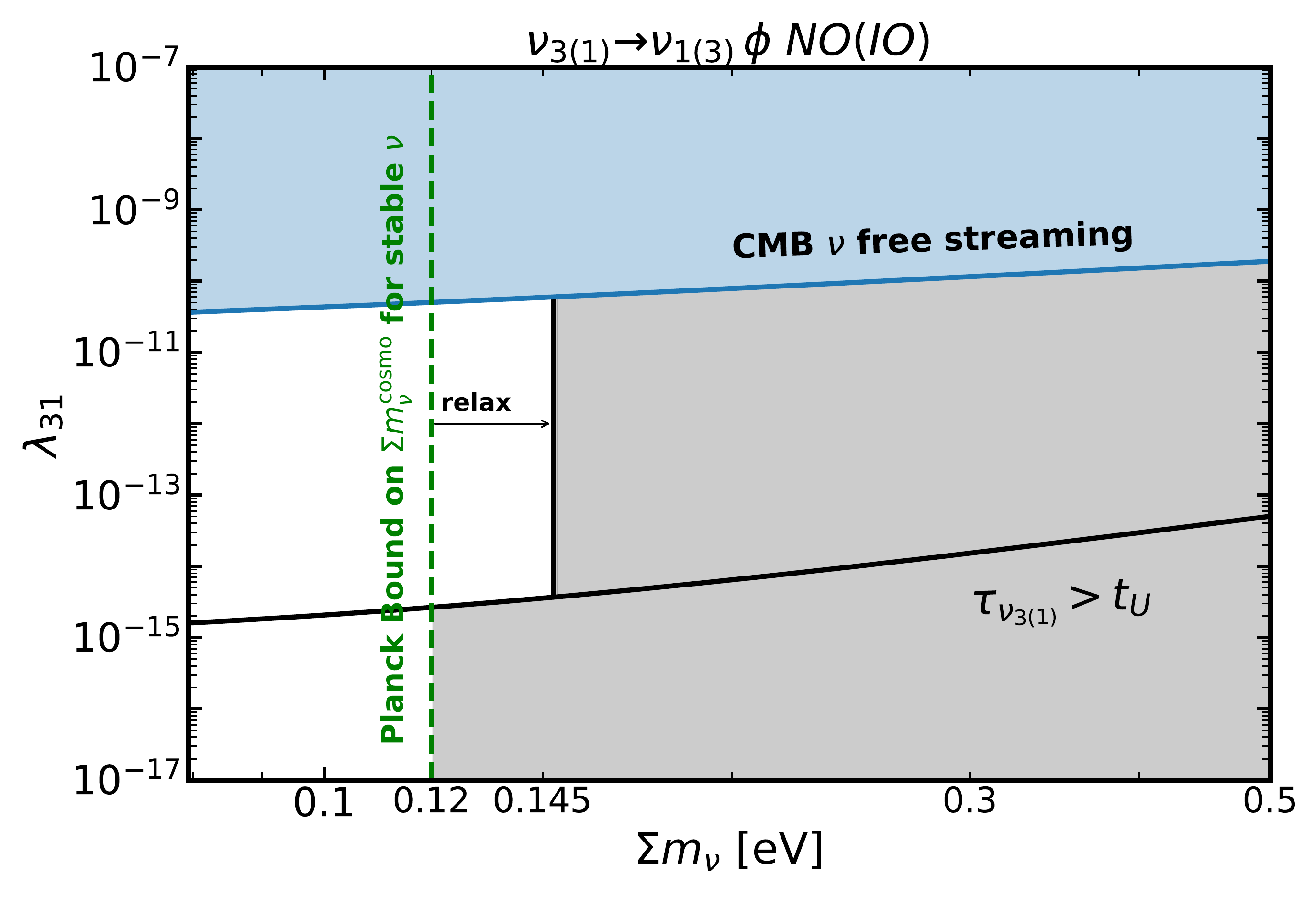} & \hspace{-0.3cm}  \includegraphics[width=0.5\textwidth]{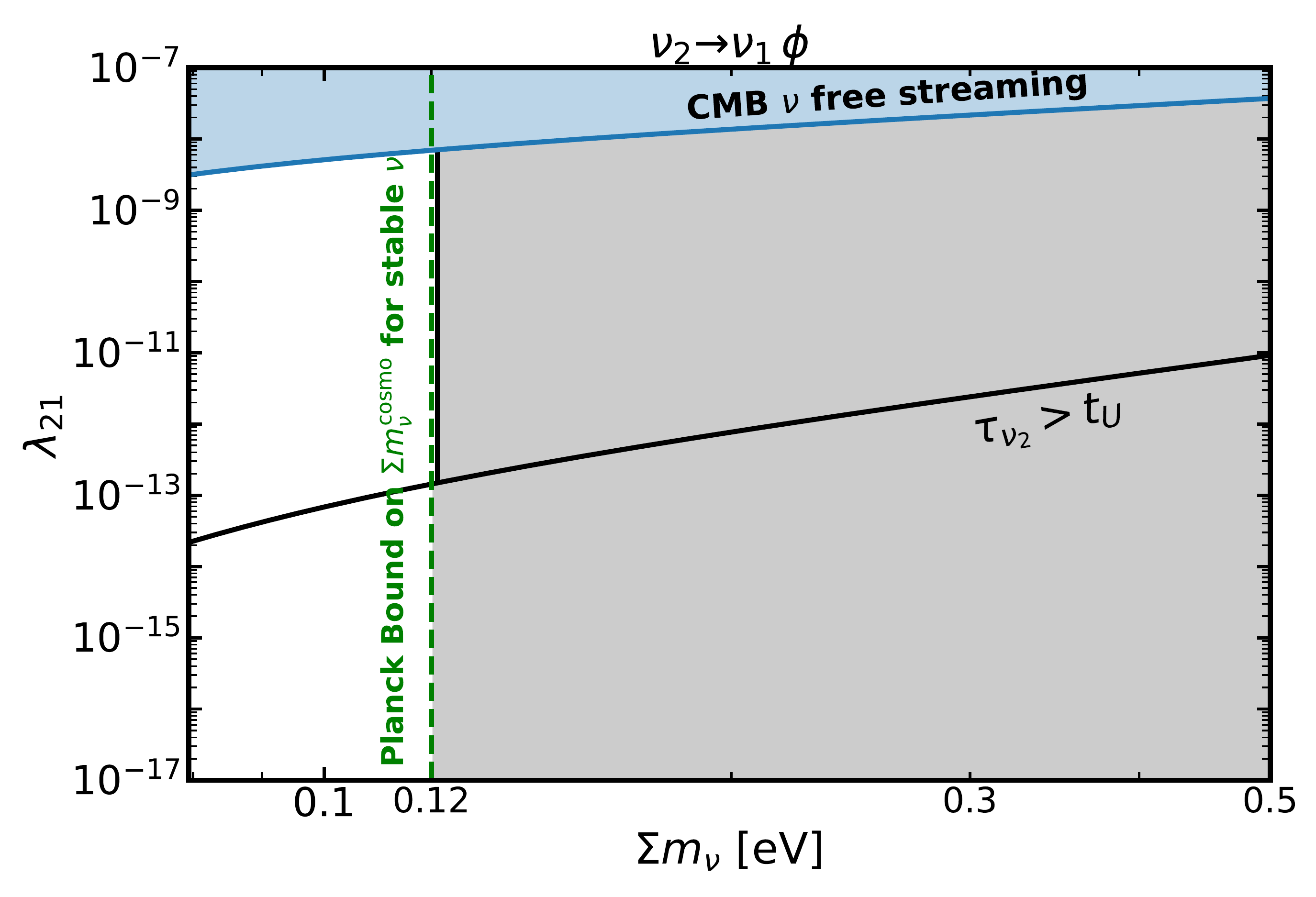}     \\
\hspace{-0.5cm}
 \includegraphics[width=0.5\textwidth]{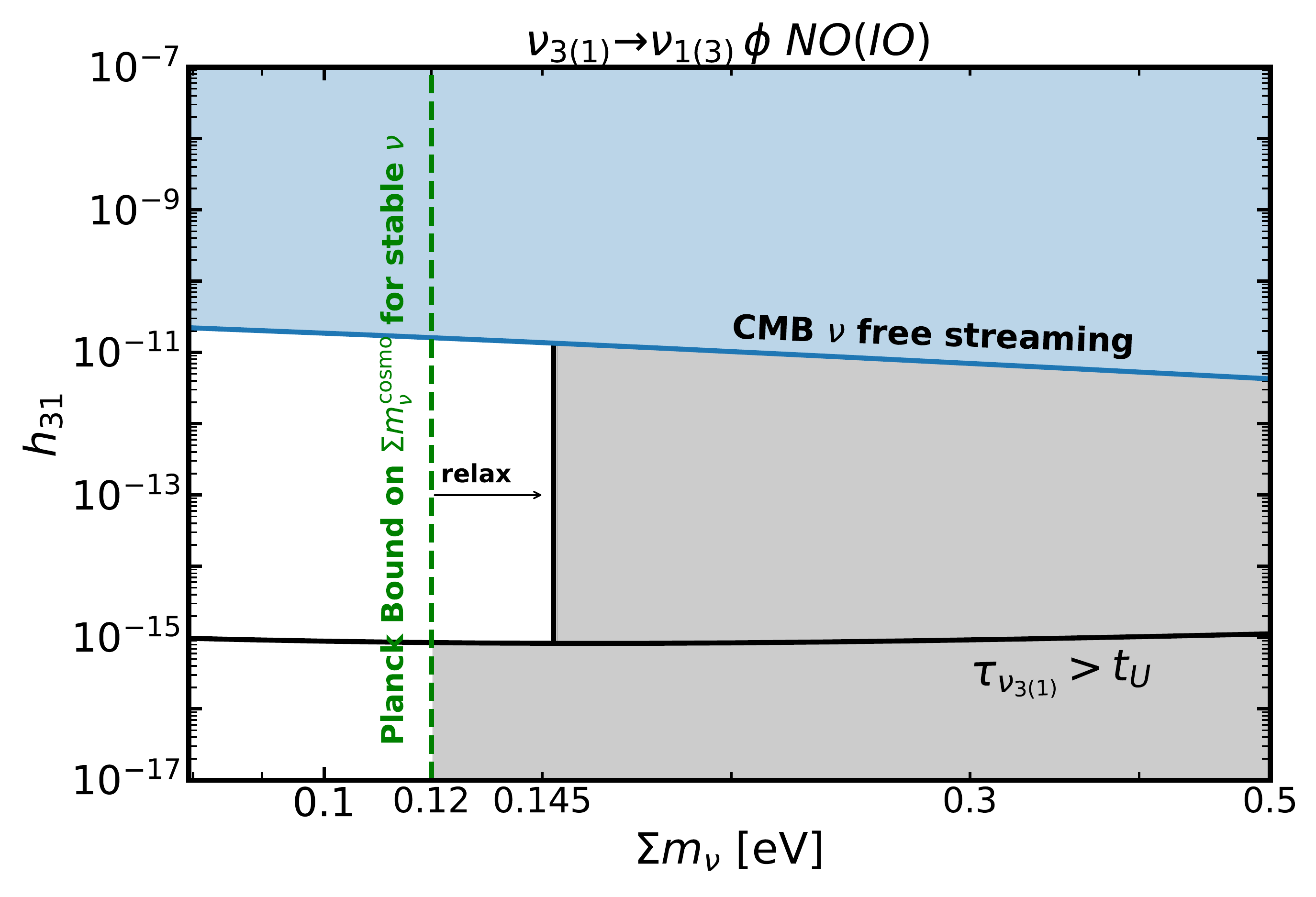} & \hspace{-0.3cm}  \includegraphics[width=0.5\textwidth]{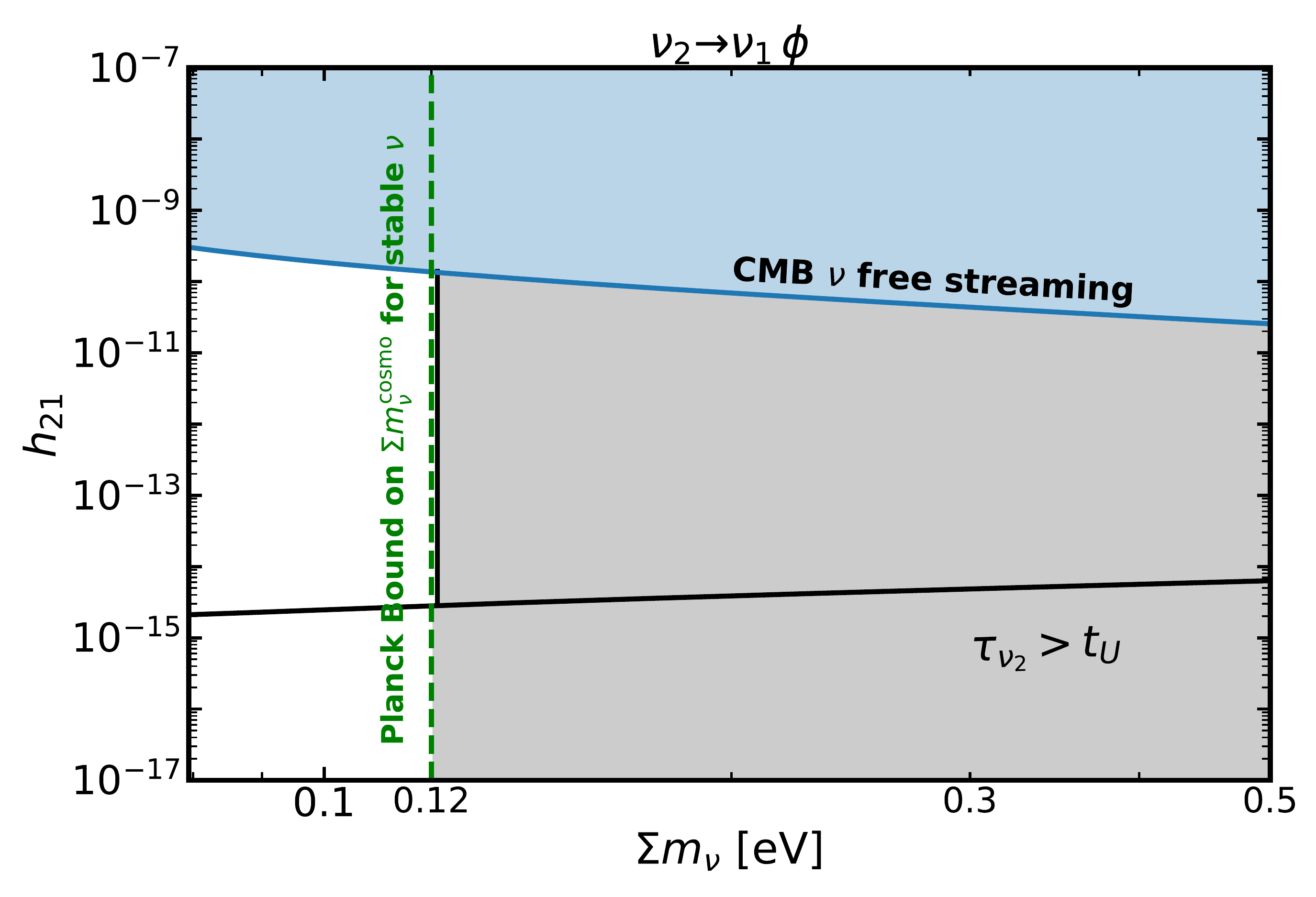}    \\
\hspace{-0.5cm}
\includegraphics[width=0.5\textwidth]{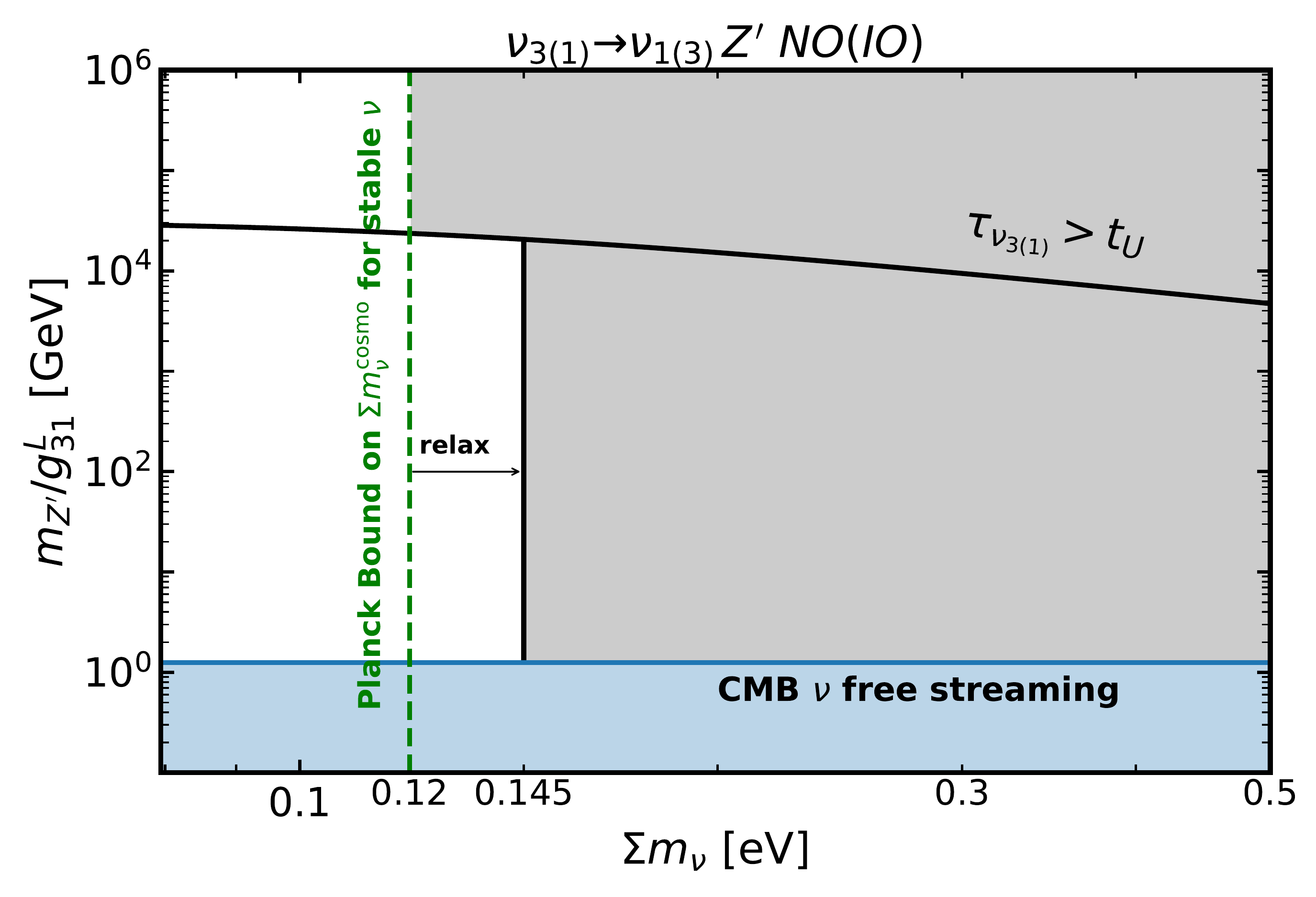}  & \hspace{-0.3cm}   \includegraphics[width=0.5\textwidth]{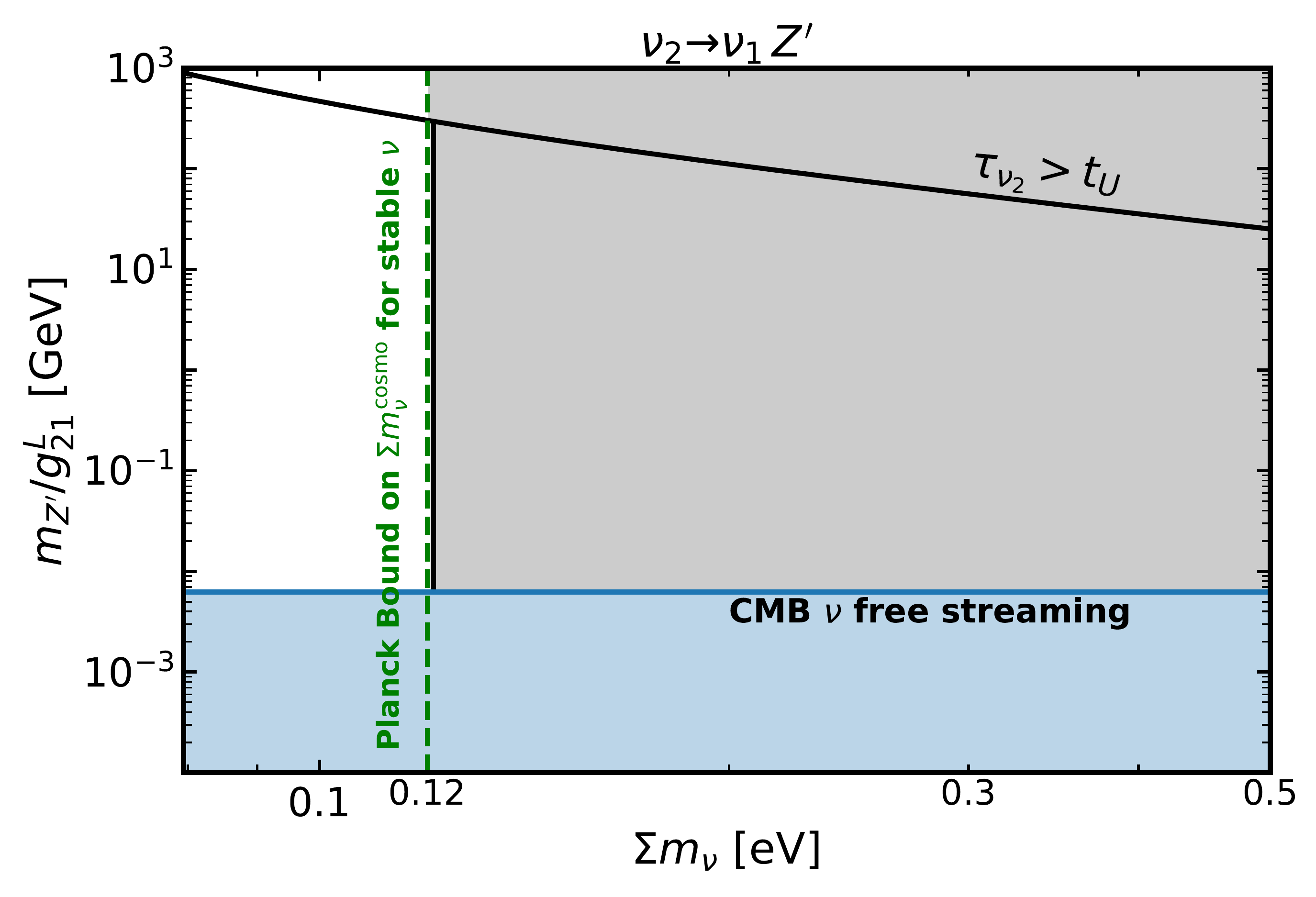}   
 \end{tabular}
\vspace{-0.4cm}
\caption{Parameter space for $2$-body invisible neutrino decays (case \textit{2a} in Figure~\ref{fig:NeutrinoDecayDiagrams}). \textit{Left panel:} $\nu_{3(1)}\to\nu_{1(3)} \,\phi/Z'$ decay modes for NO (IO). \textit{Right panel:} $\nu_2\to \nu_1 \,\phi/Z'$ decays. Note that the figure for $\nu_{3(2)} \to \nu_{2(3)} \phi$ for NO (IO) are almost identical to the left panels. The upper panels correspond to neutrino decays via pseudo-scalar couplings, the middle panels correspond to scalar couplings, and the lower panels to neutrino decays with a vector boson in the final state. For concreteness, we have considered $m_\phi,\,m_{Z'} \ll m_{\nu}$. We highlight in blue the bound on invisible neutrino decays from Planck CMB observations~\cite{Escudero:2019gfk} (see Section~\ref{sec:parameterpsace}). In dashed green we highlight the current bound on $\sum m_\nu$ within $\Lambda$CDM~\cite{Aghanim:2018eyx}, and show with an arrow the extent to which it can be relaxed. We shade in grey regions of parameter space that are be excluded by the current Planck bound $\sum m_\nu^{\rm cosmo} < 0.12\,\text{eV}$.  
}
\label{appfig:2acase}
\end{figure}

In Figure~\ref{appfig:2acase} we show the parameter space for decays of the type $\nu_{i}\to\nu_{j}\,\phi/Z'$ when every decay mode is individually considered while the rest are assumed to be switched off. According to the discussion in Section~\ref{sec:parameterpsace}, we estimate that the maximal relaxation of the cosmological mass bounds is given by the mass difference between the mother and daughter particle. Hence, it is obvious that, if the only opened decay channel is $\nu_{2}\to\nu_{1}\,\phi/Z'$, cosmological neutrino mass bounds can not substantially be modified because $\Delta m_{21}^2 \approx 7.4\times10^{-5}\,\text{eV}^2$, which leads to a maximal shift of roughly $10^{-3}\,\text{eV}$ for both NO and IO.
The situation is different for the decay modes $\nu_{3}\to\nu_{1}\,\phi/Z'$ and $\nu_{3}\to\nu_{2}\,\phi/Z'$ ($\nu_{1}\to\nu_{3}\,\phi/Z'$ and $\nu_{2}\to\nu_{3}\,\phi/Z'$) for NO (IO), since the relevant mass difference involved is $|\Delta m_{31}^2|\sim |\Delta m_{32}^2|= 2.5\times10^{-3}\,\text{eV}^2 \gg \Delta m_{21}^2$. Each of these four channels can potentially relax the present Planck neutrino mass bound up to roughly $0.14\,\text{eV}$. It is important to remark that the excluded grey region in the figures can be considered as a conservative bound since: \textit{i)} we have considered the different decay modes individually; \textit{ii)} a detailed analysis of all the potential invisible neutrino decay effects in cosmology would likely lead to a stronger constraint.

\subsection{Annihilation and decay rates of neutrinophilic bosons}

In this appendix we outline the annihilation rates as relevant for BBN and energy losses in supernovae. 
Given the effective Lagrangians presented in Equations~(\ref{eq:Lag_scalar}) and (\ref{eq:Lag_vector}), the relevant rates are
\begin{align}
\sigma(s)_{\bar{\nu}_i \nu_i \to \phi \phi } &= \frac{1}{2\pi}  \left(h_{ii}^2 + \lambda_{ii}^2 \right)^2 \frac{1}{s} \log\left(\frac{s}{m_{\nu_i}^2}\right)\,,\qquad \text{(mediated by} \,\nu_i\,\text{only)}\,,\\
\sigma(s)_{\bar{\nu}_i \nu_i \to \phi \phi } &= \frac{1}{\pi}  \left(h_{ij}^2 + \lambda_{ij}^2 \right)^2 \frac{1}{s} \log\left(\frac{s}{m_{\nu_i}^2}\right)\,,\qquad \text{(mediated by} \,\nu_j\,\text{only)}\,,\\
\sigma(s)_{\bar{\nu}_i \nu_i \to \phi \phi } &= \frac{1}{16\pi}  \left(h_{i4}^2 + \lambda_{i4}^2 \right)^2 \frac{1}{s} \log\left(\frac{s}{m_{\nu_i}^2}\right)\,,\qquad \text{(mediated by} \,\nu_4\,\text{only)}\,,\\
\sigma(s)_{\bar{\nu}_4 \nu_4 \to \phi \phi } &= \frac{1}{2\pi} \left(h_{44}^2 + \lambda_{44}^2 \right)^2 \frac{1}{s} \log\left(\frac{s}{m_{\nu_4}^2}\right)\,,\qquad \text{(mediated by} \,\nu_4\,\text{only)}\,,\\
\sigma(s)_{\bar{\nu}_i \nu_i \to \bar{\nu}_4 \nu_4 } &= \frac{1}{32\pi}  \left( 5 \lambda_{i4}^4+ 8 \lambda_{11} \lambda_{44} \lambda_{i4}^2 +16 \lambda_{11}^2 \lambda_{44}^2  \right) \frac{1}{s}\,,\qquad \text{taking }\, h_{ij} = 0\,,\\
\sigma(s)_{\bar{\nu}_i \nu_i \to \bar{\nu}_4 \nu_4 } &= \frac{1}{32\pi} \left( 5 h_{i4}^4+ 8 h_{ii} h_{44} h_{i4}^2 +16 h_{ii}^2 h_{44}^2  \right) \frac{1}{s}\,,\qquad \text{taking }\, \lambda_{ij} = 0\,,
\end{align}
where we have assumed that all particles are substantially lighter than the centre of mass energy $\sqrt{s}$.

The rates in the opposite direction are given by
\begin{align}
\label{appeq:anni1}
\sigma(s)_{ \phi \phi \to \bar{\nu}_i \nu_i  } &= \frac{4}{\pi}  \left(h_{ii}^2 + \lambda_{ii}^2 \right)^2 \frac{1}{s} \log\left(\frac{s}{m_{\phi}^2}\right)\,,\qquad \text{(mediated by} \,\nu_i\,\text{only)}\,,\\
\label{appeq:anni2}
\sigma(s)_{ \phi \phi  \to \bar{\nu}_i \nu_i } &= \frac{8}{\pi}  \left(h_{ij}^2 + \lambda_{ij}^2 \right)^2 \frac{1}{s} \log\left(\frac{s}{m_{\phi}^2}\right)\,,\qquad \text{(mediated by} \,\nu_j\,\text{only)}\,,\\
\label{appeq:anni3}
\sigma(s)_{ \phi \phi \to \bar{\nu}_i \nu_i } &= \frac{1}{2\pi}  \left(h_{i4}^2 + \lambda_{i4}^2 \right)^2 \frac{1}{s} \log\left(\frac{s}{m_{\phi}^2}\right)\,,\qquad \text{(mediated by} \,\nu_4\,\text{only)}\,,\\
\sigma(s)_{\phi \phi \to \bar{\nu}_4 \nu_4 } &= \frac{1}{\pi} \left(h_{44}^2 + \lambda_{44}^2 \right)^2 \frac{1}{s} \log\left(\frac{s}{m_{\phi}^2}\right)\,,\qquad \text{(mediated by} \,\nu_4\,\text{only)}\,.
\end{align}

Finally, the decay rate for the process $\phi \to \nu_{i} \bar{\nu}_{j}$ in limit $m_{\phi} \gg m_{i,j}$, is given by
\begin{align}
\label{appeq:inversedecay}
    \Gamma_{\phi\to \nu_{i} \bar{\nu}_{j}} &= \frac{m_{\phi}}{8\pi}(h_{ij}^2 + \lambda_{ij}^2)\,.
    \end{align}

\subsection{Constraints on sub-MeV neutrinophilic bosons}\label{sec:Bosonsconstraints}

Here we provide a derivation of the BBN and SN1987A bound that we use in the main text to set constraints on sub-MeV neutrinophilic mediators. Since vector mediators couple to charged leptons, and thus suffer from stronger constraints~\cite{Croon:2020lrf,Escudero:2019gzq}, we will focus on the  scalar mediator ($\phi$) case. 

\textbf{Constraints from SN1987A:} The main constraints on neutrinophilic bosons arises from supernova cooling considerations since, within the Standard Model, neutrinos are expected to carry away most of the gravitational energy released from a supernova~\cite{Raffelt:1996wa}. The SN1987A neutrino signal is indeed compatible with this expectation and the rate at which neutrinos cool the supernova, $\Delta t \sim 10\,\text{s}$. Thus, in order to be compatible with SN1987A observations, neutrinophilic bosons cannot be copiously produced inside a supernova and escape it. Requiring that the total luminosity emitted in these BSM particles is less or equal to the one measured from neutrinos of SN1987A one can robustly exclude $\nu-\phi$ interaction strengths of $3\times 10^{-7}\lesssim \lambda \lesssim 2\times 10^{-5} $ for $m_{\phi} \lesssim 10\,\text{MeV}$~\cite{Kachelriess:2000qc,Farzan:2002wx} and $2 \times 10^{-9}\frac{\text{MeV}}{m_{\phi}} < \lambda < 2 \times 10^{-6}\frac{\text{MeV}}{m_{\phi}}$~\cite{Heurtier:2016otg,Brune:2018sab} for $m_{\phi}\lesssim 200\,\text{MeV}$. The latter bound only applies provided $\lambda <2\times 10^{-5}$. 

Here we explicitly revaluate the SN1987A bound from supernova cooling, but considering all possible production channels of $\phi$'s for $m_{\phi} \neq 0$. In order to set the constraint we consider a supernova core of  temperature $T_{\rm SN}=30\,\text{MeV}$ and radius $R_{\rm SN}=10\,\text{km}$. We also take into account that there is a large number of electrons neutrinos in the SN core~\cite{Mirizzi:2015eza} and choose an associated chemical potential to be $\mu_{\nu_e} = 200\,\text{MeV}$ (while a negligible one for $\nu_\mu$ and $\nu_\tau$ ones). These values for the temperature, radius and neutrino chemical potentials are typically found in supernova simulations~\cite{Raffelt:1996wa,Mirizzi:2015eza} and, in particular, are the same values considered in Ref.~\cite{Heurtier:2016otg}. In order to set a constraint we require that the luminosity of emitted $\phi$ species does not exceed the typical binding energy of a neutron star $E_b = 3\times 10^{53}\,\text{erg}$.
We take into account production via $\bar{\nu}\nu \to \phi$ and $\bar{\nu}\nu \to \phi \phi$ and account for the fact that if the produced bosons decay or annihilate back into neutrinos within $R<R_{\rm SN}$ there is no cooling. Explicitly, this means
\begin{align}\label{eq:snbound}
\Delta E_{\text{SN}} &= (V \Delta t)  \frac{T}{64\pi^4}  \int_{4m_\phi^2}^{\infty} \text{d}s \left[ s^2 \sigma_{\bar{\nu} \nu \to \phi \phi}(s) K_{2} \left(\frac{\sqrt{s}}{T} \right) e^{\frac{\mu}{T}} e^{-\left[n_\phi \sigma_{\phi \phi\to \bar{\nu} \nu }(s) + \Gamma_{\phi\to \bar{\nu}\nu} \frac{m_\phi}{\sqrt{s}/2}\right]R_{\rm SN}} \right]  +\\
 &+ (V \Delta t) \frac{3m_\phi}{2\pi^2}  \Gamma_{\phi \to \bar{\nu}\nu}  \int_{0}^{\infty} \text{d}p \left[  p^2 \,{e}^{-\frac{E_\phi(p) - \mu}{T}} e^{-\left[n_\phi \sigma_{\phi \phi\to \bar{\nu} \nu }(s = 4E_\phi^2)+\Gamma_{\phi \to \bar{\nu}\nu}  \frac{m_\phi}{E_\phi(p)} \right]R_{\rm SN} } \right]\,,\nonumber
\end{align}
where $V = \frac{4\pi}{3} R_{\rm SN}^3$, $\Delta t = 10\,\text{s}$, $K_{2}(x)$ is a Bessel function of the second kind, $\mu$ is the chemical potential of a given neutrino, and $\Gamma_{\phi \to \bar{\nu}\nu}$ is the decay rate at rest of $\phi$ into neutrinos. The number density of the neutrinophilic boson can be approximated by its thermal equilibrium value, $n_\phi \simeq 1.2\,T^3 / \pi^2$. 

The physical interpretation of Equation~\eqref{eq:snbound} is the following: the first line accounts for emission of $\phi$'s via $\bar{\nu}\nu \to \phi \phi$. It is the very same expression as for the energy density rate for this process, see e.g. Equation 2.15 of~\cite{Escudero:2020dfa}, but modulated by an exponential that accounts for the fact that the $\phi$'s will become trapped in the core if $\phi \phi \to \bar{\nu}\nu$ and $\phi \to \bar{\nu}\nu$ processes have a mean free path shorter than $R_{\rm SN}$.  The second line in~\eqref{eq:snbound} represents the same as the first line but for production via $\bar{\nu}\nu \to \phi$ inverse decays. 

\begin{figure}[t]
\centering
\begin{tabular}{cc}
\hspace{-0.5cm} \includegraphics[width=0.5\textwidth]{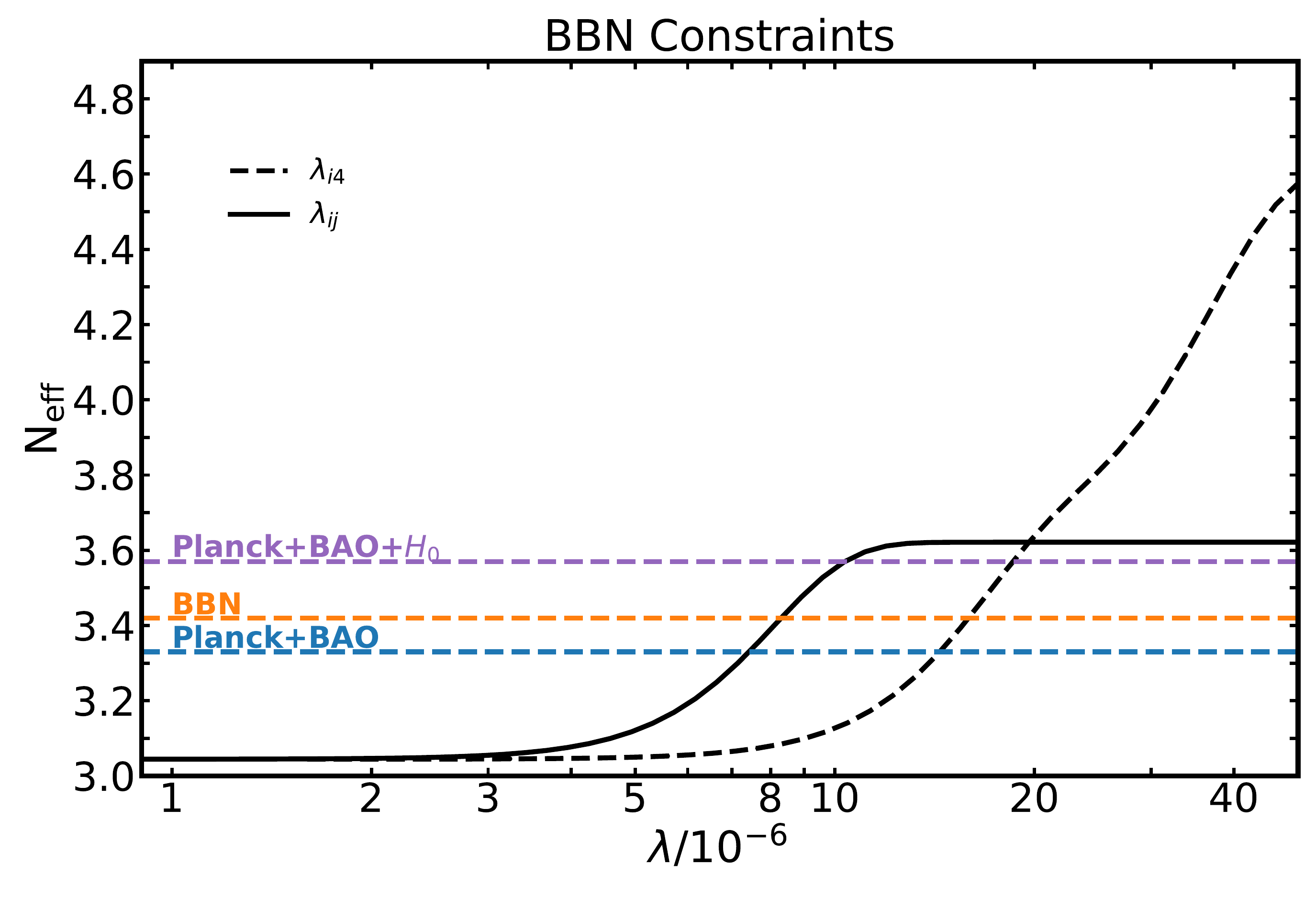} & \hspace{-0.55cm}   \includegraphics[width=0.5\textwidth]{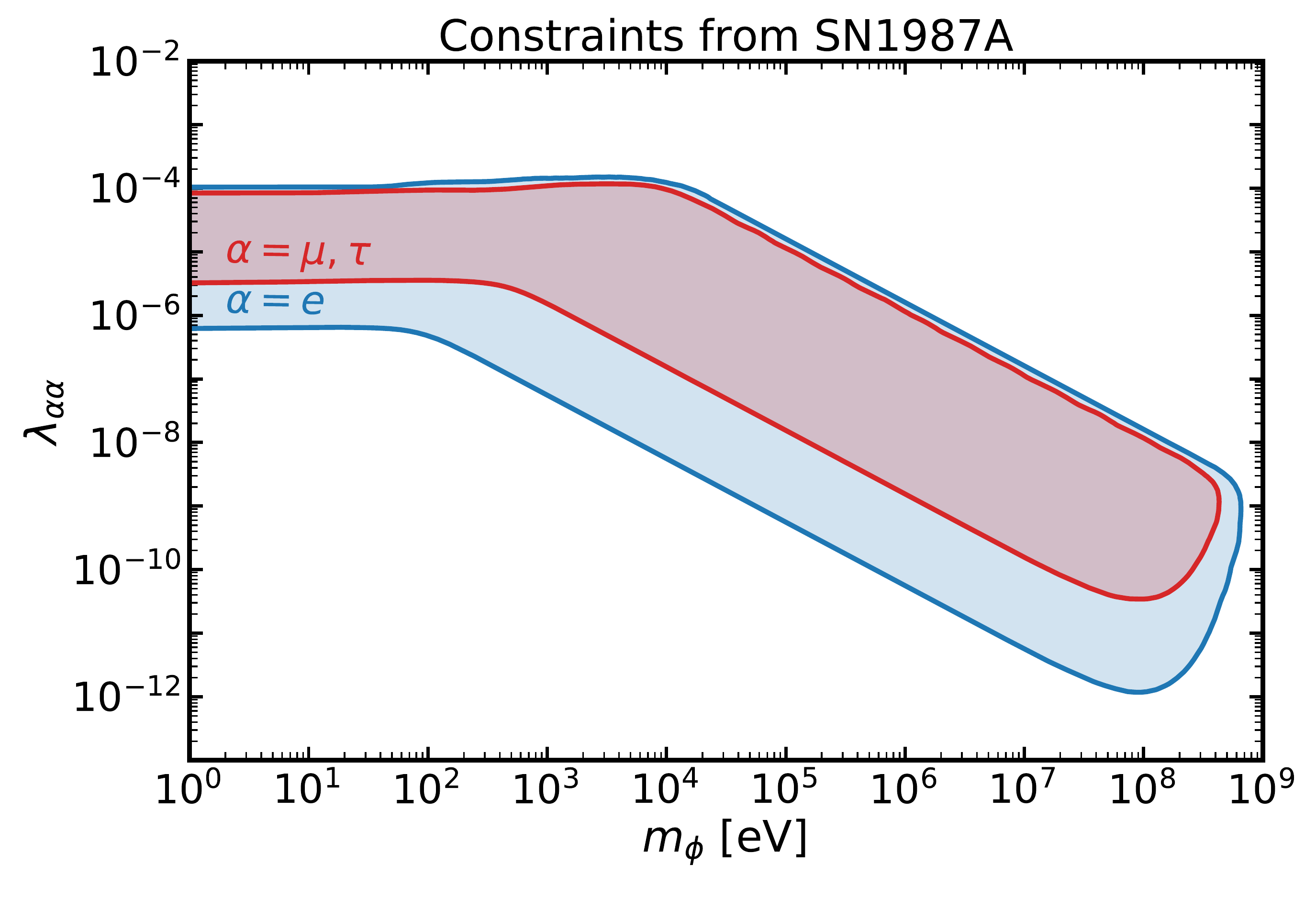}   \\
 \end{tabular}
\vspace{-0.4cm}
\caption{\textit{Left panel:} $N_{\rm eff}$ constraints on neutrinophilic mediators. Both $\phi$ and $\nu_4$ species are produced via the $\lambda_{i4}$ coupling. Via $\lambda_{ij}$ only $\phi$ can be produced. We have assummed $m_\phi < 1\,\text{MeV}$. \textit{Right panel:} Constraints from the supernova SN198A observation. The constraint is given in the flavor basis as there are different chemical potentials involved for the three SM flavors. The translation to the mass basis via the PMNS matrix will only induce $\mathcal{O}(1)$ corrections, such that $\lambda_{ij} \simeq \lambda_{ee}$.}
\label{fig:BBN_constraint}
\end{figure}

Finally, we obtain our constrain on the relevant coupling strengths by taking the annihilation and decay rates from Equations~\eqref{appeq:inversedecay},~\eqref{appeq:anni1}, and~\eqref{appeq:anni2}, using~\eqref{eq:snbound} and by requiring $\Delta E_{\rm SN} < E_b \simeq 3 \times 10^{53}\,\text{erg}$. For masses below $m_\phi \sim 100\,\text{MeV}$ the SN1987A bound for electron neutrinos is given by
\begin{align}\label{eq:ee_SN}
\left[\left(5 \! \times \! 10^{-11}\frac{\text{MeV}}{m_{\phi}}\right)^{-1} + (5 \! \times \! 10^{-7})^{-1}\right]^{-1}<  \lambda_{ee} < \left[\left(10^{-6} \frac{\text{MeV}}{m_{\phi}}\right)^{-1} + (10^{-4})^{-1}\right]^{-1}\,,
\end{align}
while for $\mu$ and $\tau$ neutrinos it leads to 
\begin{align}\label{eq:mumu_SN}
\left[\left(1 \! \times \! 10^{-9}\frac{\text{MeV}}{m_{\phi}}\right)^{-1} + (5 \! \times \! 10^{-6})^{-1}\right]^{-1}<  \lambda_{\alpha \alpha} < \left[\left(10^{-6} \frac{\text{MeV}}{m_{\phi}}\right)^{-1} + (10^{-4})^{-1}\right]^{-1}\,.
\end{align}
In the right panel of Figure~\ref{fig:BBN_constraint} we show the bounds from SN1987A for masses up to $1\,\text{GeV}$. The $\lambda_{ij}$ couplings (mass basis) are related to $\lambda_{ee}$ (flavor basis) via a rotation with the PMNS matrix and thus the constraint on $\lambda_{ij}$ is similar to the $\lambda_{ee}$. As a result, in the main text, and in particular in the left panel of Figure~\ref{fig:3a1}, we choose to show the bound on the $\lambda_{ee}$ coupling. 

We can also use SN1987A data to constrain the coupling among the scalar field, the active neutrino and the sterile state. In this case, the main constraint arises due to processes such as $\bar{\nu}_i\nu_j \to \phi \phi $ as mediated by a sterile neutrino. In this case, we assume $\lambda_{44} \gg \lambda_{i4}$ and that $\phi$ decays to $\bar{\nu}_4 \nu_4$. Given these processes, SN1987A excludes
\begin{align}\label{eq:ij_SN}
3 \! \times \!10^{-6} <  \lambda_{i4} <3 \! \times \!10^{-4} \,, \qquad \text{for} \qquad m_\phi \lesssim 10\,\text{MeV}\,\,\text{and}\,\,\, m_{\nu_4} \ll \,\text{MeV}\,,
\end{align}

Constraints on $\nu$-$\phi$ interactions from supernova deleptonization have also been derived restricting $\lambda \lesssim 10^{-4}$~\cite{Kachelriess:2000qc}. These bounds are however subject to the details of the supernova explosion and, thus, have not being considered in the present paper. We also note that recently, bounds on $\nu$-$\phi$ interactions have been derived based on neutrino shockwave considerations within a supernova: $\lambda \gtrsim 10^{-2}$, but extending up to $m_{\phi/Z'}\lesssim 1 \,\text{GeV}$~\cite{Shalgar:2019rqe}.

\textbf{Big Bang Nucleosynthesis:}
We follow~\cite{Escudero:2020dfa,Escudero:2018mvt} and use~\href{https://github.com/MiguelEA/nudec_BSM}{\texttt{NUDEC\_BSM}} to model the physics of neutrino decoupling in the presence of $\phi$ bosons and $\nu_4$ neutrinos. We write down evolution equations for the temperature of the electromagnetic plasma, active neutrinos, $\nu_4$, and $\phi$ bosons including all relevant interactions among them. For $2\leftrightarrow 2$ processes we take $m_{\nu_i} = m_{\nu_4} = m_\phi = 0$ since these masses are negligible for temperatures that can impact $N_{\rm eff}$, namely $T > T_\nu^{\rm dec} \simeq 2 \,\text{MeV}$. We assume that there is no primordial population of $\phi$ and $\nu_4$ particles for $T\gtrsim 10\,\text{MeV}$ and that for $T\lesssim 10\,\text{MeV}$ these species are only produced via their interactions with active neutrinos. We note that the bounds we derive below would become more stringent had we considered a primordial population of such species or additional production channels.

We show the resulting values of $N_{\rm eff}$ as a function of the relevant coupling constants in the left panel of Figure~\ref{fig:BBN_constraint}. The current BBN bound, $N_{\rm eff} < 3.4$~\cite{Fields:2019pfx,Pitrou:2018cgg} at 95\% CL, is indicated by the orange dashed line. We set conservative bounds on individual neutrino-boson interactions by assuming that the rest of the couplings vanish -- otherwise constraints will be more stringent. By doing so we find that successful BBN requires
\begin{align}
\lambda_{ij} < 8\times 10^{-6}\,,\,\,\,\,  \lambda_{i4} < 1.6\times 10^{-5}\,,\qquad \text{(BBN)}
\end{align}
where these couplings are defined in Lagrangian~\eqref{eq:Lag_scalar}. Note that these bounds strengthen by a factor $\sqrt[4]{3}\simeq 1.3$ if three neutrinos simultaneously interact with $\phi$.

In the limit in which $m_\phi$ is not negligible, $\phi$ species can also be produced in the early Universe via $\bar{\nu}_i \nu_j
 \to \phi$ processes prior to neutrino decoupling and therefore modify the expansion history of the Universe. If the process $\bar{\nu}_i \nu_j \to \phi$ thermalizes prior to $T_\nu^{\rm dec}$ it would lead to $N_{\rm eff} \simeq 3.62$. Including this type of processes, we find that for $N_{\rm eff} < 3.4$ the interaction couplings are required to be
 \begin{align}
\lambda_{ij} < 1.2 \times 10^{-9} \,\text{MeV}/m_\phi \,,\qquad \text{(BBN)}
\end{align}
which roughly corresponds to requiring $\left<\Gamma(\bar{\nu}_i \nu_j \to \phi)\right> < H(T = 4 \,\text{MeV})$ and applies when $ 2\,m_\nu \lesssim m_\phi < 1\,\text{MeV}$.

\subsection{BBN bounds on very light sterile neutrinos}\label{sec:BBN_steriles}
 
In this appendix we place BBN bounds on the mixings of mostly sterile neutrinos with active neutrinos as relevant for invisible neutrino decays. There are very sophisticated studies dealing with cosmological constraints on eV-scale sterile neutrinos~\cite{Gariazzo:2019gyi,Hagstotz:2020ukm,Hasegawa:2020ctq}. Previous studies, however, have focused on the case of sterile neutrinos that are heavier than active neutrinos, $\Delta  m^2 \equiv m_{\nu_4}^2-m_{\nu_i}^2> 0$. We, on the other hand, are interested in sterile neutrinos that are \textit{lighter} than active neutrinos since they could represent the active neutrino decay final state, $\Delta m^2 < 0$. The early Universe phenomenology of both $\Delta m^2 >0$ and $\Delta m^2 <0$ is fairly analogous but with a relevant difference. In our case of interest, $\Delta m^2 <0$, there is resonant sterile neutrino production via a MSW-like effect. This leads to significantly more stringent constraints on $|\theta_{\alpha 4}|$ for $\Delta m^2 < 0$ than for  $\Delta m^2 > 0$.

\begin{figure}[t]
\begin{tabular}{cc}
\hspace{-0.5cm} \includegraphics[width=0.515\textwidth]{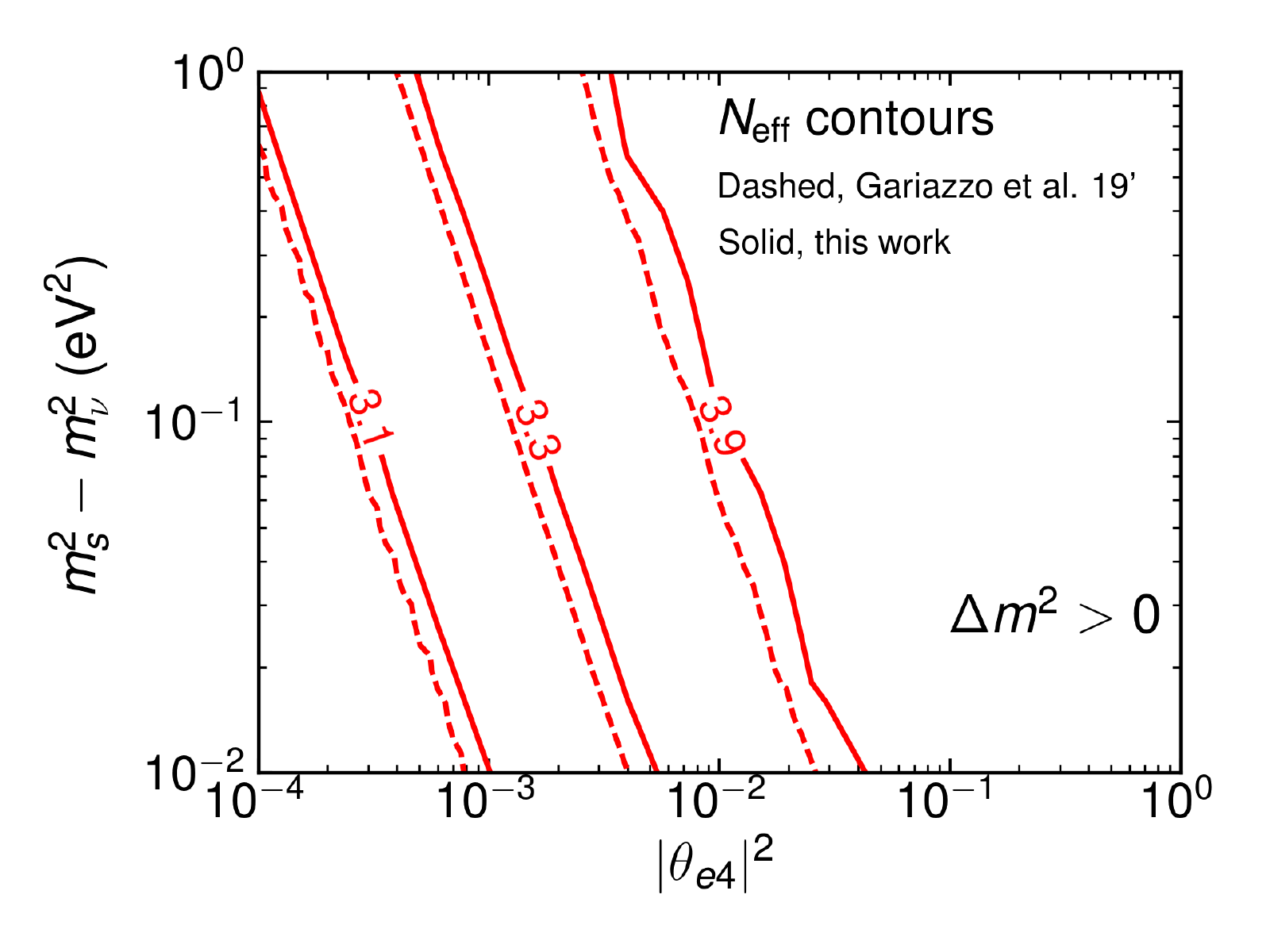} & \hspace{-0.55cm}   \includegraphics[width=0.515\textwidth]{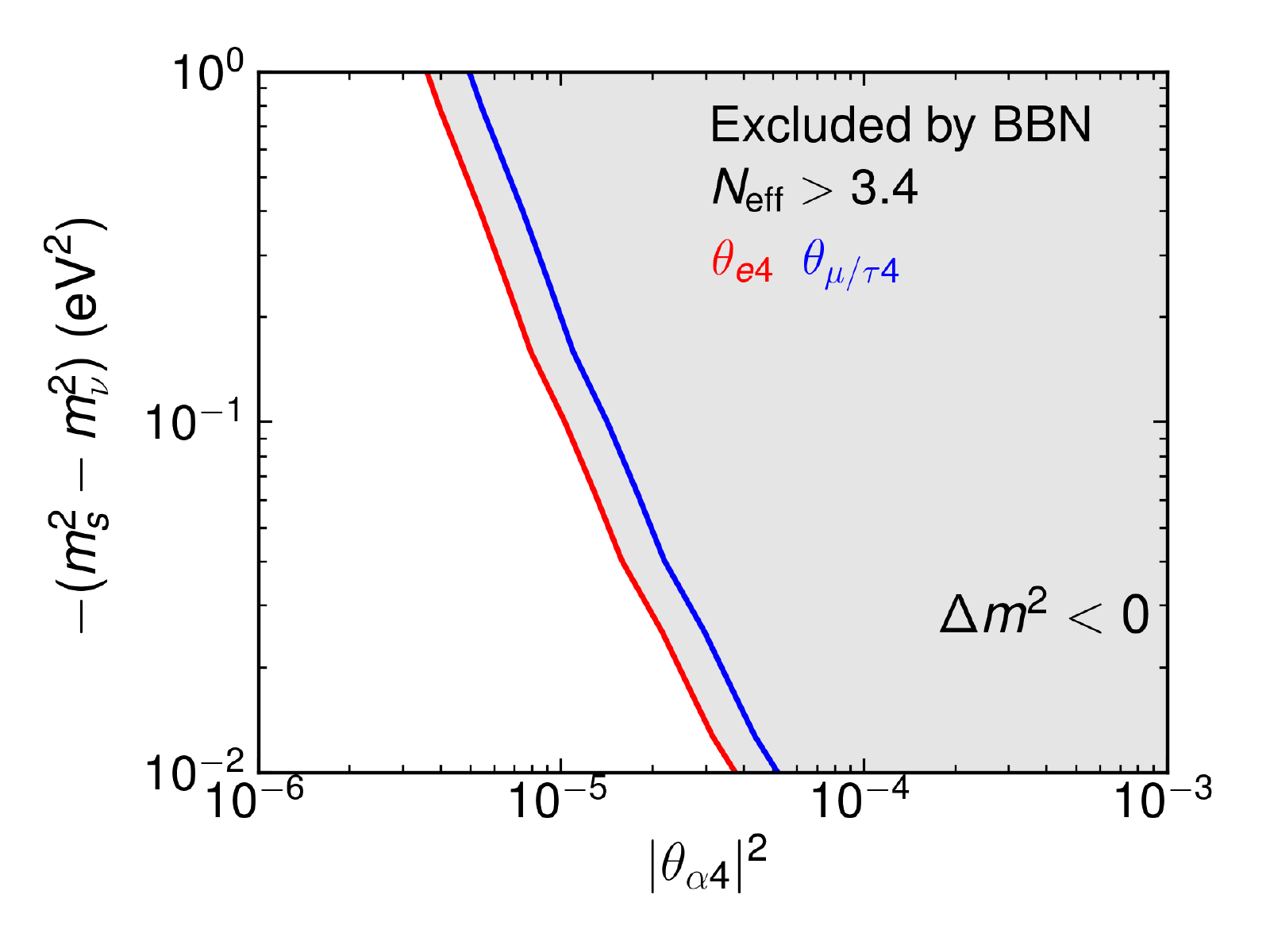}   \\
 \end{tabular}
\vspace{-0.5cm}
\caption{\textit{Left panel:} $N_{\rm eff}$ as a function of $|\theta_{4\alpha}|$ for eV-scale sterile neutrinos with $\Delta  m^2 \equiv m_{\nu_4}^2-m_{\nu_i}^2> 0$. Solid lines correspond to our calculation using~\href{https://github.com/MiguelEA/nudec_BSM}{\texttt{NUDEC\_BSM}} and dashed lines correspond to the results of~\cite{Gariazzo:2019gyi}. We appreciate the excellent agreement. We note that a similar agreement is found for $|\theta_{4\mu/\tau}|$.  \textit{Right panel:} 95\% CL BBN exclusion contour for $|\theta_{4\alpha}|$ for sterile neutrinos \textit{lighter} than the mainly active neutrinos ($\Delta m^2 < 0$) as computed using~\href{https://github.com/MiguelEA/nudec_BSM}{\texttt{NUDEC\_BSM}}. }
\label{fig:BBN_constraint_nu4}
\end{figure}

In the early Universe, sterile neutrinos are produced via scatterings and oscillations of active neutrinos. The production of eV-scale sterile neutrinos in the thermal plasma roughly peaks at $T\sim 4\,\text{MeV} \,(\sqrt{|\Delta m^2|}/0.1\,\text{eV})^{1/3}$~\cite{Abazajian:2005gj}. Active neutrinos decouple from electrons and positrons at $T \sim 2\,\text{MeV}$ and therefore any sterile neutrinos produced at $T \gtrsim 2\,\text{MeV}$ will impact $N_{\rm eff}$ as relevant for BBN and CMB observations. In this appendix we use~\href{https://github.com/MiguelEA/nudec_BSM}{\texttt{NUDEC\_BSM}}~\cite{Escudero:2020dfa,Escudero:2018mvt} to model neutrino decoupling in the presence of an eV-scale sterile-neutrino. As in~\cite{Escudero:2020dfa,Escudero:2018mvt}, we assume all relevant species can be described by thermal equilibrium distribution functions with an evolving temperature and negligible chemical potentials. We then solve for the temperatures of photons, active neutrinos and sterile neutrinos including all relevant interactions among them. We account for the active-sterile neutrino interactions by using the collision term described in Equation 1 of~\cite{Abazajian:2005gj} and assuming there is no asymmetry between particles and antiparticles in the early Universe. We validate our method for $\Delta m^2 > 0$ by comparing our results against state-of-the-art references in the left panel of Figure~\ref{fig:BBN_constraint_nu4}. There we can appreciate the excellent agreement between our calculation and that of~\cite{Gariazzo:2019gyi} for the case $\Delta m^2 > 0$. In the right panel of  Figure~\ref{fig:BBN_constraint_nu4} we show the bounds on $|\theta_{\alpha 4}|$ for $\Delta m^2 < 0$. We consider the latest BBN bound $N_{\rm eff} < 3.4$~\cite{Fields:2019pfx,Pitrou:2018cgg} at 95\% CL and find a limit on $|\theta_{\alpha 4}|$ of
\begin{align}
\label{appeq:theta_bbn}
|\theta_{ e 4}|^2 &\lesssim 3\times 10^{-6} \,\text{eV}/\sqrt{|\Delta m_{4i}^2|}\,,\,\,\,\,  m_{\nu_4}^2-m_{\nu_i}^2 < 0 \,,\qquad \text{(BBN)}\,,\\
|\theta_{\mu/\tau 4 }|^2 &\lesssim 4\times 10^{-6} \,\text{eV}/\sqrt{|\Delta m_{4i}^2|}\,,\,\,\,\,  m_{\nu_4}^2-m_{\nu_i}^2 < 0 \,,\qquad \text{(BBN)}\,.
\end{align}
We notice that this bound is roughly two orders of magnitude more stringent than for $\Delta m^2 = m_{\nu_4}^2-m_{\nu_i}^2 > 0$. 

\subsection{Implications of a relaxation of the CMB bounds on invisible neutrino decays}\label{sec:comments}

\begin{figure}[t]
\centering
\begin{tabular}{cc}
\hspace{-0.5cm}
\includegraphics[width=0.5\textwidth]{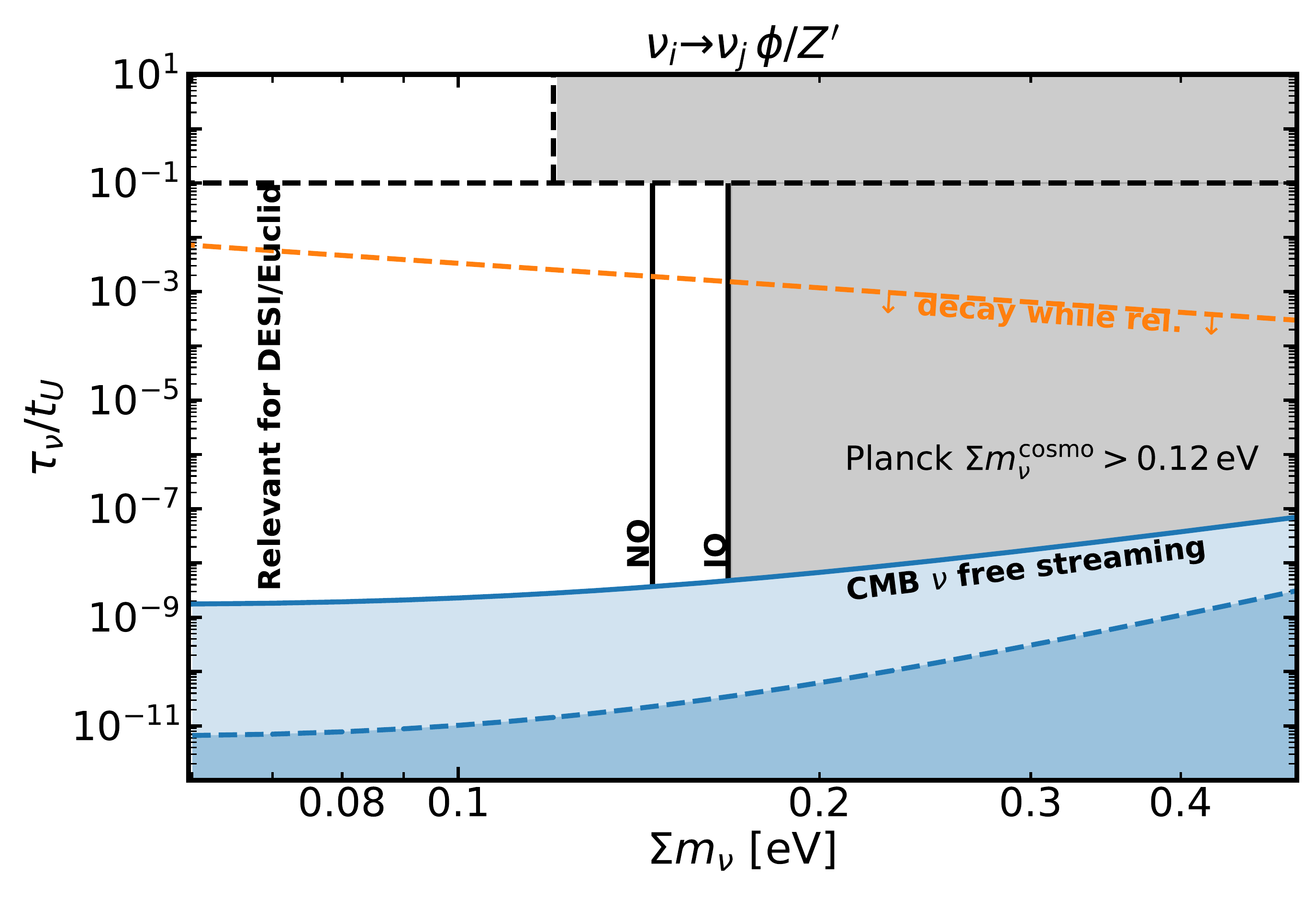} & \hspace{-0.3cm}  \includegraphics[width=0.5\textwidth]{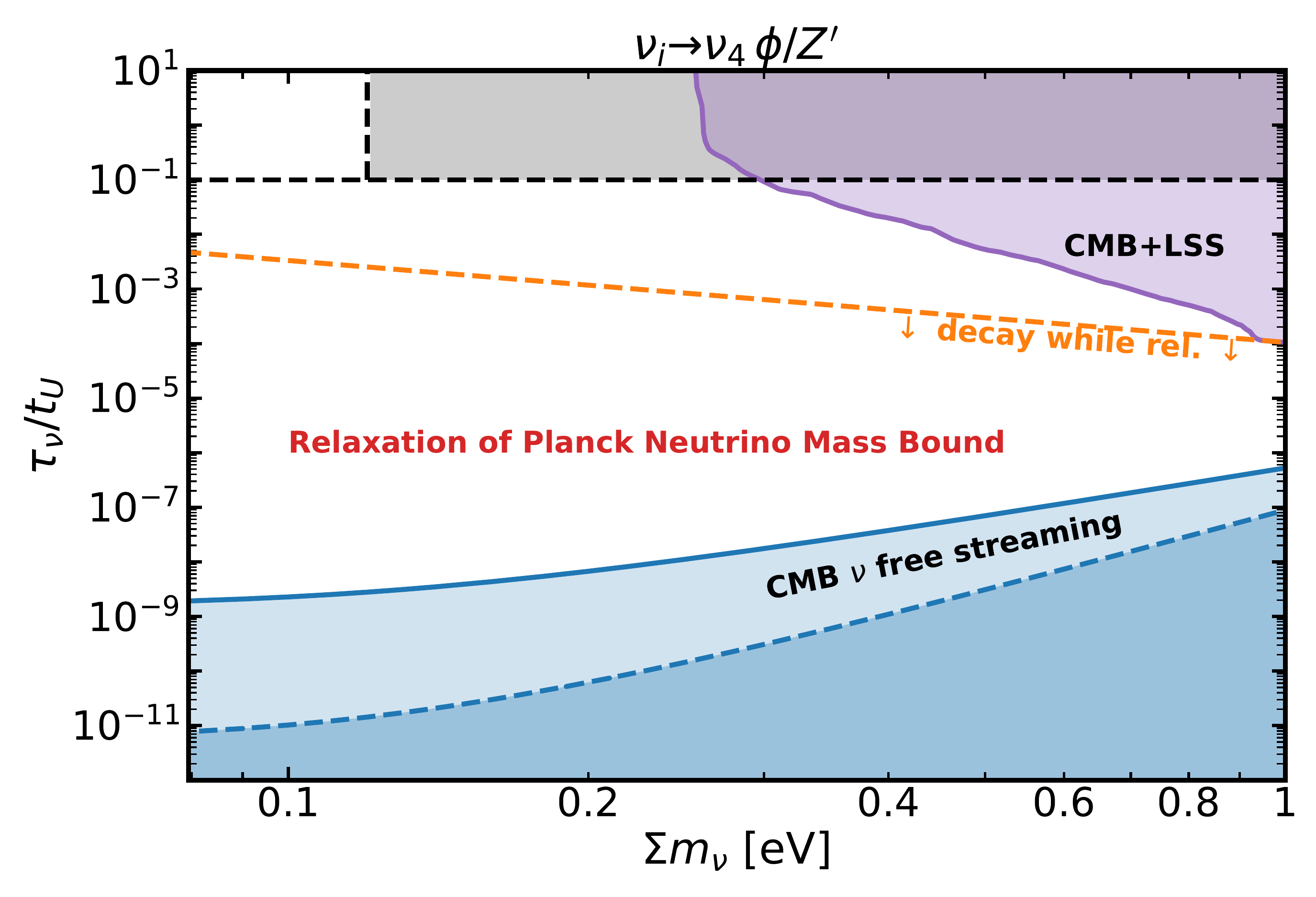}     \\
  \includegraphics[width=0.5\textwidth]{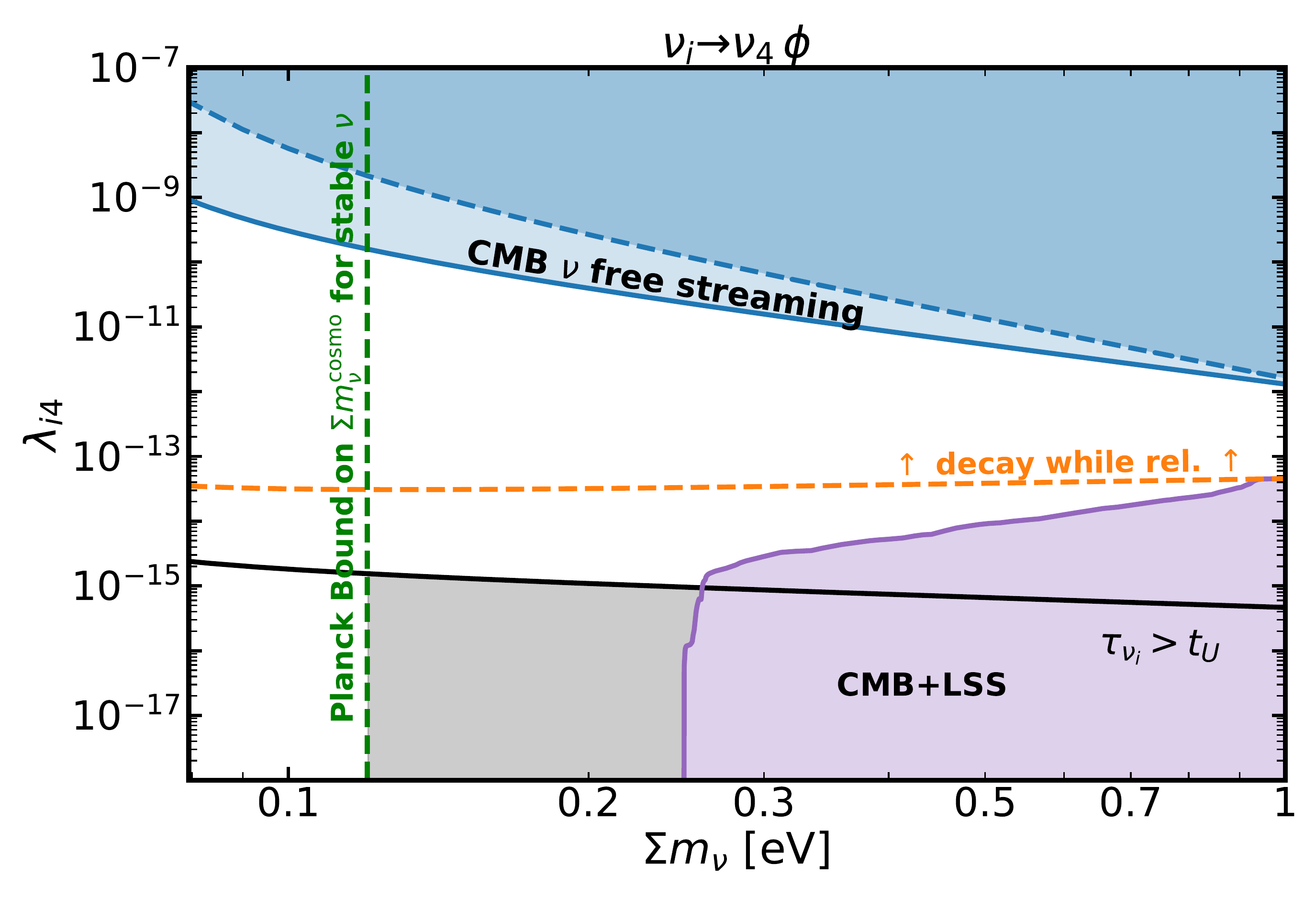} & \hspace{-0.3cm}  \includegraphics[width=0.5\textwidth]{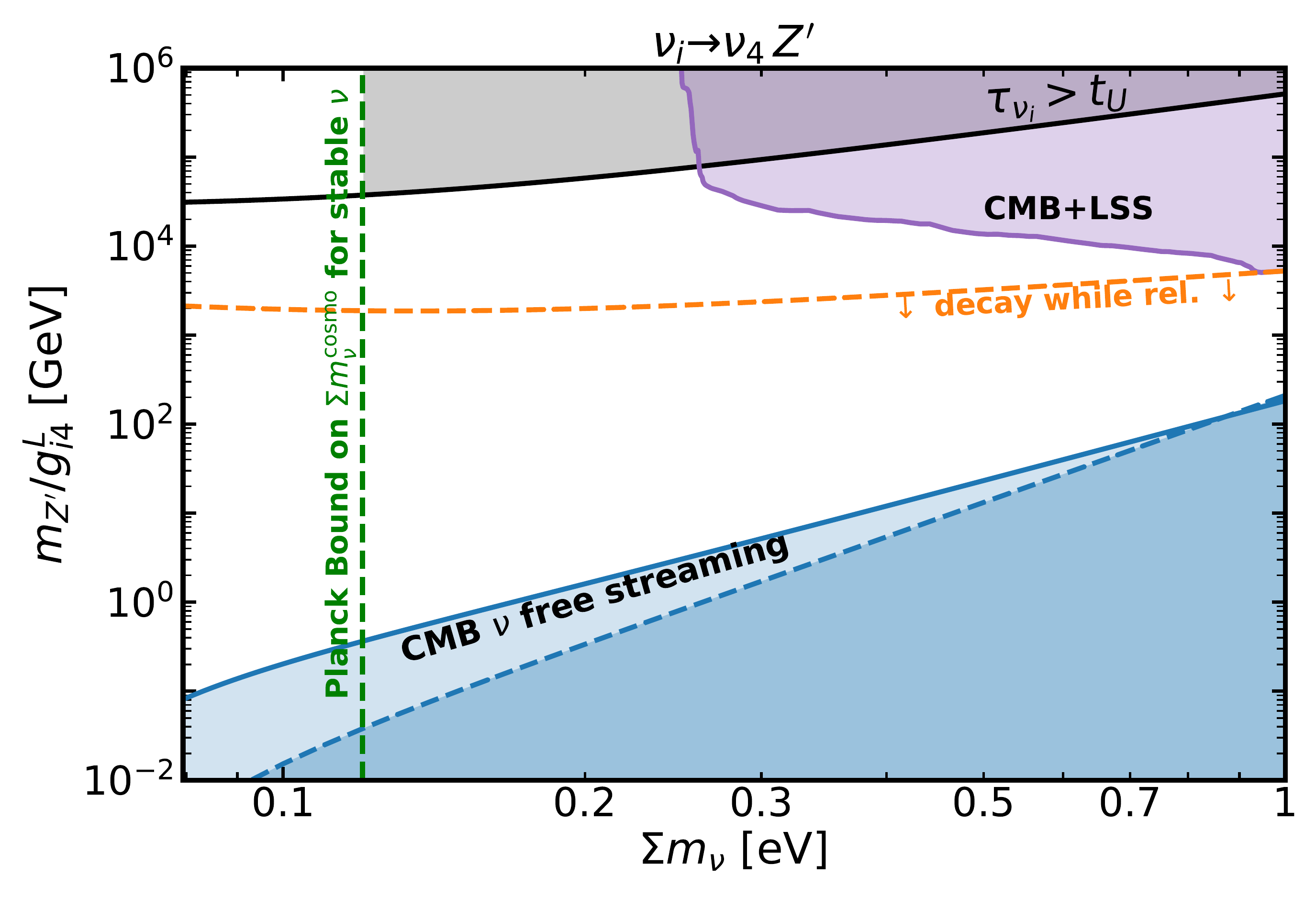}    
\hspace{-0.5cm}
 \end{tabular}
\vspace{-0.4cm}
\caption{Same as Figure~\ref{fig:lifetimes} (\textit{upper panels}) and Figure~\ref{fig:2scase} (\textit{lower panels}) but including the results from~\cite{Barenboim:2020vrr} (dashed blue lines), to be compared with the bounds presented in our analysis and derived from~\cite{Escudero:2019gfk} (solid blue lines).
}
\label{fig:appnew}
\end{figure}

Ref.~\cite{Barenboim:2020vrr} represents a substantial improvement in our understanding of the cosmological implications of neutrino decays and is the first to write down the full inhomogeneous Boltzmann equation for neutrino decays in the early Universe. Armed with this machinery, the authors argue that the bound on relativistically decaying neutrinos from CMB observations considered in our analysis (see Section~\ref{sec:Results}) $\tau_\nu > 1.3 \times  10^{9}\,\text{s}\,(m_\nu/0.05\,\text{eV})^3$~\cite{Escudero:2019gfk} is relaxed by up to $3$ orders of magnitude to $\tau_\nu > (0.4-4)\times 10^{6}\,\text{s}\,(m_\nu/0.05\,\text{eV})^5$~\cite{Barenboim:2020vrr}. In spite of this relaxation, using this new bound in our study does not significantly modify our results and conclusions. This is explicitly shown in Figure~\ref{fig:appnew} where we highlight the impact of considering the bound from Ref.~\cite{Barenboim:2020vrr} in our analysis (dashed blue lines) to be compared with our results from~\cite{Escudero:2019gfk} (solid blue lines).

In Ref.~\cite{Barenboim:2020vrr} it is also argued that the derived bound on relativistically decaying neutrinos a priori only applies for neutrino masses $m_\nu < 0.2\,\text{eV}$. However, for masses $m_{\nu} > 0.2\,\text{eV}$, decaying neutrinos will still substantially reduce the neutrino anisotropic stress energy tensor prior to recombination. Indeed, for $m_{\nu} \sim 0.5\,\text{eV}$ neutrinos become non-relativistic close to recombination which means that the actual rate that should suppress neutrino freestreaming is $\Gamma \sim \tau_\nu^{-1}$ at $T\sim m_\nu/3$. For $T\lesssim m_\nu/3$ one needs to account for the suppression arising from freeze-out inverse decays. Even though the exact constraints in this region have not been studied, we expect that the region of parameter space in which neutrinos decay at a rate $1/\tau_\nu > H(T=T_{\rm CMB})$ would be excluded by Planck data. However, for the purpose of our study we extrapolated the result of~\cite{Escudero:2019gfk} that we therefore expect to be a conservative estimate.

Finally, in Ref.~\cite{Barenboim:2020vrr} the CMB anisotropies including these new details of the neutrino decay process in the early Universe were not calculated. Thus, it remains a task to confirm the expectations from~\cite{Barenboim:2020vrr} with a full analysis of the Planck CMB data. As we discuss in Section~\ref{sec:Results} and in the previous paragraph, such analysis will finally clarify what are the precise constraints on ultrarelativistically decaying neutrinos from CMB observations.

\end{document}